\newcommand{\logg}{$\log\,g$}
 \newcommand{\teff}{$T_{\rm eff}$}

 \documentclass{aa}
 \usepackage{txfonts}
 \usepackage{graphicx}
 \begin{document}

 \title{An abundance analysis from the STIS-HST UV spectrum of the non-magnetic Bp star HR\,6000
 }

 \author{
 F.\, Castelli
 \inst{1}
 \and
 C. R. \,Cowley
 \inst{2}
\and
T.R. Ayres 
\inst{3}
\and
G.\, Catanzaro
\inst{4}
\and
F.\, Leone
\inst{4,}
\inst{5}
 }

 \offprints{F.\,Castelli}

 \institute{
 Istituto Nazionale di Astrofisica,
 Osservatorio Astronomico di Trieste, Via Tiepolo 11,
 I-34143 Trieste, Italy\\
 \email{castelli@oats.inaf.it}
  \and
 Department of Astronomy, University of Michigan, Ann Arbor, MI 48109-1042, USA\\\email{cowley@umich.edu}
\and
Center for Astrophysics and Space Astronomy, University of Colorado, Boulder, CO 80309-0389, USA\\\email{Thomas.Ayres@Colorado.EDU}
\and
Istituto Nazionale di Astrofisica,
Osservatorio Astronomico di Catania, via S. Sofia 78,I-95123 Catania,Italy\\\email{gcatanzaro@ct.astro.it}  
\and
Universita' di Catania, Dipartimento di Fisica e Astronomia, Sezione Astrofisica, Via S. Sofia 78,
I-95123 Catania, Italy\\\email{fleone@ct.astro.it}
 }

 \date{}

 \abstract
{The sharp-line spectrum of the non-magnetic, main-sequence Bp star HR\,6000
 has peculiarities that distinguish it from those of the HgMn stars with which it is sometimes
 associated. The position of the star close to the center of the Lupus\,3 molecular cloud, 
 whose estimated age is on the order of $~$9.1$\pm$2.1\, Myr,
 has lead to the hypothesis that the anomalous peculiarities of HR\,6000
 can be explained by the young age of the star.  
 }
 {Observational material from the Hubble Space Telescope
 (HST) provides the opportunity to extend the abundance analysis previously performed
 for the optical region and clarify the properties of this
 remarkable peculiar star. Our aim was to obtain the atmospheric
 abundances  for all the elements observed in a broad 
 region from 1250 to 10000\,\AA.  
 }
 {An LTE synthetic spectrum was compared with a high-resolution  spectrum 
 observed with the Space Telescope Imaging Spectrograph (STIS) equipment in the 1250-3040\,\AA\ interval.
 Abundances were changed until the
 synthetic spectrum fit the observed spectrum. The assumed model is
 an LTE, plane-parallel, line-blanketed ATLAS12 model already used for
 the abundance analysis of an high-resolution optical spectrum observed
 at ESO with the Ultraviolet and Visual Echelle Spectrograph (UVES).
 The stellar parameters are $T_{eff}=13450$K, $\log{g} = 4.3$, and zero microturbulent velocity. 
 }
 {Abundances for 28 elements and 7 upper limits were derived from
 the ultraviolet spectrum. Adding results from previous work, we have now
 quantitative results for 37 elements, some of which show striking contrasts
 with those of a broad sample of HgMn stars.
 The analysis has pointed out numerous abundance anomalies, such as ionization
 anomalies and line-to-line variation in the derived abundances,
 in particular for silicon. The inferred discrepancies could be explained
 by non-LTE effects and with the occurrence of diffusion and
 vertical abundance stratification. 
 In the framework of the last hypothesis, we obtained, by means of trial and error,
 empirical step functions of abundance versus optical depth $\log(\tau_{5000})$ 
 for carbon, nitrogen, silicon, manganese, and gold, while we failed
 to find such a function for phosphorous. The poor results for carbon,
 and mostly for phosphorus, suggest the possible importance in this star
 of NLTE effects to be investigated in future works.
 }
{}


 \maketitle{}

 \section{Introduction}

 The ASTRAL spectral library for hot stars (Ayres 2014) includes observations
 of the peculiar star HR\,6000 (HD\,144667), giving us the possibility of
 studying at high resolution the ultraviolet spectrum of this special 
 peculiar star  for the first time after the International Ultraviolet
 Explorer (IUE) era. 
 The star is anomalous because it does not fit into any of the typical 
 CP subclasses, but seems to combine abundance anomalies from several 
 Bp subtypes (Andersen et al 1984). 
 A remarkable characteristic is the extreme deficiency of silicon, which makes 
 HR\,6000 the most deficient silicon star of the HgMn group, with which
 it is usually associated.

 Peculiarities in the spectrum of HR 6000 were first noted
 by  Bessell \& Eggen (1972), who  classified  it 
 as a peculiar main-sequence B star with strong \ion{P}{ii} lines
 and weak helium (see also van den Ancker et al. 1996).
 This star is the brighter, more massive component of the visual binary 
 system $\Delta$199 or\,Dunlop 199 (Dunlop 1829). 
The second component is the Herbig Ae star
 HR\,5999 (HD\,144668) (Eggen 1975). Because the $\Delta$199 
double system is located close
 to the center of the Lupus\,3 molecular cloud, which is populated by
 numerous TTauri stars, it has been assumed that the system is the
 same age as the cloud, i.e., about (9.1$\pm$2.1)\,Myr  
 (James et al. 2006). In this case, HR\,6000 is a very young peculiar star
 and its anomalous peculiarity could be explained  by its young age. 
 The spectrum of HR\,6000 also seems to be polluted by lines of some
 cool star. The identification of an unshifted \ion{Li}{i} line at 6707.7\,\AA\
 led Castelli\& Hubrig (2007) to support the hypothesis from van den
 Ancker et al. (1996) of
  the presence of a faint TTauri companion, which could be
 either physically related to HR\,6000 to form a close binary
 system or located in the foreground of HR\,6000. 

 After recognition of its peculiar nature,
 the first high-resolution  studies of the 
 chemical composition of HR\,6000 were of an ultraviolet spectrum obtained with
  IUE in the far and near ultraviolet on March 1979 (Castelli et al. 1981;
  Castelli et al. 1985) and an optical spectrum observed at ESO with a dispersion
  of 12\,\AA\,mm$^{-1}$ covering the spectral range 
  $\lambda\lambda$ 3323$-$5316\,\AA\  (Andersen \& Jaschek 1984).

  While very significant qualitative results on the nature of the star
  were provided from the optical spectra by Andersen et al. (1984),
  the quantitative  analysis from the IUE spectra (Castelli et al. 1985) has given
  abundances for almost all the  elements observed in the IUE range
  (1258-1960\,\AA\ and 2050-3150\,\AA), but with large uncertainties.
   The quantitative analysis of the IUE spectra was 
  a very difficult task, in particular owing to the rather low resolving power 
  and  poor S/N ratio,  severe line crowding, and  
  poor quality of the line lists
  available at that time for computing synthetic 
  spectra.
  However, the underabundance
  of several elements (Be, C, N, Mg, Al, Si, S, Sc, Co, Ni, Cu, and Zn) and
  overabundance of others elements (B, P,  Ca, Mn, Fe, and Ga ) were determined. 
  For a few elements, spurious high abundances were assigned owing to 
 several unresolved blended features and unidentified components.

  Later on, in a series of papers, Smith (1993; 1994; 1997) and Smith \& Dworetsky (1993) 
  published  the
  abundances of selected elements obtained from IUE spectra of
  a sample of HgMn stars, including HR\,6000, which they noted
  did not fit the typical abundance pattern.  
   They pointed out three additional stars, 33 Gem, 36 Lyn, and 46 Aql,
   with
   abundances that depart  from the more general behavior observed in 
   the majority of the HgMn class members.
 
   A comparison of the Smith and Smith\& Dworetsky results with those from
   Castelli et al.(1985) for the elements in common ( \ion{Mg}{ii}, 
   \ion{Cr}{ii}, \ion{Mn}{ii}, \ion{Fe}{ii}, \ion{Ni}{ii}, \ion{Zn}{ii})
   has shown agreement within the error limits for all
   the elements except for \ion{Zn}{ii} whose abundance was estimated 
    $-$8.3\,dex by Castelli et al (1985) and $-$9.2\,dex by Smith (1994).
    We currently adopt an upper limit of $-8.84$ for Zn.

    Several years later, high-resolution optical spectra 
   became available to study  HR 6000 in the visual range.
   Catanzaro et al. (2004) analyzed the 3800-7000\,\AA\ region on FEROS spectra 
   observed with 48000 resolving power, while
   Castelli \& Hubrig (2007) and Castelli et al. (2009)
   studied the 3040-10000\,\AA\ region on a  
   spectrum observed at ESO with the Ultraviolet and Visual Echelle Spectrograph (UVES)
   at a resolving power ranging from 80000 to 110000. 
   The comparison of the abundances derived from the  FEROS and UVES spectra
   were in rather good agreement, with discrepancies larger than  0.3\,dex  only for
   \ion{He}{i}, \ion{Mg}{ii}, and \ion{Mn}{ii}. For iron the difference was 
   0.3\,dex. In both analyses there are only upper limits for a few elements.
   Discordant abundances from the visual and ultraviolet spectra 
   were found for \ion{O}{i}, \ion{S}{ii}, and \ion{Ti}{ii}.
   In Castelli \& Hubrig (2007)  several studies on the nature of 
   the star are quoted, included those on variability. While a
   weak photometric variability was pointed out by van den Ancker et al. (1996)
   and by Kurtz \& Marang (1995),  no spectroscopic variability was evident
   from the different spectra taken by Andersen \& Jaschek (1984)
   and  Castelli \& Hubrig (2007) at different epochs and with different
   instruments.   
   
   The Space Telescope Imaging Spectrograph (STIS) ultraviolet  spectrum of the ASTRAL library
   gives  us  the opportunity to extend the  UVES observations to the
   ultraviolet and supersede the low quality IUE data.
   The entire range from 1250 to 10000\,\AA\  is now covered by  high-resolution
   spectra of HR\,6000. The presence of elements not observed in the
optical region and elements that are present in the ultraviolet with more 
ionization states and more lines allow us to improve  the investigation
of effects related to the atomic
diffusion (Michaud 1970), in particular, vertical abundance stratification.

 The first theoretical attempt to point out the effect of abundance
stratification on line profiles is from Alecian \& Artru (1988) who
analyzed the case of the \ion{Ga}{ii} resonance line at 1414.401\,\AA\
in the atmospheres of Ap stars. Babel (1994) explained the \ion{Ca}{ii} profile
at 3933\,\AA\ observed in Ap stars with the help of a step function of the
calcium abundance 
versus  optical depth. Leone \& Lanzafame (1997) and Leone (1998)
introduced the use of the contribution functions  to associate abundances
with atmospheric layers. In addition, Leone et al. (1997) and
Catanzaro et al. (2016) extended  the search for 
vertical abundance stratification to the ultraviolet.

All the previous studies tell us that 
in stars with low rotational velocity and no convective motions, such as in HR\,6000,
vertical abundance stratification for certain elements
may be present near the surface.
Currently several studies have been performed with the 
aim of detecting vertical stratification
of the abundances. In this context, of particular interest is
the VeSE1kA project (Le Blanc et al. 2015). This paper could add some more
information related to the abundance peculiarities observed in HgMn stars.

   \section{Observations}

   HR\,6000 is one of the targets included in the Hot Stars part
   of the HST Cycle 21 ``Advanced Spectral Library (ASTRAL)''
   Project (Ayres 2014; Ayres 2014: GO-13346). The star was observed several days in
   October 2014 with moderate and high-resolution echelle settings of STIS. 
   The final spectrum covers the range 1150-3045\,\AA. The nominal resolving 
   power R ranges from about  30000 to 110000; the S/N ratio typically
   is greater than 100, approaching 200 in selected regions.
   We used the final  spectrum that resulted from the calibration and 
   merging of the overlapping echelle spectra observed in the different 
   wavelength intervals, as performed by the ASTRAL Science Team\footnote{
   http://casa.colorado.edu/$\sim$ayres/ASTRAL/}. 

   For this work we analyzed the whole region from 1250 to 3045\,\AA,
   using the IRAF tool ``continuum'' to normalize the observed spectrum to the
   continuum level. When we compared observed and computed spectra, we
   tentatively fixed different resolving powers for the different intervals
   corresponding roughly to major switches in the echelle modes.
   While the theoretical resolving powers of the 
   moderate-resolution far-ultraviolet E140M and moderate-resolution near-UV E230M
   echelle modes are 45000 and 30000, respectively (Ayres 2010), measurements of
   narrow features in the HR\,6000 spectrum suggest lower effective resolutions
   of 30000 and 25000, respectively. However, at the longer wavelengths ($>$ 2300\,\AA)
   recorded   with the NUV high-resolution echelle E230H, there appears to be little
   degradation of the theoretical resolving power of 110000. The cause of the
   resolution decrease for the medium-resolution spectra is unknown, but currently
   is under investigation.  The wavelength scale of the STIS
   spectra  are provided in  vacuum for the whole observed region.
   We converted the wavelength scale of the final spectrum
   from vacuum to air for the 2000-3040\,\AA\ interval.

   \section{The analysis}

   Abundances were determined with the synthetic spectrum method by changing
   the abundance of a given element until agreement between observed and computed profiles
   is obtained. We used the SYNTHE code (Kurucz 2005) to compute  LTE
   synthetic spectra when the adopted abundance is constant with depth and a modified
   version of SYNTHE (Kurucz 2015, private communication) when a vertical abundance
   stratification is considered for a given element.
   To be consistent with a previous abundance analysis performed on an UVES optical
   spectrum by Castelli et al. (2009), we adopted the same model atmosphere without any modification. 
   It is a plane-parallel, LTE model atmosphere  with parameters
   \teff=13450\,K, \logg=4.3, and microturbulent velocity $\xi$=0.0\,km\,s$^{-1}$,
   computed with the ATLAS12 code (Kurucz 2005) for the individual abundances
   of HR\,6000, as derived from the optical spectrum and tabulated in Castelli et al. (2009).
   In all the computations the continuous opacity includes the scattering contribution
   from electron scattering and  Rayleigh
   scattering from neutral hydrogen, neutral helium, and molecular hydrogen H$_{2}$.
   Hydrogen Rayleigh scattering, which is an important opacity source for  modeling
   Lyman$_{\alpha}$ wings,  was computed according to Gavrila (1967). The
   other scattering sources are described in Kurucz (1970).
   No scattering was considered in the line source function, which was
     approximated with the Planck function.

   The model parameters  are updated values
   from those  derived by Castelli \& Hubrig (2007) from both Balmer profiles and the
   \ion{Fe}{i} and \ion{Fe}{ii} ionization equilibrium. In Castelli et al. (2009)
   \teff\ and \logg\ were obtained from Str\"omgren photometry, from the requirement
   that there is no correlation between the \ion{Fe}{ii} abundances derived from
   high- and low-excitation lines, and from the constraint of \ion{Fe}{i}$-$\ion{Fe}{ii}
   ionization equilibrium. The only element used to fix the model parameters was  iron
   because other elements were not observed in different ionization degrees in
   the optical spectrum, except for phosphorous, which was present 
   with \ion{P}{ii} and \ion{P}{iii} lines.   Owing to the larger number of lines
   and  more trustworthy atomic data for both \ion{Fe}{i} and \ion{Fe}{ii} lines
   than for phosphorous, in particular \ion{P}{iii}, we
   preferred to use iron to phosphorous for the determination of the model parameters.
   
   As discussed in Castelli et al. (2009), the Balmer lines were
   no longer used owing to the several  problems  related
   with the UVES spectra in the regions of the Balmer lines.
   
   The computed spectrum was broadened for a rotational velocity
   v{\it sini}=1.5\,km\,s$^{-1}$ and a Gaussian instrumental profile corresponding to  
   different resolving powers for different intervals, as indicated in Sect.\,2.
   In order to superimpose the observed profiles
   on the computed profiles the observed spectrum was shifted by a velocity 
   ranging from 0.0\,km\,s$^{-1}$ at 1250\,\AA\ to 2.0\,km\,s$^{-1}$ at 3040\,\AA.

   \subsection{The line lists}

   To compute the synthetic spectrum we adopted as  main line list the 
   file gfall05jun16.dat, available on the Kurucz website on
   June 2016 (K16) (Kurucz 2016)\footnote{http://kurucz.harvard.edu/linelists/gfnew/}. 
   All the lines arising from observed levels 
   of elements from hydrogen to zinc up to six or more ionization degrees
   are listed with wavelength, $\log\,gf$ value, excitation potential of upper
   and lower levels and broadening parameters. 
   The $\log\,gf$ values are taken from the literature,
   mostly when experimental data are available,  otherwise the values are those
   computed by Kurucz with a semi-empirical method. 
   
   The list also includes lines of \ion{Fe}{ii} (Castelli\& Kurucz 2010),
   \ion{Fe}{i} (Peterson \& Kurucz 2015), and \ion{Mn}{ii}  (Castelli et al. 2015)
   observed in high-resolution stellar spectra but neither measured or classified in
   laboratory work. They arise  from energy levels predicted (approximately) 
   by the theory. 
   The method takes note of coincidences of unidentified stellar spectral 
   lines with the lines  arising from a given predicted energy level. 
   After an appropriate  
   energy shift (correction) of the levels, the lines may be considered to be
   identified, and used in spectral synthesis.
   
   For elements heavier than zinc the 
   Kurucz line lists contain lines taken from  the literature. 
   They are mostly lines of neutral and first ionized atoms. 

   We changed and implemented the line data in the Kurucz line list.  
   The big file gfall05jun16.dat was divided into 
   several smaller files that we  then modified.  The main changes were made on 
   \ion{Cr}{ii}, \ion{Cr}{iii}, \ion{Mn}{ii}, \ion{Mn}{iii}, and \ion{Fe}{ii}.
   For numerous lines of \ion{Cr}{ii} and \ion{Fe}{ii},  
   the Kurucz $\log\,gf$ values were replaced by  those given in the
   \ion{Fe}{ii} and \ion{Cr}{ii} line lists from Raassen \& Uylings (1998), 
   which were available some time ago on a website cited by  the authors, but it is 
   no longer online.
   We changed a $\log\,gf$ value from Kurucz only when that from Raassen \& Uylings (1998)
   improved the agreement between the  observed and computed spectra.
   For a few \ion{Cr}{iii} lines and several \ion{Mn}{ii}, \ion{Mn}{iii},
   and lines we lowered the Kurucz $\log\,gf$ value when 
   it gave rise to a very strong computed, but  
   unobserved line.   We checked these changes by comparing synthetic spectra of 
   the HgMn star HD\,175640 and of the normal star $\iota$\,Her
   with their STIS spectra.

   For other elements we modified the Kurucz line lists by 
   using data taken from the
   atomic database (version 5)\footnote{https://www.nist.gov/PhysRefData/ASD/lines-from.html}
   (Kramida et al. 2015) of the National Institute of Standards and Technology (NIST),  or
   from more recent literature.
   We both modified several $\log\,gf$ values and added lines for \ion{Ga}{i},
   \ion{Ga}{ii}, \ion{Ge}{ii}, \ion{Zr}{ii}, \ion{Ru}{ii}, \ion{Rh}{ii},
   \ion{Pd}{ii}, \ion{Ag}{ii}, \ion{Cd}{ii}, \ion{Sn}{ii}, \ion{Os}{ii},
   \ion{Ir}{ii}, \ion{Hg}{i}, and \ion{Hg}{ii}. For \ion{In}{ii} we only 
   replaced by the NIST values the atomic data of the Kurucz line list 
   for the 1250-3000\,\AA\ region.
   We added lines of \ion{Ga}{iii}, \ion{As}{ii}, \ion{Y}{iii},
   \ion{Zr}{iii}, \ion{Xe}{i}, \ion{Xe}{ii}, \ion{Yb}{iii}, \ion{Pt}{ii},
   \ion{Pt}{iii}, \ion{Au}{ii}, \ion{Au}{iii}, \ion{Hg}{iii}, and \ion{Tl}{ii}.
   At the moment these ions are left out  of the Kurucz line list. 
   References for the modified or added $\log\,gf$ sources of the ions  
   relevant for this paper are given in Table\,1.
   The complete modified list may be obtained at the
   Castelli's website\footnote{http://wwwuser.oats.inaf.it/castelli/linelists.html}.

\begin{table}[]
 \caption[ ]{$\log\,gf$ sources for the heavy elements relevant for this paper.}
 \font\grande=cmr7
 \grande
 \begin{flushleft}
 \begin{tabular}{llllrrllllllll}
 \hline\noalign{\smallskip}
 \multicolumn{1}{c}{Elem}&
 \multicolumn{1}{l}{$\log\,gf$ sources$^{a}$}\\
 \hline\noalign{\smallskip}
 \ion{Ga}{i} & NIST5 (Shirai et al. 2007), K16\\
 \ion{Ga}{ii}& NIST5 (Shirai et al. 2007), K16, Ryabchikova\\
             & \& Smirnov (1994), Nielssen et al. (2005), HD\,175640\\ 
 \ion{Ga}{iii} & NIST5 (Shirai et al. 2007), guessed$^{b}$, HD\,175640\\
 \ion{Ge}{ii}  & NIST5 (Wiese \& Martin 1980),  Morton 2000; K16\\ 
 \ion{As}{ii}  & Biemont et al. 1998\\
 \ion{Y}{iii}  & Biemont et al. 2011\\
 \ion{Zr}{ii}  & Ljung et al. 2006, K16\\
 \ion{Zr}{iii} & NIST5 (Reader \& Acquista 1997)\\
 \ion{Cd}{ii}  & NIST5 (Wiese \& Martin 1980) , K16, guessed$^{b}$\\
 \ion{In}{ii}  & NIST5 (Curtis et al. (2000); Jonsson and Andersson (2007);\\
               & Ansbacher et al. (1986) (see the comment in NIST5)\\
 \ion{Sn}{ii}  & NIST5 (Oliver \& Hibbert 2010; Haris et al. 2014;\\
               & Alonso-Medina et al. 2000) \\
 \ion{Xe}{i}   & NIST5 (Fuhr \& Wiese 2005)\\
 \ion{Xe}{ii}  & Yuce et al. 2011\\ 
 \ion{Au}{ii}  & Fivet et al. 2006; Rosberg et al. (1997), Biemont et al. (2007)\\ 
 \ion{Au}{iii}  & Enzonga Yoca et al. (2008), Wyart et al. (1996)\\ 
 \ion{Hg}{ii}  & NIST5 (Sansonetti \& Reader 2001) , Proffitt et al. (1999)\\ 
 \ion{Hg}{iii}  & Proffitt et al. (1999)\\ 
 \hline \noalign{\smallskip}
 \end{tabular}
 \end{flushleft}
 $^{a}$: K16: Kurucz(2016); NIST5 is the version 5 of the NIST database (Kramida et al. 2015).
 In parenthesis are given the original sources quoted in the NIST database.\\
 $^{b}$: guessed $\log\,gf$ values are estimated values on the basis of\\
 laboratory intensities and excitation energies.\\
\end{table}

\subsection{The abundances}

A short description of the elements observed in the ultraviolet spectrum
of HR\,6000 together with a list of the analyzed lines with the atomic data and
the corresponding abundances is given in Appendix\,A. 
The average abundance for each element 
is given in col.\,6 of Table\,2, which also displays in successive columns 
the abundances for HR\,6000 determined from spectra
   observed with different instruments in different ranges and by different researchers,
   as described in the Introduction.   Column 2  of Table\,2 lists the abundances
   derived from the IUE spectra by Castelli et al. (1985) (CCHM85)  and col.\,3
   those obtained from IUE spectra  by Smith\& Dworetsky (1993) (SD) and
   Smith (1993; 1994; 1997) (S). 
   Columns 4 and 5 list the abundances derived from
   the optical spectra by Catanzaro et al. (2004) (CLD04) and by Castelli et
   al. (2009) (CKH).  In column\,6 the abundances obtained in this paper from the STIS spectrum
   are given (CCACL16).
   The final abundances from both UVES and STIS spectra are collected in col.\,7.
   Columns 8 and 9 list the solar abundances from Asplund et al. (2009), 
   Scott et al. (2015a),   Scott et al. (2015b), and 
   Grevesse et al. (2015) and their differences with the
   HR\,6000 abundances. 
   Advancing from column\,2 to column\,7 one can note the increasing number
   of elements for which abundances became available with time and  
   the decreasing number of upper limits.

   \begin{table*}[]
   \caption[ ]{Abundances $\log(N_{elem}/N_{tot})$ in HR 6000 from different studies.}
   \font\grande=cmr7
   \grande
   \begin{flushleft}
   \begin{tabular}{llllll|llllllll}
   \hline\noalign{\smallskip}
   \multicolumn{1}{c}{Elem}&
   \multicolumn{1}{c}{CCHM85}&
   \multicolumn{1}{c}{S,SD}&
   \multicolumn{1}{c}{CLD04}&
   \multicolumn{1}{c}{CKH}&
   \multicolumn{1}{c}{CCACL16}&
   \multicolumn{1}{c}{Final}&
   \multicolumn{1}{l}{Sun}&
   \multicolumn{1}{l}{HR-Sun}\\
   \multicolumn{1}{c}{}&
   \multicolumn{1}{c}{IUE}&
   \multicolumn{1}{c}{IUE}&
   \multicolumn{1}{c}{FEROS}&
   \multicolumn{1}{c}{UVES}&
   \multicolumn{1}{c}{STIS}&
   \multicolumn{1}{c}{}&
   \multicolumn{1}{l}{}\\
   \hline\noalign{\smallskip}
    \ion{He}{i}$^{a}$  &              &        &  $-$2.50$\pm$0.10      & $-$2.11         &                                  & $-$2.11 & $-$1.11 & $-$1.0\\
    \ion{Be}{ii}$^{a}$ & $\le$ $-$9.5 &        &                        & $-$9.78          &                                  & $-$9.78  & $-$10.66 & $+$0.88\\
    \ion{B}{ii}$^{a}$  & $-$10.1      &        &                        &                 & $-$10.1                          & $-$10.1 & $-$9.34 & $-$0.76 \\
    \ion{C}{i}   & $\ge$$-$5.4  &        &                        &                 & $-$6.1/$-$4.9                    & $-$5.52$\pm$0.30 & $-$3.61 & $-$1.91\\
    \ion{C}{ii}$^{a}$  & $\ge$$-$5.4  &        & $\le$ $-$5.60$\pm$0.10 & $-$5.50         & no fit                           & $-$5.50          &         & $-$1.89\\
    \ion{N}{i}   & $\ge$$-$6.5: &        &                        & $\le$$-$5.82    & $-$6.8/ $-$5.6                   & $-$6.01$\pm$0.47 & $-$4.21 & $-$1.80\\
    \ion{N}{ii}$^{a}$  &              &        &                        &                 & $-$5.2                           & $-$5.23$\pm$0.04 &         & $-$1.02\\
    \ion{O}{i}   & $-$3.3       &         &  $-$3.80 $\pm$0.10    & $-$3.68$\pm$0.4 & $-$3.68                          & $-$3.68$\pm$0.04 & $-$3.35 & $-$0.33\\
    \ion{Ne}{i}$^{a}$ &              &         &                        & $-$5.34    &                                 &$-$5.34      & $-$4.11 & $-$1.23 \\          
    \ion{Na}{i}  &              &        &  $\le$$-$4.64$\pm$0.15 &$\le$ $-$5.71    &                                  & $\le$$-$5.71     & $-$5.83 & $\le$$+$0.12\\
    \ion{Mg}{ii} & $-$5.6 & -5.3 $\pm$ 0.13 &  $-$6.00$\pm$0.10   & $-$5.66         & $-$5.42$\pm$0.05                    & $-$5.45$\pm$0.09 & $-$4.45 & $-$1.00 \\
    \ion{Al}{ii}$^{a}$ & $-$7.7       &        &  $\le$$-$6.75$\pm$0.11 & $\le$ $-$7.30   & $-$7.8 $\pm$0.3                  & $-$7.8$\pm$0.3   & $-$5.61 & $-$2.19\\
    \ion{Al}{iii}& $-$7.7       &        &                        &                 & $-$7.7                           & $-$7.7           &         & $-$2.09\\
    \ion{Si}{ii} & $\le$$-$5.7  &        &  $\le$$-$7.15$\pm$0.37 & $-$7.33$\pm$0.26  & $-$8.35/ $-$6.00                  & $-$6.99$\pm$0.64 & $-$4.53 & $-$2.46\\
    \ion{Si}{iii}$^{a}$& $-$4.3,$-$5.7&        &                        &                & $-$8.35/ $-$6.45                  & $-$7.60$\pm$0.59 &         & $-$3.07\\
    \ion{Si}{iv} & $-$5.7:      &        &                        &                & $-$6.65                           & $-$6.65          &         & $-$2.12\\
    \ion{P}{i}   & $\le$ $-$4.4 &        &                        &                & $-$5.53$\pm$0.08                  & $-$5.53$\pm$0.08 & $-$6.63 & $+$1.10\\ 
    \ion{P}{ii}$^{a}$  & $-$3.0/$-$6.5 &       &  $-$4.64$\pm$0.23      & $-$4.57$\pm$0.10$^{1}$ & $-$4.54$\pm$0.11          & $-$4.55$\pm$0.10  &         & $+$2.08\\
    \ion{P}{ii}$^{a}$  &               &       &                   &      $-$4.32$\pm$0.09$^{2}$ &                           & $-$4.32$\pm$0.09  &         & $+$2.31\\
    \ion{P}{iii} & $-$5.0 &      &                               & $-$4.69$\pm$0.09  & $\le$$-$5.22$\pm$0.32      & $-$4.76$\pm$0.14  &         & $+$1.87\\
    \ion{S}{ii}$^{a}$  & $-$5.5 &      &       $\le$ $-$5.85$\pm$0.10   & $-$6.36                & $-$6.17$\pm$0.23          & $-$6.36  & $-$4.92 & $-$1.44\\
    \ion{Cl}{i}  & $-$3.3/$-$6.3   &        &                         & $\le$$-$7.74       &                           & $\le$$-$7.74  & $-$6.54 & $\le$$-$1.20 \\
    \ion{Ca}{ii} & $-$5.2        &        &                           & $-$5.68            & $-$5.70$\pm$0.09          & $-$5.70$\pm$0.09  & $-$5.72 & $-$0.02\\
    \ion{Sc}{ii} & $\le$ $-$11.0 &         & $\le$$-$9.15$\pm$0.10    &  $\le$$-$9.50 \\
    \ion{Sc}{iii}$^{a}$&               &         &                          &                    &$-$10.0$\pm$0.28           & $-$10.0$\pm$0.28    & $-$8.88 & $-$1.12\\
    \ion{Ti}{ii} &  $-$6.0  &        & $-$6.56$\pm$0.10               & $-$6.47$\pm$0.13   &                           & $-$6.47$\pm$0.13  & $-$7.11 & $+$0.64\\
    \ion{Ti}{iii}$^{a}$ & $-$5.8/$-$6.5  &        &                         &                    & $-$6.37$\pm$0.41          & $-$6.37$\pm$0.41  &         & $+$0.74 \\
    \ion{V}{ii} & $\le$ $-$6.7         &        &                  & $\le$$-$9.2           &$-$9.23$\pm$0.12           & $-$9.23$\pm$0.12  & $-$8.15 & $-$1.08 \\
    \ion{Cr}{ii} & $-$6.2 & -6.4$\pm$ 0.4  & $-$6.25 $\pm$0.08 &   $-$6.1$\pm$0.09&   $-$5.9$\pm$0.2                   & $-$6.10$\pm$0.25  & $-$6.42 & $+$0.32 \\
    \ion{Cr}{iii}$^{a}$ &$-$5.2/ $-$6.2 &   &                            &                 &  $-$6.22$\pm$-0.25                & $-$6.22$\pm$0.25  &         & $+$0.20\\
    \ion{Mn}{ii} & $-$5.3/$-$6.3 & -5.3 $\pm$ 0.1  & $-$5.6$\pm$0.10&   $-$5.18$\pm$0.32&  $-$5.29$\pm$0.21              & $-$5.29$\pm$0.21  & $-$6.62 & $+$1.33\\
   \ion{Mn}{iii}$^{a}$ & $-$5.3: &    &                &                   &  $-$5.79$\pm$0.36                               & $-$5.79$\pm$0.36  &         & $+$0.83\\
   \ion{Fe}{i}   &         &    & $-$3.95$\pm$0.03&  $-$3.65$\pm$0.07 &                                                & $-$3.65$\pm$0.07  & $-$4.57 & $+$0.92\\
   \ion{Fe}{ii} & $-$3.91$\pm$0.27 & -3.7$\pm$0.05  & $-$3.98$\pm$0.20&  $-$3.65$\pm$0.09 & $-$3.68$\pm$0.05           & $-$3.65$\pm$0.09  &         & $+$0.92\\
   \ion{Fe}{iii}$^{a}$ & $-$3.73$\pm$0.42 &   & $-$3.86$\pm$0.20&            & $-$3.78$\pm$0.14                       & $-$3.78$\pm$0.14  &         & $+$0.79 \\
   \ion{Co}{ii}$^{a}$ & $\ge$ $-$9.3 & $\le$$-$9.0    &               & $\le$ $-$8.42  &$\le$$-$10.12                   & $\le$$-$10.12 & $-$7.11 & $\le$$-$3.01\\
   \ion{Ni}{ii}$^{a}$ &$-$5.5/ $-$7.0 & -6.3 $\pm$0.2 & $\le$ $-$6.00 $\pm$0.10&  $-$6.24 &$-$6.24                           & $-$6.24           & $-$5.84 & $-$0.40 \\
   \ion{Ni}{iii} &$-$6.5/ $-$7.0 &               &                       &          &$-$6.64$\pm$0.14                        & $-$6.64$\pm$0.14  &         & $-$0.80 \\
   \ion{Cu}{ii}$^{a}$ & $\ge$ $-$9.3 &    -7.5       & -10.53  & $\le$ $-$7.83 & $-$10.53                                    & $-$10.53          & $-$7.86 & $-$2.67\\ 
   \ion{Zn}{ii}$^{a}$ & $-$8.3 &    -9.2 &           &         &              $\le$$-$8.84                                   & $\le$$-$8.84       & $-$7.48 & $\le$$-$1.36\\
   \ion{Ga}{ii}$^{a}$ & $-$8.4 &         &           &         &                $-$8.85                                      & $-$8.85           & $-$9.02 & $+$0.36\\
   \ion{Ga}{iii} &       &         &           &         &                $-$8.15                                      & $-$8.15           &         & $+$0.86\\
   \ion{Ge}{ii} &       &         &           &         &                $\le$$-$10.64                                 & $\le$$-$10.64     & $-$8.41 &$\le$ $-$2.23\\
   \ion{As}{ii}$^{a}$ &       &         &           &         &                $-$9.74                                       & $-$9.74           & $-$9.74 & 0.0 \\
   \ion{Sr}{ii} &        &                 &      & $\le$ $-$10.67 &                                                   & $\le$$-$10.7      & $-$9.21 & $\le$$-$1.49\\    
   \ion{Y}{ii}  &        &                 &      & $-$8.60        &                                                   & $-$8.60           & $-$9.83 & $+$1.23   \\
   \ion{Y}{iii}$^{a}$  &        &                 &      &        &$-$7.6                                                    & $-$7.6            &         & $+$2.23 \\
   \ion{Zr}{iii}$^{a}$  &        &                 &      &        &$\le$$-$10.24                                            & $\le$$-$10.24     & $-$9.45 & $\le$$-$0.79\\
   \ion{Cd}{ii}$^{a}$  &         &                &        &       &$-$7.00                                                  & $-$7.00           & $-$10.27& $+$3.27\\
   \ion{In}{ii}$^{a}$  &         &                &        &       &$-$10.24:                                                & $-$10.24:         & $-$11.24& $+$1.00\\
   \ion{Sn}{ii}  &         &                &        &       &$-$8.23$\pm$0.64                                         & $-$8.23$\pm$0.64  & $-$10.02 & $+$1.79\\
   \ion{Xe}{i} &        &         &         &        & $-$5.55$\pm$0.45                                                & $-$5.55$\pm$0.25  & $-$9.80 & $+$4.25\\
   \ion{Xe}{ii}$^{a}$ &        &         &            &$-$5.25$\pm$0.17 &                                                    & $-$5.25$\pm$0.17  &         & $+$4.55\\
   \ion{Au}{ii}$^{a}$ &        &         &                  &          & $-$7.82$\pm$0.55                                    & $-$7.82$\pm$0.55  & $-$11.13& $+$3.31\\
   \ion{Au}{iii}&        &         &                  &          & $-$8.80$\pm$0.28                                    & $-$8.80$\pm$0.28  &         & $+$2.33\\
   \ion{Hg}{ii}$^{a}$ & $--$   & $-$9.25/$-$7.25 & $\le$$-$8.00$\pm$0.10&  $-$8.20 & $-$9.60                               & $-$9.60            & $-$10.87& $+$1.27\\

   \hline
   \noalign{\smallskip}
   \end{tabular}
   \end{flushleft}
   \ion{P}{ii}$^{1}$: average  with lines with $\lambda$ within 4044\,\AA\ and 5200\,$\AA$.
   \ion{P}{ii}$^{2}$: average with lines with $\lambda$$\ge$5200\,$\AA$\\
   $^{a}$ dominant ionization state.

   \end{table*}

   In addition to the non-solar abundances for almost all the observed elements, 
   we point out two other types of abundance anomalies as follows:
\begin{enumerate} 
    \item Ionization anomalies,
   i.e., LTE abundances for some elements that are markedly inconsistent among 
   different ionization states.  
   The most significant abundance differences are pointed out in Table\,3.
   \item The considerable line-to-line variation in the derived abundances for
   \ion{C}{i}, \ion{N}{i}, \ion{Al}{ii}, \ion{Si}{ii}, \ion{Si}{iii}, \ion{Cl}{i},
   \ion{Ti}{iii}, \ion{Mn}{iii}, \ion{Sn}{ii}, and \ion{Au}{ii}. The large
   scatter in the abundances can be inferred from the large mean square 
   error associated with the
   average abundance listed in Table\,2 for the considered ion.
\end{enumerate}

   \begin{figure*}
   \centering
   \resizebox{8.0in}{!}{\rotatebox{90}{\includegraphics[0,0][460,900]
   {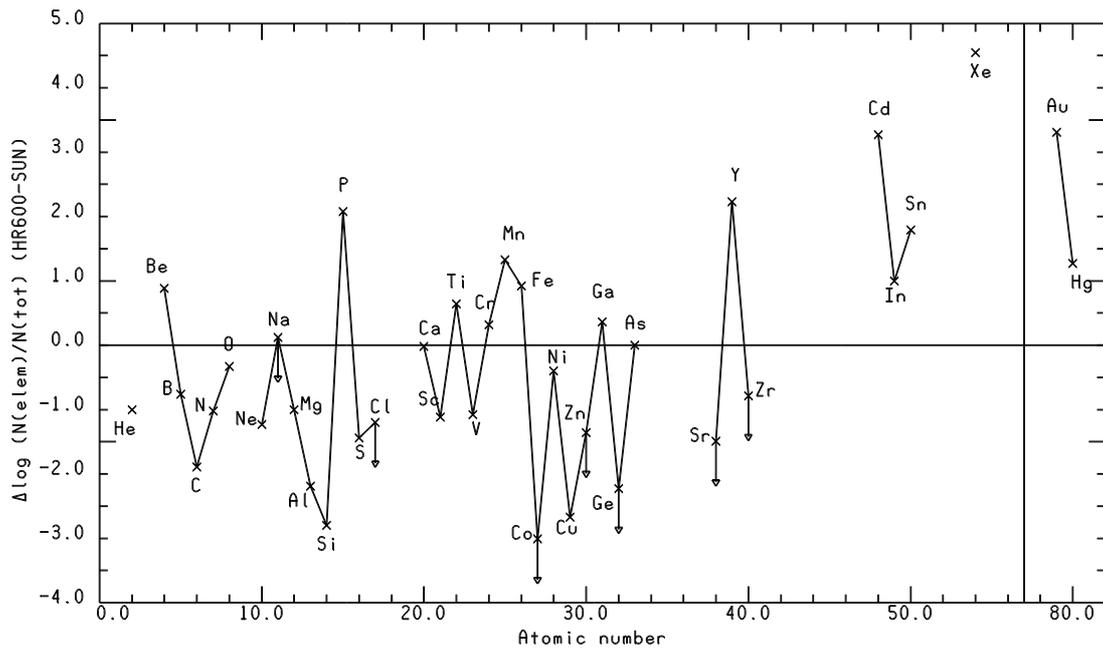}}}
   \vskip -0.75cm
   \caption{Differences between the abundance from a selected ion of a given element in HR\,6000 and the solar abundance,
     as they are listed in col.\,3 of Table\,4. The arrow for Na, Cl, Co, Zn, Ge, Sr, and Zr
     indicates an upper abundance limit.  The vertical full line 
   indicates the jump from Z=57 to Z=77, a range in atomic numbers for which no
   elements were observed in HR\,6000. The errors can be inferred from col.\,1 of Table\,4}
   \end{figure*}

   In the final spectrum synthesis calculation, it was
   necessary to adopt single abundances for elements with
   different abundances for different ionization stages. This is also needed
   for abundance comparisons with other stars such as in Table\,4.
   For the light and heavy elements we chose the 
   abundances of the dominant ionization state, except for silicon
   for which we preferred to adopt the value $-$7.35\,dex obtained from the \ion{Si}{ii}
   lines at 3862.595, 4128.074, and 4130.894\,\AA. 
   For the iron-group elements we adopted the abundances derived from the first
   ionization state, if observed, because the $\log\,gf$ values 
   available for this state are usually more critically evaluated than  
   those for the second ionization degree. 
   The difference of these abundances with respect to the solar values are listed in
   Col.\,3 of Table\,4 and are plotted in Fig.\,1. Upper limits  for Na, Cl, Co, Zn,
   Ge, Sr, and Zr are indicated with an arrow.
   The figure shows a tendency
   for the elements from He to Zr to be more or less depleted relative to the Sun, except for
   Be, P, Ti,Cr, Mn, Fe, Ga, and Y. Instead, the observed heavy elements with atomic number
   Z $\ge$ 50 are overabundant, in particular Cd, Xe, and  Au, which show enhancements from 
   an order of magnitude of 3 to 4.

   The adopted abundances were used to compare the
   chemical composition of HR\,6000 with that of other HgMn stars.
   Ghazaryan \& Alecian (2016) presented a compilation of the abundances 
   taken from the most recent literature for a sample of 104 HgMn stars.
   Table\,4 compares the over- and underabundances of HR\,6000 with the minimum and maximum
   abundances of a given element from the Ghazaryan \& Alecian (2016)
   compilation. 
   The table shows that for HR\,6000 the most impressive deviation from the abundance 
   peculiarities observed in the other HgMn stars is the underabundance of silicon, 
   followed by the underabundances of cobalt, copper, strontium, and zirconium. 
   We note that the silicon underabundance $-$1.975\,dex and $-$2.065\,dex for the 
   two stars of the HgMn sample from Ghazaryan \& Alecian (2016), HD\,158704 and HD\,35548, 
   are based on 
   the wrong $\log\,gf$ values $-$0.360 and $-$1.30 for the two \ion{Si}{ii} lines at
   4028.465\.\AA\ and 4035.278\,\AA, respectively (Hubrig et al. 1999). The updated $\log\,gf$ values
   $-$2.322 and $-$3.238 (Kurucz 2016) increase the silicon underabundance to [0.0] and to [$-$0.5], respectively.
     The underabundance of carbon in HR\,6000 is at the lower limit of the carbon underabundances
   given for the sample. 
   The overabundances of phosphorous and iron are at the upper limit of the
   overabundances of these elements in the sample of HgMn stars,  while the
   overabundance of gallium is slightly below the lower limit of the 
   gallium overabundance
   in the other HgMn stars.
   The other elements are all fully included within the abundance ranges spanned by the
   HgMn stars of the sample.

   \begin{table}[]
   \caption[ ]{Elements with strong ionization abundance anomalies.}
   \font\grande=cmr7
   \grande
   \begin{flushleft}
   \begin{tabular}{llrrrcclcrr}
   \hline\noalign{\smallskip}
   \multicolumn{1}{c}{ion1}&
   \multicolumn{1}{c}{ion2}&
   \multicolumn{1}{c}{$\log\epsilon$1-$\log\epsilon$2}
   \\
   \hline\noalign{\smallskip}
   \ion{N}{i} &\ion{N}{ii} & $-$0.88    \\
   \ion{Si}{ii} & \ion{Si}{iii} & $+$0.60\\
   \ion{Si}{iii} & \ion{Si}{iv} & $-$1.00\\
   \ion{P}{i}&\ion{P}{ii} & $-$1.08    \\
   \ion{Mn}{ii}&\ion{Mn}{iii} & $+$0.50 \\
   \ion{Ga}{ii}&\ion{Ga}{iii} & $-$0.65 \\
   \ion{Y}{ii}&\ion{Y}{iii} & $-$1.00 \\
   \ion{Au}{ii}&\ion{Au}{iii} & $-$0.98 \\

   \hline
   \noalign{\smallskip}
   \end{tabular}
   \end{flushleft}
   \end{table}

 \begin{table*}[]
   \caption[ ]
    {Adopted abundances compared with solar values, as well as with 
     extreme (maximum and minimum) values
    for stars from the Ghazaryan \& Alecian (2016) compilation.}

   \font\grande=cmr7
   \grande
   \begin{flushleft}
   \begin{tabular}{lllllllcllllll}
   \hline\noalign{\smallskip}
   \multicolumn{1}{c}{element}&
   \multicolumn{1}{c}{HR\,6000}&
   \multicolumn{1}{c}{HR\,6000-Sun}&
   \multicolumn{1}{c}{lower limit}&
   \multicolumn{1}{c}{Star}&
   \multicolumn{1}{c}{upper limit}&
   \multicolumn{1}{c}{Star}&
   \multicolumn{1}{l}{number of stars}\\
   \hline\noalign{\smallskip}
   \ion{He}{i} & $-$2.11& $-$1.00 & $-$1.55$\pm$0.13 & HR2676     & $+$0.79$\pm$0.12 & Platais 1 No1 & 64\\
   \ion{Be}{ii}& $-$9.78& $+$0.88 & $-$0.135         & HD71066    & $+$0.015         & HD\,175640 & 2\\ 
   \ion{B}{ii} & $-$10.1& $-$0.76 & $-$2.48          & $\chi$Lupi & $-$2.48          & $\chi$Lupi & 1\\
   \ion{C}{ii} & $-$5.50& $-$1.89 & $-$1.79          & 112 Her A  & $+$0.47$\pm$0.05 & HR\,8118   & 49\\ 
   \ion{N}{ii} & $-$5.23$\pm$0.04& $-$1.02 & $-$2.265$\pm$0.35& $\nu$Her   & $+$0.47          & 21\,Aql    & 14\\
   \ion{O}{i}  & $-$3.68$\pm$0.04& $-$0.33 & $-$0.4           & HR\,8118; 46\,Dra\,A& $+$0.72 & HR\,8349   & 35\\           
   \ion{Ne}{i} & $-$5.34& $-$1.23 &  $-$1.4     & $\nu$\,Her & $+$0.69          & $\kappa$Cnc\,A & 27\\       
   \ion{Na}{i} & $\le$$-$5.71 &$\le$$+$0.12 & $+$0.285$\pm$0.09 & HD\,71066 & $+$0.865    & HR\,7775   & 6\\       
   \ion{Mg}{ii}& $-$5.45$\pm$0.09 &$-$1.00 & $-$1.76  & HD\,55362  & $+$0.30$\pm$0.25 & HR\,2844   & 67\\
   \ion{Al}{ii}& $-$7.8$\pm$0.3& $-$2.19  & $-$2.75$\pm$0.37& 112\,Her\,A & $+$0.39  & HD\,173673 & 43\\
   \ion{Si}{ii}& $-$7.33$\pm$0.26 & $-$2.80 & $-$1.49$\pm$0.03 & HD\,55362& $+$0.355 & HR\,7775 & 71\\
   \ion{P}{ii} &  $-$4.55$\pm$0.10& $+$2.08 &   $+$0.14 & 53\,Tau & $+$2.235 & 74\,Aqr\,A & 38\\
   \ion{S}{ii} & $-$6.36 & $-$1.44  & $-$1.94$\pm$0.51 & HD\,55362 & $+$0.47$\pm$0.25 & HR\,7018& 51\\
   \ion{Cl}{i} & $\le$$-$7.74 & $\le$ $-$1.20 & $-$1.40     &$\chi$Lupi\,A &  $-$0.36   &HD\,46866 &2\\
   \ion{Ca}{ii}& $-$5.70$\pm$0.09 & $-$0.02 & $-$0.805$\pm$0.21& HD\,71066 & $+$1.215 &HD\,158704 &53\\
   \ion{Sc}{iii}&$-$10$\pm$0.28 & $-$1.12 & $-$1.90 &$\phi$Phe & $+$1.83 &HR\,8118 &37\\
   \ion{Ti}{ii}& $-$6.47$\pm$0.13 & $+$0.64 & $-$0.34$\pm$0.15 & 21\,Aql & $+$1.465 &1\,Cen &60\\
   \ion{V}{ii}& $-$9.23$\pm$0.12& $-$1.08 & $-$1.895 & HD\,71066 & $+$1.155 &$\mu$\,Lep &14\\
   \ion{Cr}{ii}& $-$6.10$\pm$0.25& $+$0.32 & $-$1.44$\pm$0.2 & 46\,Aql & $+$1.20$\pm$0.26 &$\phi$\,Her\,A &86\\
   \ion{Mn}{ii}&$-$5.29$\pm$0.21& $+$1.33 & $-$0.23$\pm$0.16 & 36\,Lyn & $+$2.87$\pm$0.40 &BD\,0984 &67\\
   \ion{Fe}{ii}&$-$3.65$\pm$0.09& $+$0.92 &$-$1.10$\pm$0.20  & 87\,Psc & $+$0.92          & 112\,Her\,A & 84\\
   \ion{Co}{ii}&$\le$$-$10.12& $\le$$-$3.01     &$-$2.49 & several           & $+$2.945 &$\iota$CrB\,A& 29   \\
   \ion{Ni}{ii}&$-$6.24 & $-$0.40          & $-$2.085 & HD\,71066       & $+$0.66$\pm$0.13 & HD\,49886 & 49   \\
   \ion{Cu}{ii}&$-$10.53& $-$2.67          & $-$0.89$\pm$0.36 & 12\,Her\,A & $+$2.21$\pm$0.15 & HR\,7361 & 28  \\
   \ion{Zn}{ii}&$\le$$-$8.84& $\le$ $-$1.36     & $-$2.56 & $\phi$Phe            &$+$1.86 & $\phi$Her\,A & 30     \\
   \ion{Ga}{ii}&$-$8.66& $+$0.36     & $+$0.81$\pm$0.30 & 46\,Aql & $+$4.21 & $\alpha$And &  35    \\
   \ion{Ge}{ii}&$\le$$-$10.64& $\le$$-$2.23 & $-$1.65 & $\chi$Lupi\,A & $-$1.65 &$\chi$Lupi\,A &  1    \\
   \ion{As}{ii}&$-$9.74&  0.00 & $+$2.33 & $\chi$Lupi\,A & $+$3.365 & HD\,71066 &  2    \\
   \ion{Sr}{ii}&$\le$$-$10.7& $\le$$-$1.49 & $-$1.17 & 112\,Her\,A & $+$2.940 & HR\,7775 &  44    \\
   \ion{Y}{iii}&$-$7.60& $+$2.23 & $+$0.67 & HR\,2676 & $+$4.3  & HD\,2844 &  53    \\
   \ion{Zr}{iii}& $\le$$-$10.24& $\le$ $-$0.79 & $+$4.95$\pm$0.15 & HR\,7775 & $+$2.85  & AV\,Scl &  34    \\
   \ion{Cd}{ii}& $-$7.00& $+$3.27 & $+$0.56 & $\chi$Lupi\,A & $+$0.56  & $\chi$Lupi\,A &  1    \\
   \ion{In}{ii}& $-$10.24& $+$1.00 & $--$ & $--$ & $--$  & $--$ &  0    \\
   \ion{Sn}{ii}& $-$8.23$\pm$0.64& $+$1.79 & $+$1.38 & $\chi$Lupi\,A & $+$1.38  & $\chi$Lupi\,A &  1    \\
   \ion{Xe}{ii}& $-$5.25$\pm$0.17& $+$4.55 & $+$2.98$\pm$0.11 & $\phi$Her\,A & $+$4.89$\pm$0.14  & $\kappa$Cnc\,A & 29    \\
   \ion{Au}{ii}& $-$7.82$\pm$0.55&  $+$3.31 & $+$3.5  & 46\,Dra\,A & $+$5.74  & 66\,Eri\,B & 11    \\
   \ion{Hg}{ii}& $-$9.60&  $+$1.27  & $+$1.06  & 53\,Tau & $+$6.61$\pm$0.20  &USNO-A2.0\,0825-03036752 & 82    \\
   \hline
   \noalign{\smallskip}
   \end{tabular}
   \end{flushleft}
   \end{table*}

   \section {NLTE and stratification}

   The classical LTE abundance analysis of the ultraviolet spectrum of HR\,6000
   described in the previous sections has
   pointed out the presence of numerous abundance anomalies and inhomogeneities.
   We can therefore state {\it a posteriori} that a more refined theory is needed
   to predict a spectrum that fits the observations better than the spectrum we computed
   fits. Possible explanations for 
   the observed abundance anomalies can be  NLTE-effects and the occurrence of
   diffusion and vertical abundance stratification.

   There are few NLTE studies available for stars in the temperature range from
   10000 to 15000\,K and these studies are mostly devoted to investigate lines in the
   optical and infrared regions. In fact,  to explain the
   \ion{Si}{ii}$-$\ion{Si}{iii} anomaly observed in a set of late B-type stars,
   Baily \& Landstreet (2013) performed  a rather elaborate test to conclude
   that NLTE-effects in Si could be potentially
   important well below the conventional limit of \teff\ $=$ 15000\,K.
   However, they demonstrate that a vertical abundance stratification of
   silicon can also partly provide an explanation of the observed silicon anomaly.
   
   Hempel \& Holweger (2003) concluded from
   a NLTE abundance analysis in the optical region of 27 optically bright B5-B9 main-sequence stars
   that elemental stratification due to diffusion is a common property of these stars.

   A NLTE analysis of \ion{C}{i} lines at 1657\,\AA\
   and \ion{C}{ii} lines at 1334-1335\,\AA\ performed by Cugier \& Hardorp (1988) for main-sequence stars
   of spectral types A0 to B3, has shown  remarkable NLTE effects
   on \ion{C}{i} lines in $\pi$ Cet, a star with stellar parameters 
   close to those of HR\,6000.
   The \ion{C}{i} lines computed in NLTE are  weaker and narrower than the
   LTE lines, so that a NLTE synthetic spectrum would improve the agreement
   between observations and computations  in the case of HR6000.
   Instead, NLTE \ion{C}{ii} lines have a narrower core than the LTE lines, but
   the difference in the wings is negligible.
   Therefore NLTE effects do not explain the very broad wings
   observed in HR\,6000. More in general, Cugier \& Hardorp (1988) have
   demonstrated that the resonance lines of \ion{C}{ii} at 1335\,\AA\ are
   largely insensitive to effective temperature, microturbulence, and NLTE effects
   in late B-type stars.

   Given the difficulty of discussing NLTE-effects in HR\,6000 owing to
   the scarce sources available in the literaure for  star of this spectral type,
   and because we do not have the tools to carry out non-LTE synthesis
   for this study,  we consider here the hypothesis of vertical abundance
   stratification in more detail.  We hope that this paper encourages investigators
   to perform
   studies on the NLTE effects on ultraviolet lines in HgMn and related stars. 

   Ryabchikova et al. (2003) discuss  spectroscopic observational
   evidence for the vertical abundance stratification in stellar atmospheres.
   In HR\,6000 all these observational signs are present, and in particular, we find:
 \\
\begin{enumerate}
\item  The impossibility to fit the wings and core of strong spectral
   lines with the same abundance.
   It was observed for \ion{He}{i} in the optical region (Castelli \& Hubrig 2007),
   for \ion{C}{i} and for the strong \ion{C}{ii} and  \ion{Si}{ii} lines 
   in the ultraviolet.   
 \\
\item  Violation of LTE ionization balance.
   This occurs in HR\,6000 for \ion{N}{i},\,\ion{N}{ii}, \ion{Si}{ii},\,
   \ion{Si}{iii}, \ion{Si}{iv},
   \ion{P}{i},\, \ion{P}{ii}, \ion{Mn}{ii},\,\ion{Mn}{iii},
   \ion{Ga}{ii},\,\ion{Ga}{iii}, \ion{Y}{ii},\,\ion{Y}{iii}, and
   \ion{Au}{ii},\, \ion{Au}{iii} (Table\,3).
   \\
\item Disagreement between the abundances derived from strong and weak lines
   of the same ion. In traditional atmospheric analysis, this trend is
   eliminated by introducing microturbulence. In the present case,
   abundances are typically lower for the strongest lines. This cannot be
   changed by taking a lower microturbulence, as we have already adopted
   a microturbulence of zero. The effect is present for all the elements 
   with average
   abundances affected by a large mean square error. They are \ion{C}{i}, \ion{N}{i},
   \ion{Si}{ii}, \ion{Si}{iii}, \ion{Cl}{i}, \ion{Ti}{iii}, \ion{Mn}{iii},
   \ion{Sn}{ii}, and \ion{Au}{ii} (Table\,2).
   \\
\item An unexpected behavior of high-excitation lines of the ionized iron
   peak elements. For $\lambda$ $>$ 5800\,\AA, emission lines 
   were observed in HR\,6000 for \ion{Cr}{ii}, \ion{Mn}{ii}, and \ion{Fe}{ii}
   (Castelli \& Hubrig 2007). Sigut (2001) showed that NLTE effects
   coupled with a stratified manganese abundance  satisfactorily accounts for
   the emissions from the \ion{Mn}{ii} lines at $\lambda\lambda$ 6122-6132\,\AA\  
  observed  in several HgMn stars, HR\,6000 included.
   \\
\item  Different abundances obtained from different lines of the same ion formed at {\it a priori}
 different optical depths, for example, before and after the Balmer jump, in the UV
 and visual spectral region. In addition to the possibility of different abundances
 for  mercury, as discussed above, this 
 anomaly was observed for \ion{Mg}{ii}.
 The ultraviolet lines are consistent with the abundance of $-$5.45$\pm$0.09\,dex,
 while the doublet at 4481\,\AA\ indicates an abundance of $-$5.7\,dex.

\end{enumerate}

Stratification may manifest itself in one or more of the ways enumerated.
We call attention to cases where stratification is indicated by the profile
of a single line (Item\,1) or several lines (Items\,2,\,3,\,4, and 5).
Stratification is indicated by the line profiles in only a few cases.
The various indicators may overlap in a given element (Silicon) or 
spectrum (\ion{Si}{ii}).

We searched in more detail the presence of stratification for carbon, nitrogen,
silicon, phosphorous, manganese, and gold.
The lines of gallium have oscillator strengths that are too uncertain to be used, while
yttrium has too few lines, none of which have wings.
 
We determined  the atmospheric layers where the optical depth  
of several points of a given profile from the center to the wings
is $\tau_{\nu}$=1. 
Each profile was computed with the WIDTH code (Kurucz 2005)
for the particular abundance determined from that line using
the synthetic spectrum (Table A1). 
Among the several outputs of the WIDTH code
there is the  mass depth value ($\rho$x) of the layer where $\tau_{\nu}$=1 
for each of the considered  line profile points, as well as  the average
mass depth  of the forming region of the line, defined as

\( log_{10}(\rho x)_{av} = {{\int_{0}^{b} {log_{10}(\rho x)_{\tau\nu=1}(1-H_{\nu}(0)/H_{c}(0))d\nu}}
\over{\int_{0}^{b} {(1-H_{\nu}(0)/H_{c}(0))d\nu}}}  \)\\  
where the integral
is performed over the frequencies from center to  wings and H$_{\nu}(0)$ 
and H$_{c}(0)$ are the line and continuum 
Eddington flux at the stellar surface, respectively. For more details  see Castelli (2005).

As further step, in order to be consistent with other papers 
dealing with stratification, the mass depth scale ($\rho$x) was converted
to the continuum optical depth scale at $\lambda$=5000\,\AA\ ($\tau_{5000}$).
Finally, the abundance derived with 
the synthetic spectrum from each line was plotted as a function of $\log(\tau_{5000}$)$_{aver}$,
the average optical depth of line formation. Figs. 2a, 4a, 7a, 9, 10a, and 12a
show these plots for carbon, nitrogen, silicon, phosphorous, manganese, and gold.

This kind of approach, which  is an indication of the presence of
stratification for a given element, is similar to that used in others studies on
this subject, although usually the optical depth in the line core is considered
(Khalack et al. 2007, Khalack et al. 2008, Thiam et al. 2010) rather than 
the average optical depth adopted here.
Qualitatively, we expect that stronger lines are formed higher in the
photosphere than weaker lines. Since the current study is based on synthesis
rather than equivalent widths, we use depth $\tau_{5000}$ where the line
profile optical depth is unity as a proxy for the line strength.
 
We  carried out a series of simple experiments for some elements.
We derived, by means of trial and error, an empirical step-like
function  of abundance  versus  $\log(\tau_{5000})$ for a given element.  
The adopted  functions reduce the observed 
abundance anomalies when  used in the synthetic spectrum computations.
This basic approach is used for both anomalous line profiles and sets
of lines with inconsistent abundances from the unstratified model.
We do not claim that these functions are unique. In fact, only a more rigorous
theoretical approach including the stratification on the temperature-pressure
structure of the atmosphere could give an adequate
answer concerning the origin of the observed abundances inhomogeneities. 

Empirical step functions for the abundance of carbon, nitrogen, silicon, manganese,
and gold are given in Figs.\,2b, 4b, 7b, 10b, and 12b.  
These elements are now discussed.

\subsection{Carbon}

Several signs of possible vertical abundance stratification are present for carbon.
Numerous \ion{C}{i} lines can be observed in the ultraviolet, but most of them
have profiles that cannot be reproduced by an unique abundance.
In fact, they have a observed core that is too weak as compared to that of the computed profile 
whose wings fit the observed spectrum.   
The best fit abundance from  \ion{C}{i} individual lines  ranges from 
$-$6.1\,dex to $-$4.9\,dex. The average abundance is $-$5.52$\pm$0.3\,dex,
which is consistent with the value $-$5.50\,dex derived by Castelli et al. (2009)
from the \ion{C}{ii} lines at 4267.001\,\AA\ and 4267.261\,\AA. 
However, a  plot of the \ion{C}{i} abundance versus $\log\,\tau_{5000}$(aver) 
does not shows any  dependence of the abundance on the optical depth
(Fig.\,2a).

For \ion{C}{ii}, it is impossible to reproduce the resonance profile at 
$\lambda\lambda$ 1335.663, 1335.708\,\AA, with any abundance that is
constant with depth. Either the core is fitted  but not the wings or 
the wings are fitted but  not the core.
The same kind of anomaly affects the strong \ion{C}{ii} profile at 1323.9\,\AA.
We note that the \ion{C}{ii} line at 1334.532\,\AA\ cannot be used for the abundance analysis
because it is blended with
a strong blue component of interstellar or circumstellar origin.

\begin{figure}
\centering
\resizebox{4.50in}{!}{\rotatebox{90}{\includegraphics[0,0][300,720]
{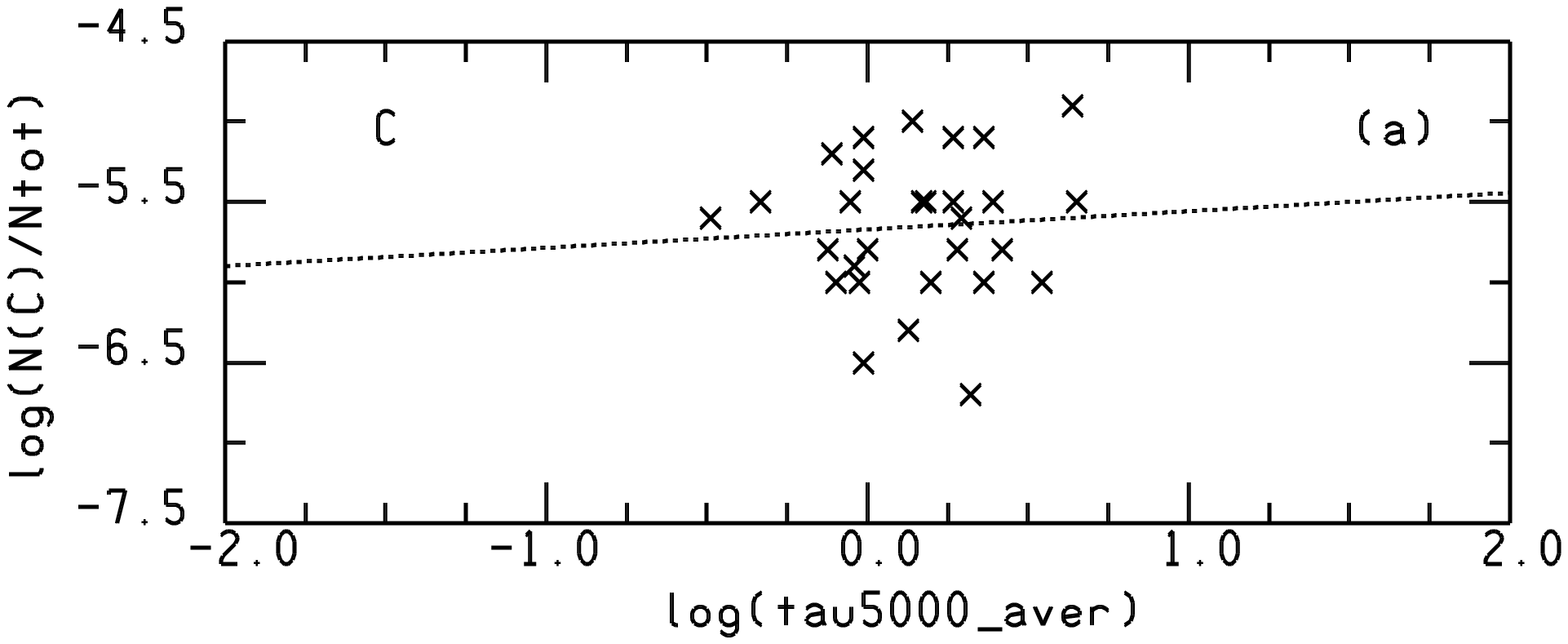}}}
\centering
\resizebox{4.50in}{!}{\rotatebox{90}{\includegraphics[0,0][350,720]
   {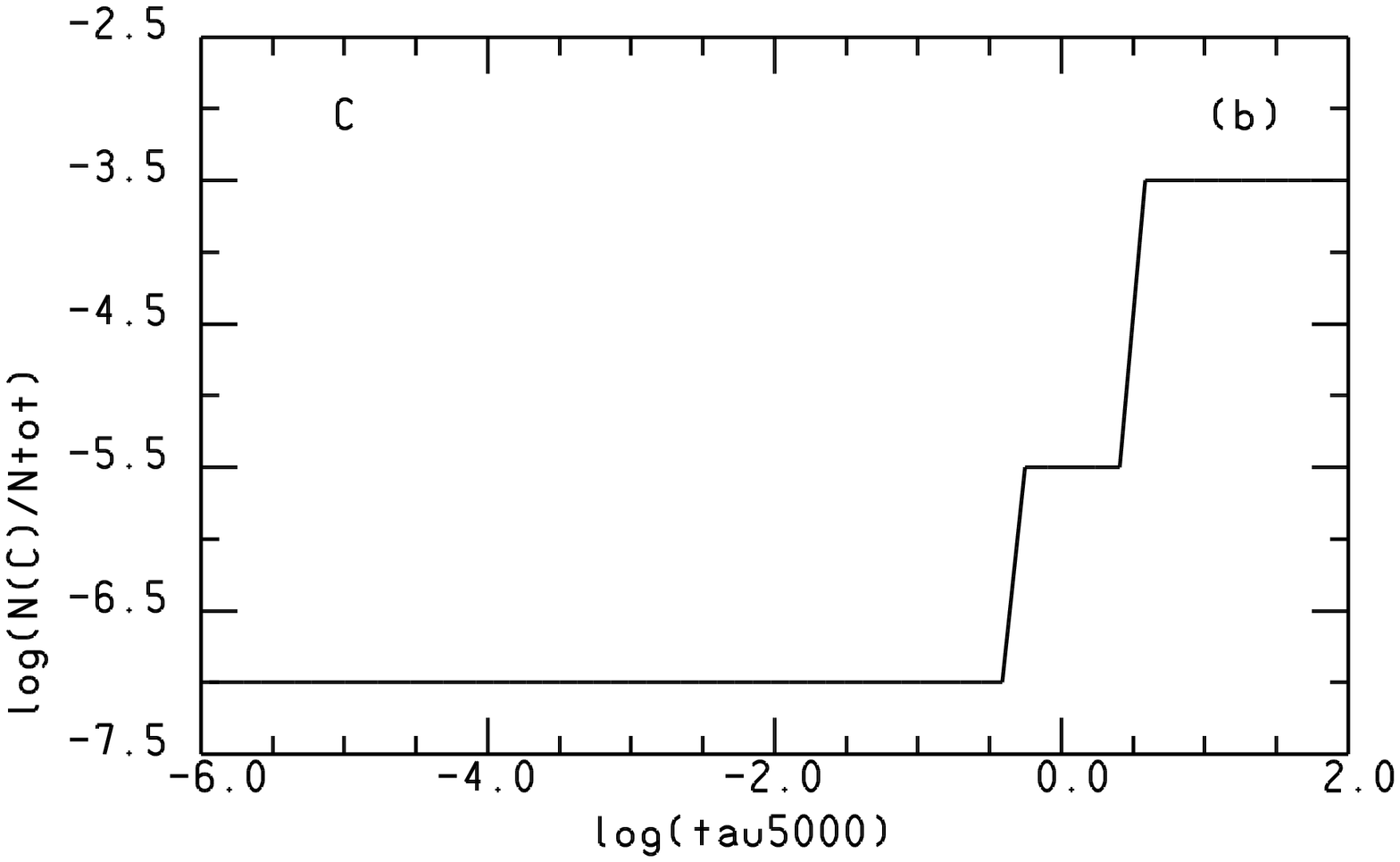}}}
\vskip -1.0cm
\caption{(a):\ion{C}{i} abundance versus $\log\tau_{5000}$(aver), average depth of 
line formation on the $\log\tau_{5000}$ scale. The dashed line is the best fit straight line to the 
plotted points. 
(b): Vertical abundance
distrbution as function of $\log\tau_{5000}$ obtained by the trial and error
applied to \ion{C}{ii} $\lambda\lambda$ 1335.663, 1335.708\,\AA.} 
\end{figure}

We were able to fit, at the same time, both the core and wings of \ion{C}{ii} at
1335.663, 1335.708\,\AA\ with the carbon abundance profile, shown in Fig.\,2b,
obtained with  trial and error.

In Fig.\,3 the \ion{C}{i} and \ion{C}{ii} lines 
computed with both the average carbon abundance  $-$5.5\,dex  and 
the  abundance step function shown in Fig.\,2b are plotted.  They are compared  with each other and
with the observed spectrum.
We point out that while the step function improves the agreement between the observed and
computed spectra, as shown in Fig.\,3, some non-negligible inconsistencies are still present. In particular,
the \ion{C}{ii} line at 1323.9\,\AA\ is still poorly reproduced.
As already discussed at the beginning of Sect.\,4, we would like to stress that a NLTE
analysis could clarify whether the \ion{C}{i} anomalies are
related with NLTE rather than a vertical abundance stratification.

\begin{figure*}
\centering
\resizebox{5.00in}{!}{\rotatebox{90}{\includegraphics[0,100][550,700]
{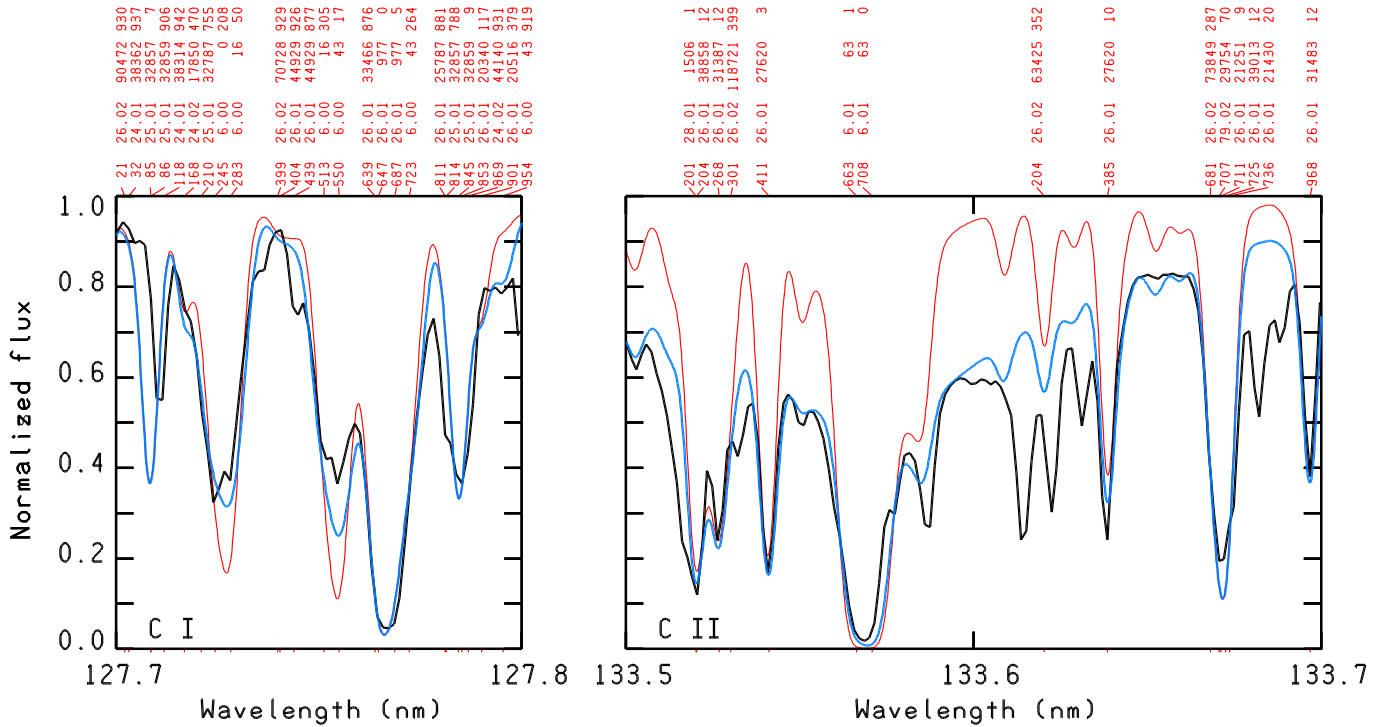}}}
\vskip -1.0cm
\caption{\ion{C}{i} lines of UV mult.\,7 ($\lambda\lambda$ 1277.245,
1277.283, 1277.513, 1277.550, 1277.273, and 1277.954\,\AA) and the
\ion{C}{ii} blend at $\lambda\lambda$ 1335.663,1335.708\,\AA\
computed with both the average carbon 
abundances of $-$5.5\,dex (red line) and  the abundance step function 
shown in Fig.\,2b (blue line) are compared each with other and with
the observed spectrum (black line). } 
\end{figure*}



\subsection{Nitrogen:} 

From the absence of \ion{N}{i} lines at 8680.282\,\AA\ and 8683.403\,\AA,
Castelli et al. (2009) determined an upper limit of $-$5.82\,dex ([$-$1.7]) 
for the nitrogen abundance. In the ultraviolet, the \ion{N}{i} lines
at 1411.9\,\AA\ (UV mul.10), 1310.5,1310.9\,\AA\ (UV mult. 13),
and 1492.625, 1492.820, and 1494.675\,\AA\ (mult.\,4) are clearly
observable, as well as the lines of \ion{N}{ii} at $\lambda\lambda$ 
1275.038, 1275.251\,\AA\ and the blend at $\lambda\lambda$ 1276.201, 1276.225\,\AA.
These lines are not all fitted by the same abundance. In fact,
the nitrogen abundance is affected by both an
ionization anomaly and line-to-line variations:
while  $-$5.2\,dex  reproduces the \ion{N}{ii} lines,
lower values,  ranging from $-$5.9\,dex to $-$6.8\,dex, 
are required to fit the \ion{N}{i} lines.
The nitrogen abundance 
versus $\log\,\tau_{5000}$(aver) 
is shown in Fig.\,4a. A steep dependence of the abundance 
on the optical depth  is evident.
The abundance inconsistencies decrease if the 
abundance profile shown in Fig.\,4b is used. 
We obtained this profile from al the \ion{N}{i} and
\ion{N}{ii} lines with the trial and error.

\begin{figure}
\centering
\resizebox{4.50in}{!}{\rotatebox{90}{\includegraphics[0,0][300,720]
{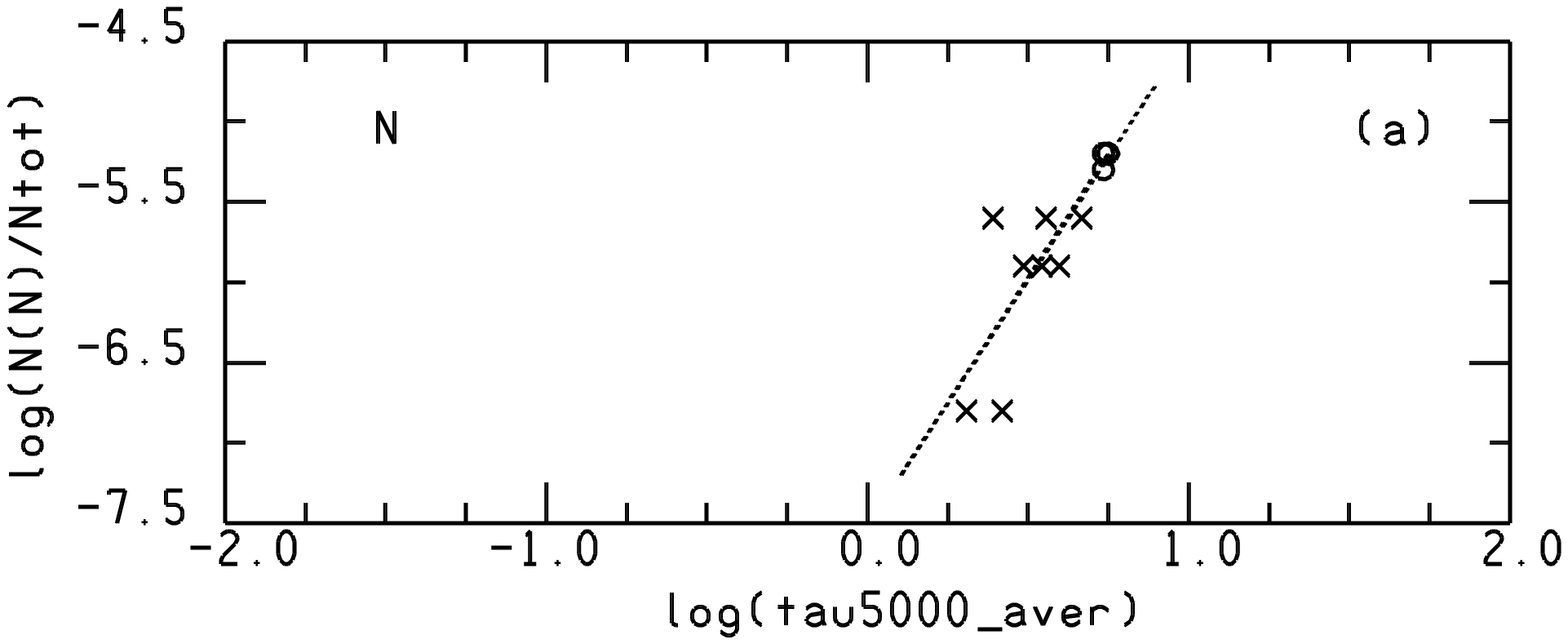}}}
\centering
\resizebox{4.50in}{!}{\rotatebox{90}{\includegraphics[0,0][350,720]
   {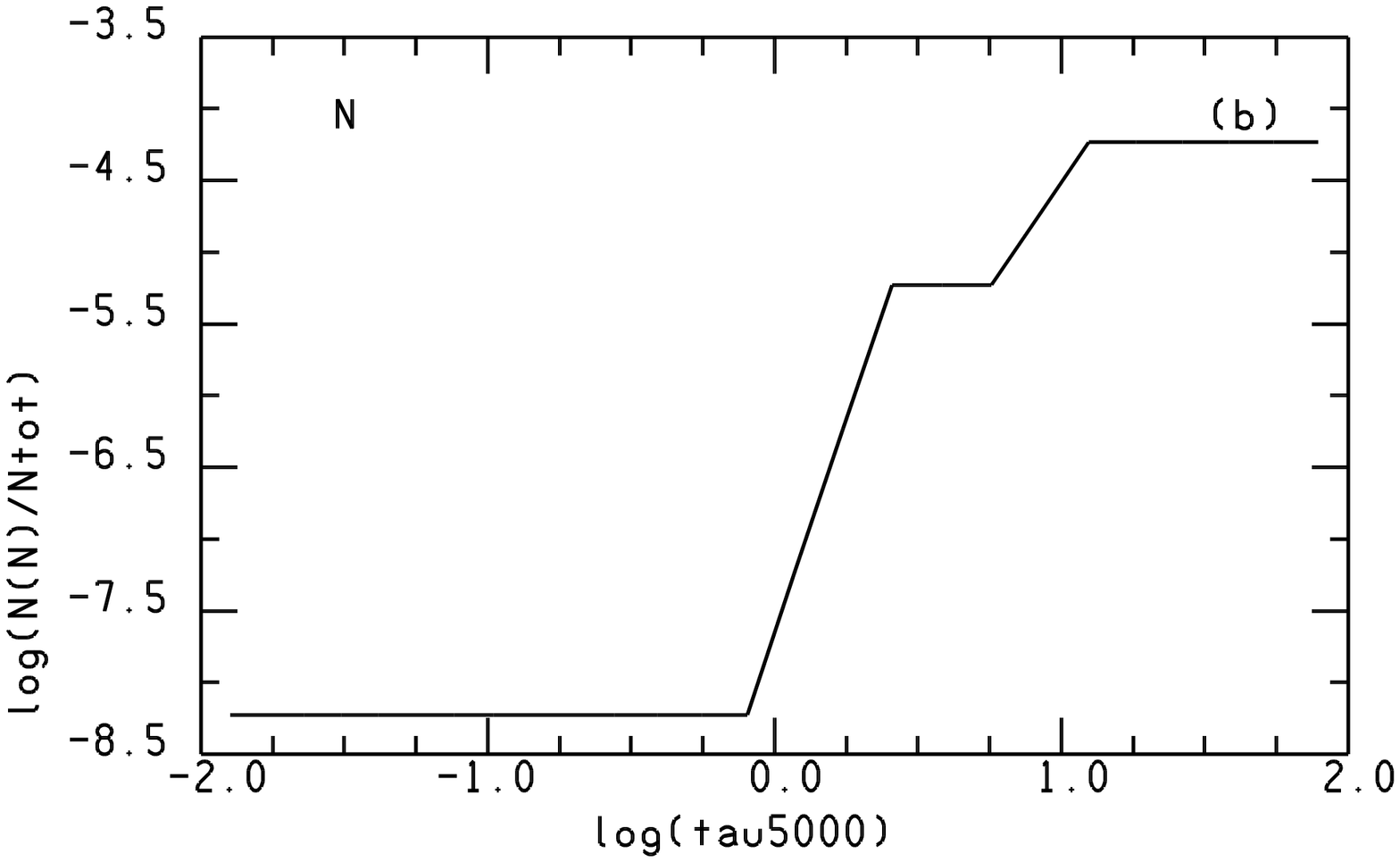}}}
\vskip -1.0cm
\caption{(a): Nitrogen abundance versus $\log\tau_{5000}$(aver);
crosses are for \ion{N}{i}, circles for \ion{N}{ii}. The dashed line is 
the best fit straight line to the plotted points. 
(b):  vertical abundance
distribution as a function of $\log\tau_{5000}$ obtained by trial and error.} 
\end{figure}

The \ion{N}{i} and \ion{N}{ii} lines
computed both with a constant  abundance  $-$5.23\,dex derived from the \ion{N}{ii} lines and 
with the  abundance step profile shown in Fig.\,4b,  are compared
with and without stratification
and with the observed spectrum in Fig.\,5 and Fig.\,6, respectively.
The effect of a depth-dependent abundance is
large for the \ion{N}{i} lines, while it does not affect
the observed \ion{N}{ii} lines.

We note that nitrogen is predominantly singly ionized in late B-star atmospheres,
so that the analysis of \ion{N}{i} could be particularly susceptible to
NLTE effects in the ionization equilibrium.  For \ion{N}{i} as well as
for \ion{C}{i}
a NLTE analysis is required to state the degree of reliability of the hypothesis
of vertical abundance stratification for this element.

\begin{figure*}
\centering
\resizebox{5.00in}{!}{\rotatebox{90}{\includegraphics[0,100][510,700]
{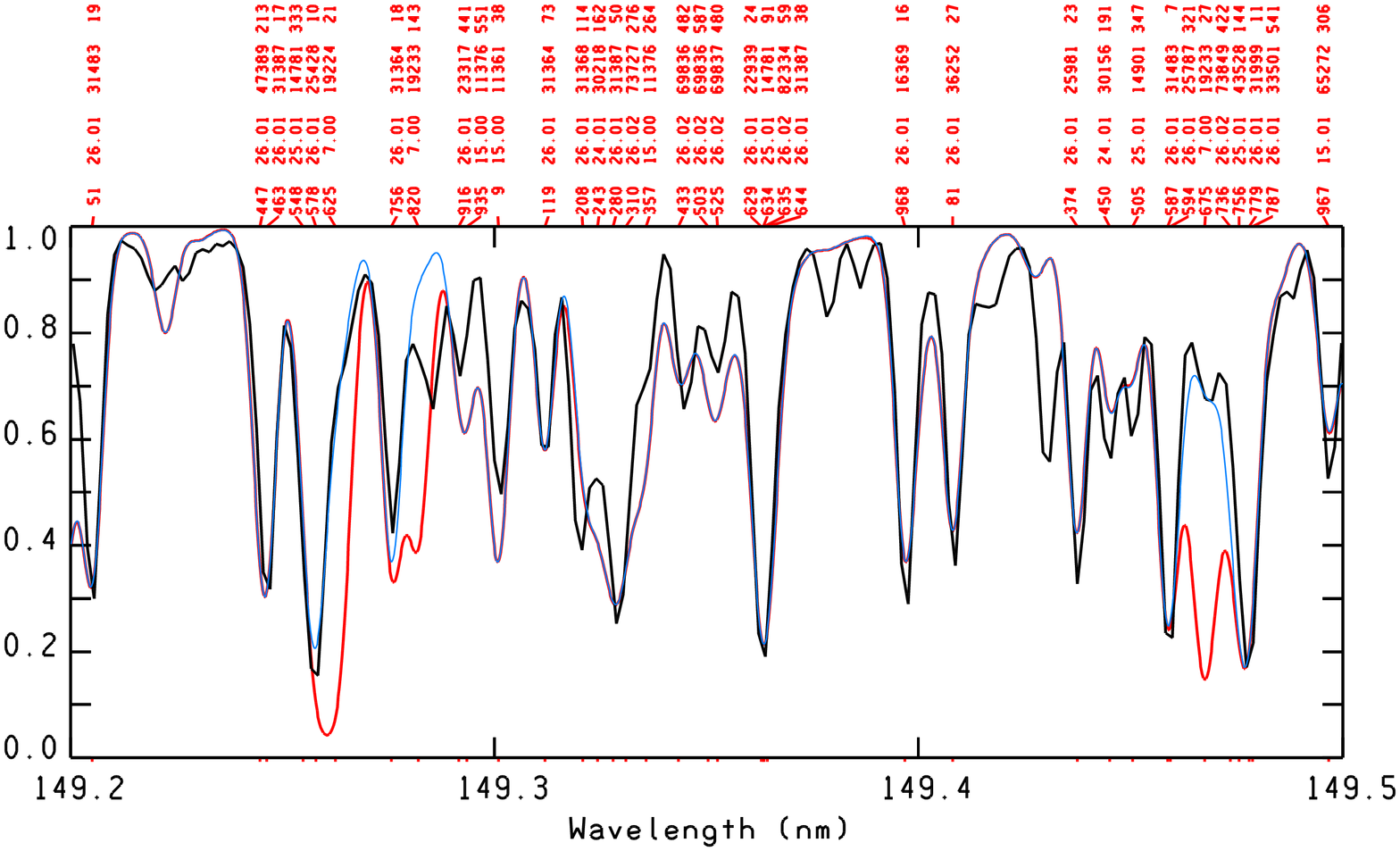}}}
\vskip -1.25cm
\caption{\ion{N}{i} lines of UV mult.\,4 at $\lambda\lambda$ 1492.625, 1492.820, and  1494.675\,\AA\
computed both with the constant nitrogen  abundance $-$5.23\,dex (red line) and with 
the abundance step function shown in Fig.\,4b (blue line). The computed spectra are superimposed to the observed spectrum.
The effect of the two different adopted abundances (constant or variable) is evident.}

\centering
\resizebox{5.00in}{!}{\rotatebox{90}{\includegraphics[0,100][525,700]
{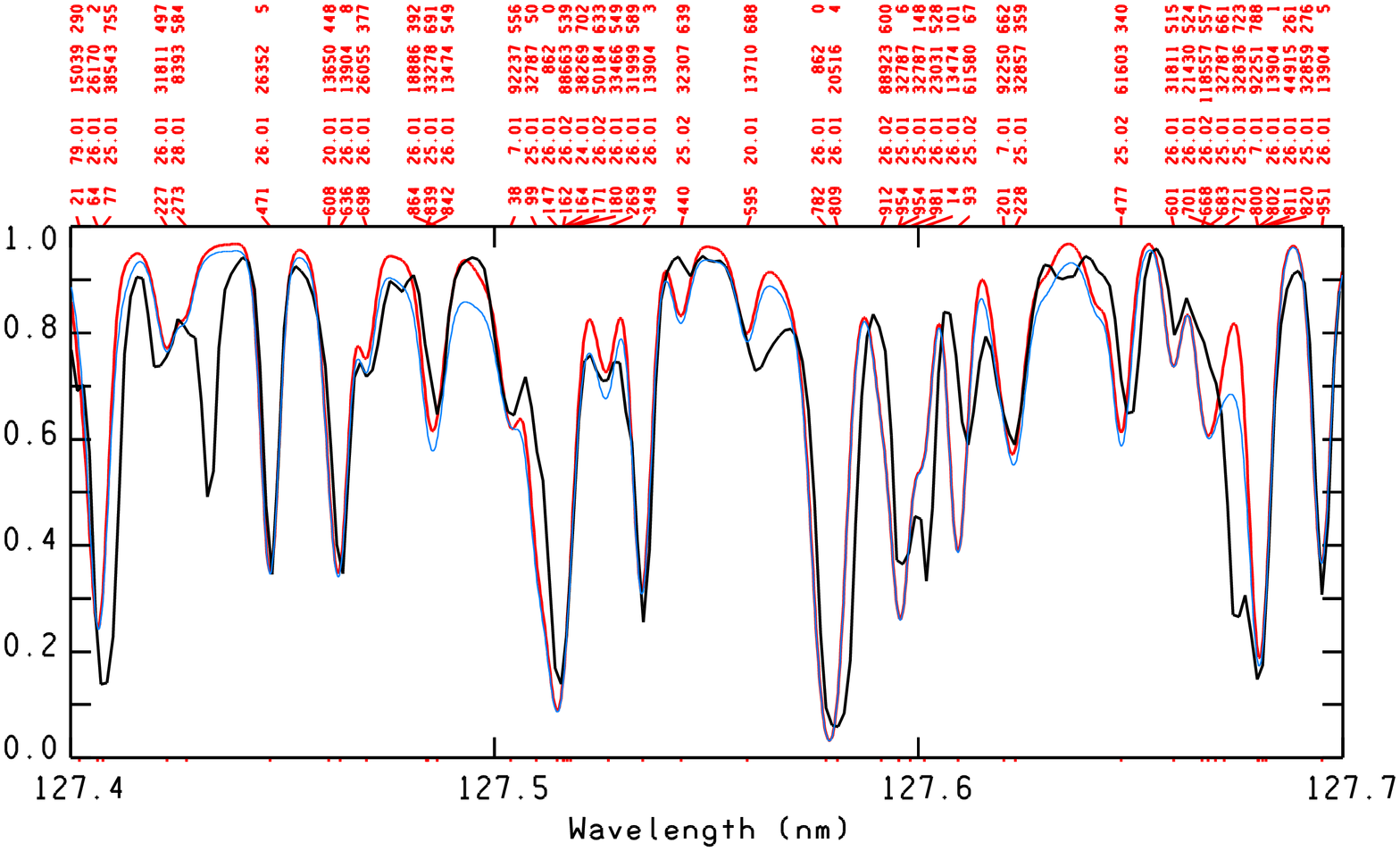}}}
\vskip -1.25cm
\caption{\ion{N}{ii} lines at $\lambda\lambda$ 1275.038, 1275.251, 1276.201, 
and 1276.225\,\AA\
computed both with the constant nitrogen  abundance $-$5.23\,dex (red line) and with 
the abundance step function shown in Fig.\,4b (blue line). 
Only the lines at 1275.038\,\AA\ and 1276.201\,\AA\ are predicted.
The computed spectra are superimposed to the observed spectrum.
The effect of the two different adopted abundances (constant or variable) is negligible.}
\end{figure*}

\subsection{Silicon}

A unique silicon abundance cannot be determined from
the analysis of  several silicon lines present such  
as  \ion{Si}{ii}, \ion{Si}{iii}, and  \ion{Si}{iv} in the ultraviolet spectrum
of HR\,6000.
In fact, the silicon abundance ranges from $-$8.35\,dex to $-$6.0\,dex. 
All the lines analyzed are collected in Table\,A.1.

A plot of the silicon abundance versus $\log\,\tau_{5000}$(aver) 
shows a  dependence of the abundance on the optical depth,
with upper layers more depleted than the deeper layers (Fig.\,7a).

\begin{figure}
\centering
\resizebox{4.50in}{!}{\rotatebox{90}{\includegraphics[0,0][350,700]
{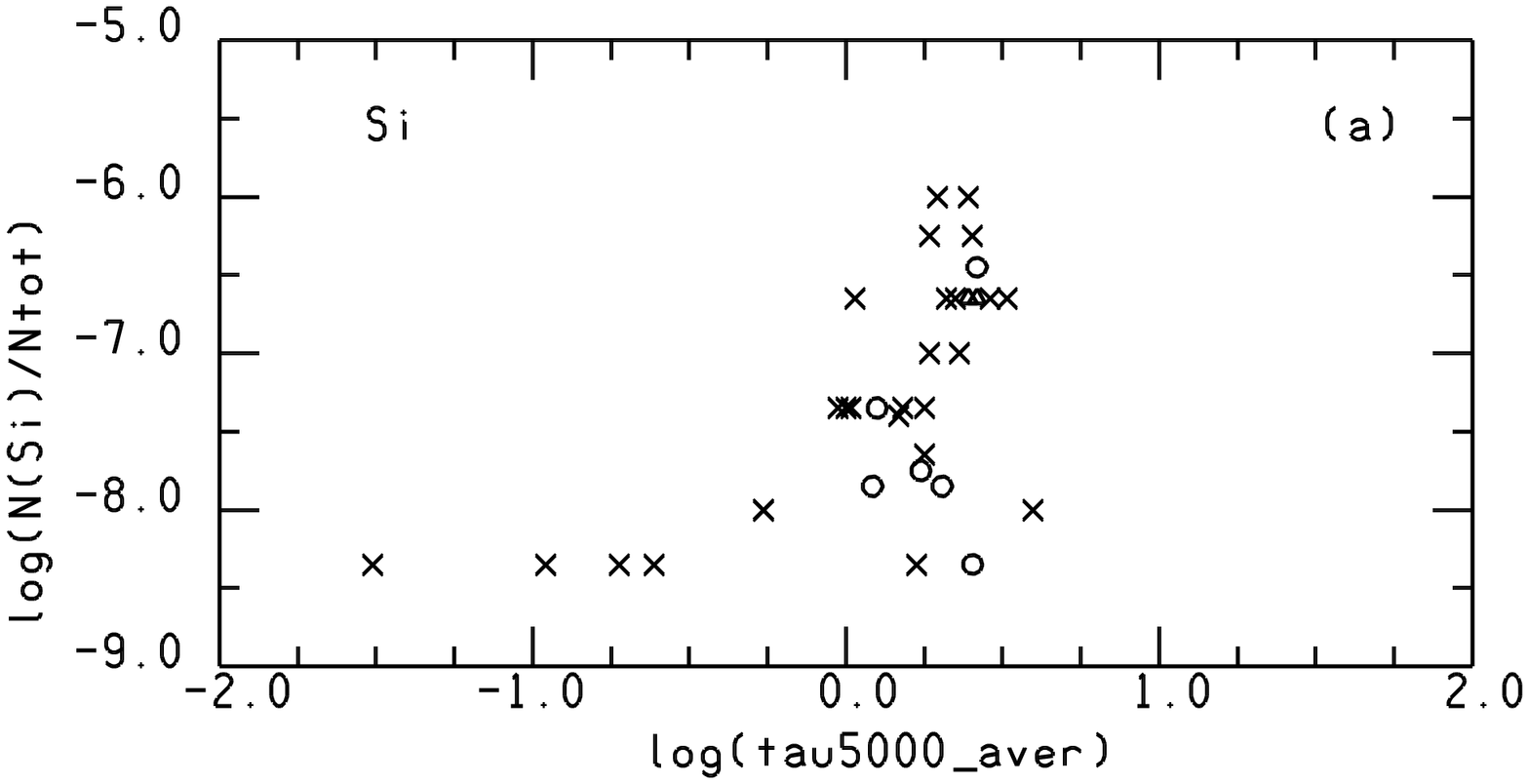}}}
\centering
\resizebox{4.50in}{!}{\rotatebox{90}{\includegraphics[0,0][350,700]
   {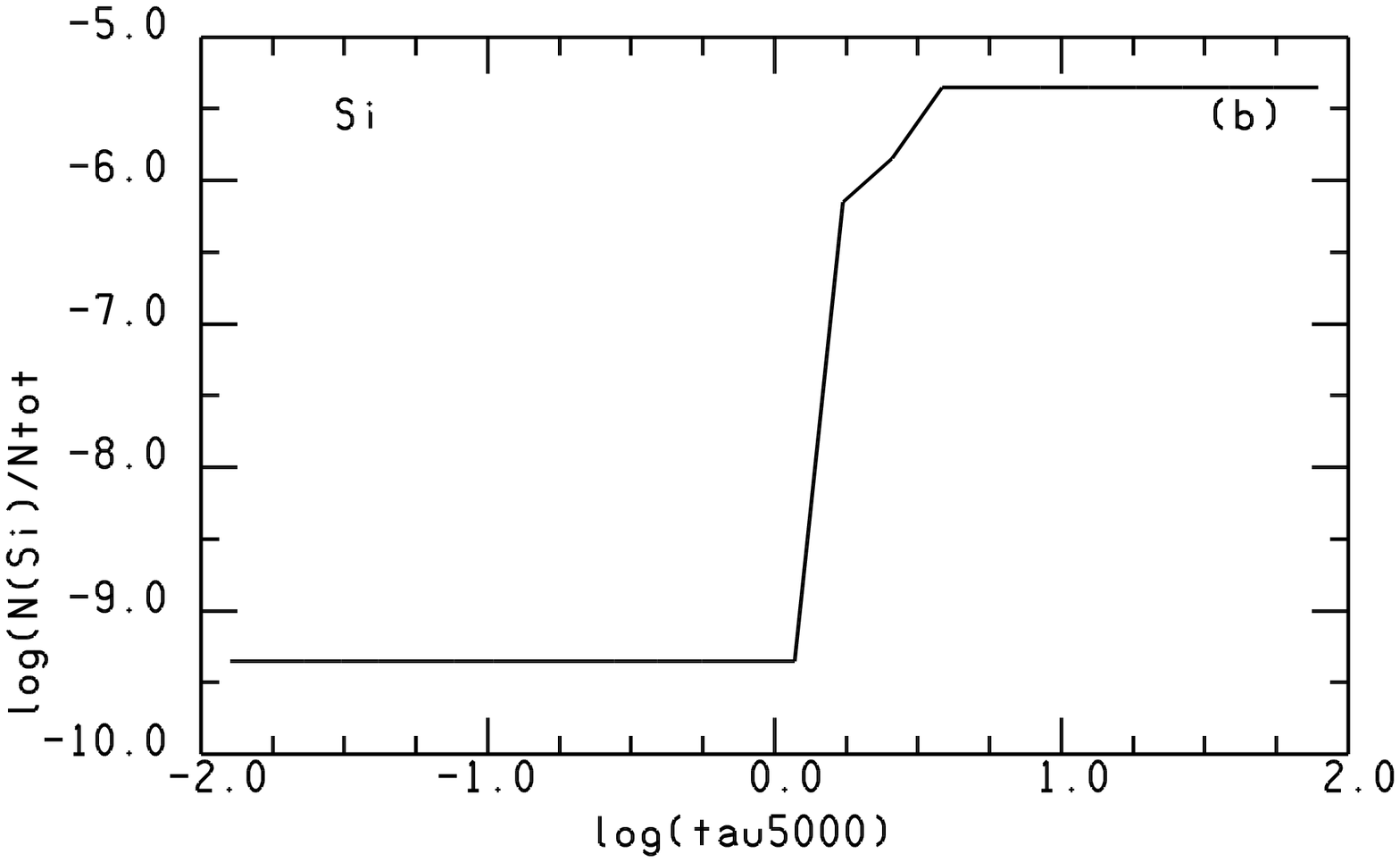}}}
\vskip -1.0cm
\caption{(a): Silicon abundance versus $\log\tau_{5000}$(aver);
crosses are for \ion{Si}{ii}, circles for \ion{Si}{iii}, triangles for\ion{Si}{iv}.
(b): Vertical abundance
distrbution as a function of $\log\tau_{5000}$ obtained by  trial and error.} 
\end{figure}

A further significant illustration of the probable silicon stratification in HR\,6000 is
given by the \ion{Si}{ii} lines at 1264.738\,\AA\ and 1265.002\,\AA\ which show a core
that is consistent with an abundance of $-$8.35\,dex and wings consistent with an
abundance of $-$6.65\,dex (Fig.\,8).
This \ion{Si}{ii} line was used  to derive, by means of  trial and error, the empirical 
stratification step profile given 
in Fig.\,7b. Its use  in the spectral synthesis allows us to 
fit numerous \ion{Si}{i} and \ion{Si}{ii} lines simultaneously.

The Alecian\& Stift (2010) computations of diffusion  of silicon in HgMn stars predict 
a stratification profile with a maximum underabundance  
between $-$0.75 dex and $-$2.0\,dex for  \teff\ between
14000\,K and 12000\,K, and \logg=4.  It occurs at $\log\tau_{5000}$=$-$2.0.
The empirical stratification profile shown in Fig.\,7b implies a slicon underabundance
that is much lower in the whole atmosphere than that predicted by Alecian \& Stift (2010).   
 
\begin{figure}
\centering
\resizebox{4.25in}{!}{\rotatebox{90}{\includegraphics[0,0][525,800]
{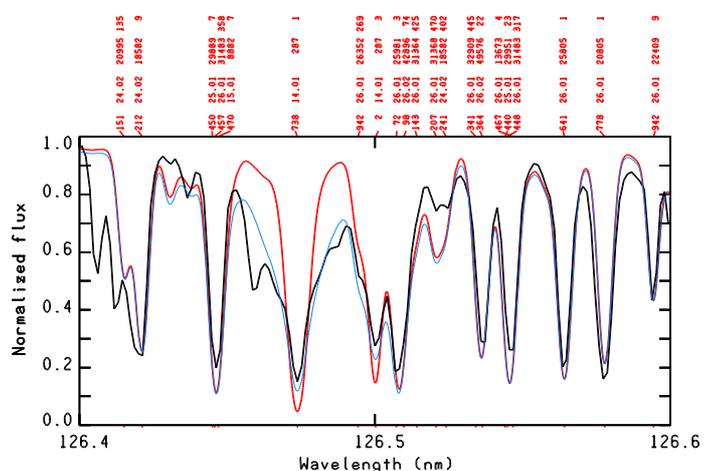}}}
\vskip -1.0cm
\caption{\ion{Si}{ii} lines at 1264.738\,\AA\ and 1265.002\,\AA\ computed 
with the average silicon abundance $-$7.35\,dex (red line) and the abundance
step function of Fig.\,7b (blue line). The computed spectra are
superimposed on the observed spectrum (black line). } 
\end{figure}

\subsection{Phosphorous}

There are numerous lines of \ion{P}{i}, \ion{P}{ii}, and \ion{P}{iii}
in the spectrum of HR\,6000, but  they are all
more or less blended in the ultraviolet, except for
\ion{P}{i} at 2136.182\,\AA,  and 2149.142\,\AA\ and  \ion{P}{ii}
at 2285.105\,\AA\ and 2497.372\,\AA. 
We used NIST $\log\,gf$ values for most \ion{P}{i} lines and
for the \ion{P}{ii} lines of
UV multiplet 2 at 1301-1310\,\AA. For the remaining \ion{P}{i},
\ion{P}{ii}, and \ion{P}{iii} lines 
we adopted $\log\,gf$ values from the Kurucz (2016) database.
For \ion{P}{ii}, they do not differ more than 0.01\,dex from
the NIST values and cover a larger number of lines than the NIST database does.

The average abundances from the \ion{P}{i} and \ion{P}{ii}
ultraviolet lines are $-$5.53$\pm$0.08\,dex and $-$4.54$\pm$0.11\,dex,
respectively, so that they differ by 1.0\,dex. 
Because all the ultraviolet  \ion{P}{iii} lines are blended, the 
\ion{P}{iii} average abundance is rather uncertain.
We assumed an upper limit of
$\log$(N$_{\ion{P}{iii}}$/N$_{tot}$)$\le$$-$5.22$\pm$0.32\,dex.

Castelli \& Hubrig (2007) did not observe  \ion{P}{i} lines in the UVES spectrum,
while several \ion{P}{ii} and \ion{P}{iii} unblended lines were measured. 
The observed cores of the strong \ion{P}{ii} lines of multiplet 5 at 
$\lambda\lambda$\,6024, 6034, and 6043\,\AA\ were so deep
that they could not  be fitted by the computations for any abundance.  
In this paper, we determined two average abundances 
from the optical \ion{P}{ii} lines depending on whether they lie shortward or longward
of 5200\,\AA. These average abundances are $-$4.57$\pm$0.1\,dex and $-$4.32$\pm$0.09\,dex, respectively. 
The average abundance from the optical \ion{P}{iii} lines  is $-$4.69$\pm$0.09.
Table\,2 summarizes all the above abundance values.

The difference of 1.0\,dex, or even larger, between 
 \ion{P}{i} and \ion{P}{ii} abundances
and the increasing of the \ion{P}{ii} abundance
from about $-$4.65\,dex at 4000\,\AA\ to about $-$4.3\,dex at 6000\,\AA,
led us to search for the presence of vertical abundance stratification
for phosphorous.

In Fig.\,9, the \ion{P}{i}, \ion{P}{ii}, and \ion{P}{iii} abundances
derived from both ultraviolet and optical lines listed in Table\,A.1
are plotted as a function of $\log(\tau_{5000})$(aver).  
While there is a weak dependence on depth for \ion{P}{ii} (circles),
no dependence is present for \ion{P}{i} (crosses), which seems
to have been formed at the same layers as \ion{P}{ii} and \ion{P}{iii},
in spite of the  abundance difference of 1\,dex. 
We may speculate that a NLTE analysis of \ion{P}{i} would explain
the large departure from the ionization equilibrium observed for  phosphorous.   

\begin{figure}
\centering
\resizebox{4.50in}{!}{\rotatebox{90}{\includegraphics[0,0][275,700]
{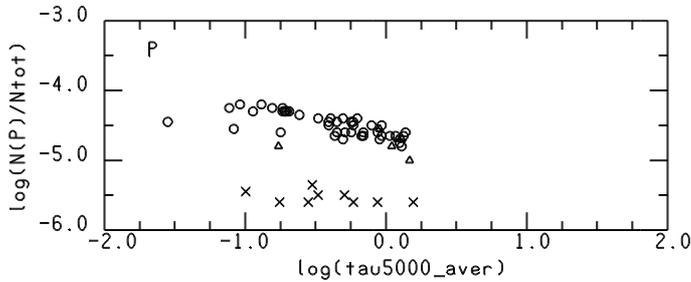}}}
\vskip -1.0cm
\caption{Phosphorous abundance versus $\log\tau_{5000}$(aver);
crosses are for \ion{P}{i}, circles for \ion{P}{ii}, triangles for
\ion{P}{iii}}.
\end{figure}

\subsection{Manganese}
  
The spectrum of HR\,6000 is very rich with  \ion{Mn}{ii} and
\ion{Mn}{iii} lines. 
The manganese abundance of $-$5.3\,dex  was obtained from the three 
\ion{Mn}{ii} lines at  $\lambda\lambda$ 2576, 2593, and 2605\,\AA,
which were also adopted by Smith \& Dworetsky (1993)
in their abundances analysis of HgMn stars 
from IUE spectra. For HR\,6000, we find the same abundance as
Smith \& Dworetsky (1993).
  
Although the average abundance $-$5.29$\pm$0.21\,dex  reproduces almost all the lines 
examined, there are  a few strong \ion{Mn}{ii} lines with a computed core that is 
less deep than the observed core. They are indicated
with the note ``core'' in the last column of Table\,A.1.
Most \ion{Mn}{ii} lines are affected by hyperfine structure. When
hyperfine structure was
taken into account in the computations 
a note ``hfs'' was added in the last column of Table\,A.1.
To compute the hyperfine components we used the hyperfine constants measured 
by  Holt et al. (1999) and by Townley-Smith et al. (2016).

In order to find a good sample of \ion{Mn}{iii} lines, we examined
the  lines with $\log\,gf$ values computed by
Uylings \& Raassen (1997) and extracted  the isolated lines 
and the main components of blends from them (Table\,A.1). The difference between the $\log\,gf$ values from
Uylings \& Raassen (1997) and those computed by Kurucz (2016) 
are on the order of 0.05\,dex, except for two lines at 1283.589\,\AA\ and 1287.584\,\AA\ for which
the difference is 0.13\,dex. The \ion{Mn}{iii} abundance equal to 
$-$5.79$\pm$0.36\,dex was derived using the Kurucz $\log\,gf$ values. 
Their general agreement with those from  Uylings \& Raassen (1997) supports the finding 
of 0.5\,dex lower  \ion{Mn}{iii} abundance compared to the \ion{Mn}{ii} value.

Figure\,10a  shows the \ion{Mn}{ii} and \ion{Mn}{iii} abundances
versus $\log\,\tau_{5000}$(aver). A  dependence of the 
\ion{Mn}{iii} abundance (circles) on the optical depth is manifest, while
the \ion{Mn}{ii} abundance (crosses) is constant with depth.
However, the abundance adopted for the strong lines did not
fit the line center. An increase of the abundance or  the microturbulent
increases the wings rather than the line core.

\begin{figure}
\centering
\resizebox{4.50in}{!}{\rotatebox{90}{\includegraphics[0,0][275,700]
{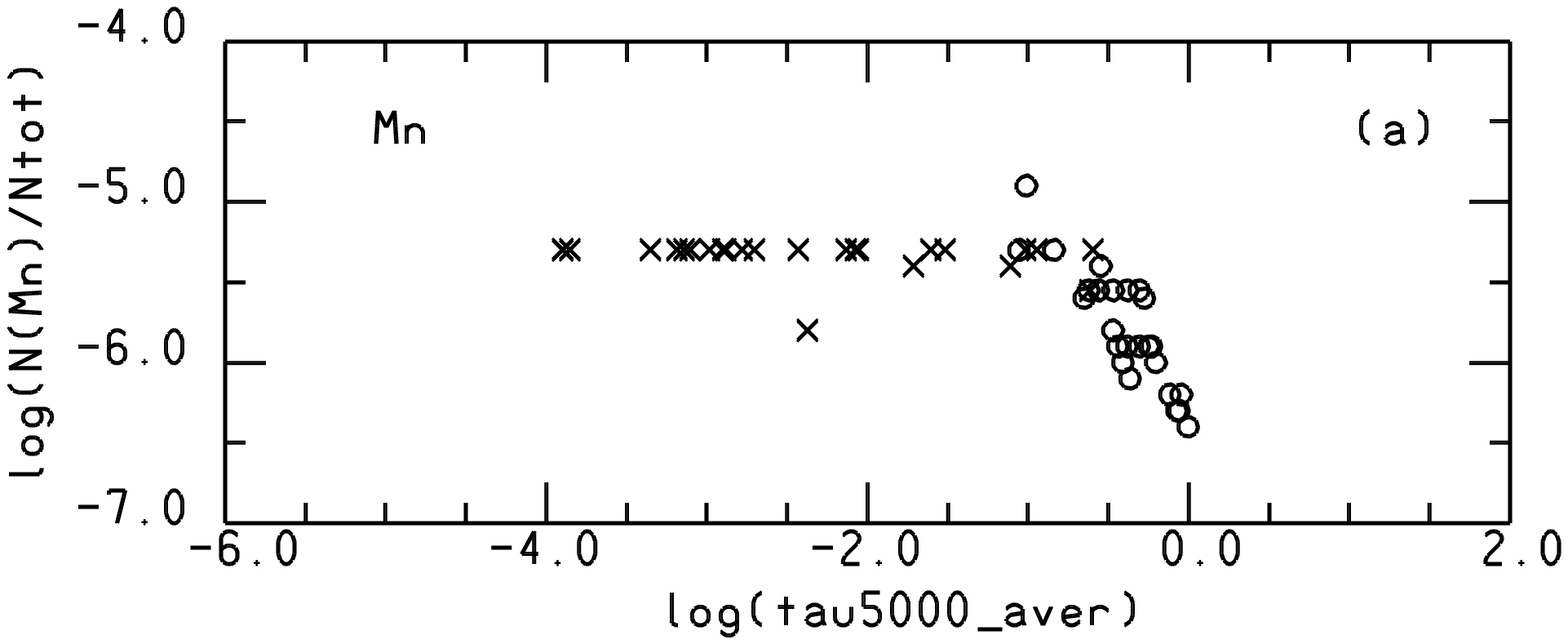}}}
\centering
\resizebox{4.50in}{!}{\rotatebox{90}{\includegraphics[0,0][400,700]
   {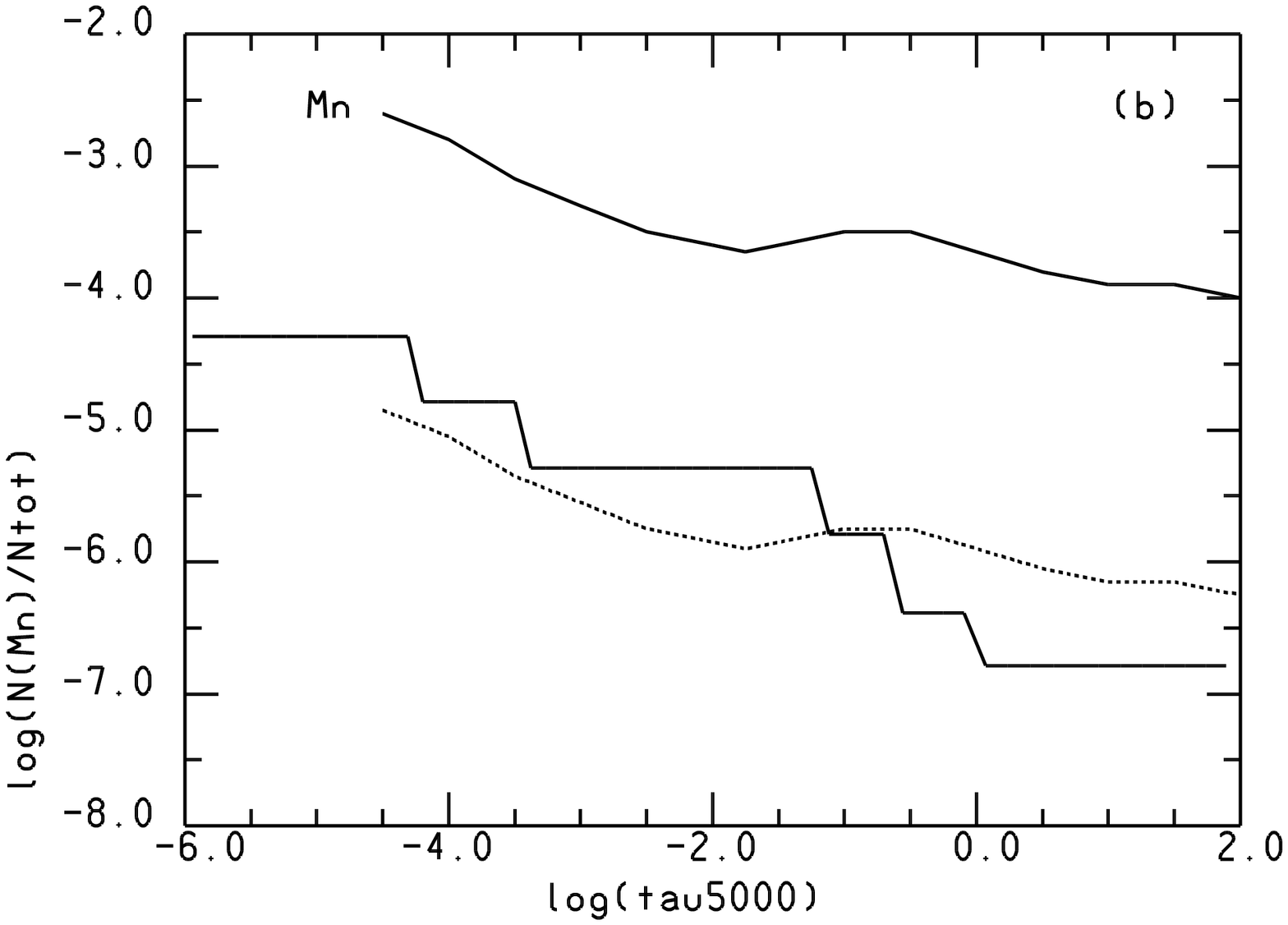}}}
\vskip -1.0cm
\caption{(a): Manganese abundance versus $\log\tau_{5000}(aver)$;
crosses are for \ion{Mn}{ii}, circles for \ion{Mn}{iii}.
(b):  Vertical abundance
distrbution as a function of $\log\tau_{5000}$ obtained by  trial and error.
The theoretical vertical stratification
for \teff=14000\,K, \logg=4.0  from  Alecian \& Stift (2010) is overplotted (full
continous line). The dashed line is the same curve shifted by $-$2.25\,dex 
in the abundance.} 
\end{figure}

The Alecian\& Stift (2010) computations of the diffusion in HgMn stars predict for 
manganese a stratification profile with a behavior similar to that we derived
for HR\,6000, but for an overabundance about 2.5\,dex larger. Fig.\,10b
compares the vertical abundance
distribution as a function of $\log\tau_{5000}$ obtained by  trial and error
with that from Alecian \& Stift (2010), both in the original form and shifted by -2.5\,dex in abundance. 

Figure\,11  is an example of the better agreement between the observed and computed core
achieved for a strong \ion{Mn}{ii} line when  the abundance step function is used in the computations.

\begin{figure}
\centering
\resizebox{4.750in}{!}{\rotatebox{90}{\includegraphics[0,0][500,800]
{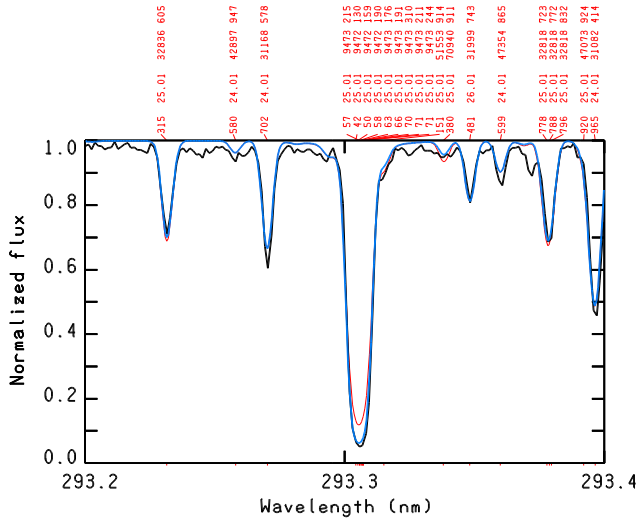}}}
\vskip -1.0cm
\caption{\ion{Mn}{ii} line at 2933.054\,\AA\ computed 
with the average manganese abundance $-$5.29\,dex (red line) and the abundance
step function of Fig.\,10b (blue line). The computed spectra are
superimposed on the observed spectrum (black line). The core of the line is well fitted 
by the abundance step function. The hyperfine structure is considered in the computations.} 
\end{figure}

\subsection{Gold}

There is a discrepancy of 1.0\,dex between the gold abundance
derived from the \ion{Au}{ii} and \ion{Au}{iii} lines.

Several \ion{Au}{ii} lines can be observed in
the spectrum of HR\,6000, but only few of them are unblended,
in particular the \ion{Au}{ii} lines at 1740.475, 1800.579,
2000.792, and 2082.074\,\AA. The abundance from the
line at 1800\,\AA\ is so high compared to that derived from
the other lines that we suspect it is blended 
with some other unknown component. If we exclude this line, the
average abundance from the three unblended lines of \ion{Au}{ii} is
$-$7.57$\pm$0.09\,dex.  However, the other blended lines
are better reproduced by a lower value equal to  $-$8.2\,dex. 
This value also nicely fits the weak unblended
line   observed  at 4052.790\,\AA\ in the UVES spectrum.
Several \ion{Au}{iii} lines, mostly blended, are predicted for
$-$8.2\,dex, but they are all stronger than the observed lines.
The abundance from the \ion{Au}{iii} lines ranges from
$-$8.5\,dex to an upper limit of $-$10\,dex.

We conclude that  gold is probably affected by the abundance
stratification observed for some other elements as well.
Figure\,12a shows the \ion{Au}{ii} and \ion{Au}{iii} abundances
versus $\log\,\tau_{5000}$(aver).
  The figure shows a  dependence of the abundance on the optical depth,
with deeper layers more depleted than the upper layers, as for manganese.
Figure\,12b shows the empirical abundance stratification that
we derived from \ion{Au}{ii} and \ion{Au}{iii} lines with  trial and error.
The abundance stratification  reduces the gold abundance inhomogeneities observed in the spectrum. 

\begin{figure}
\centering
\resizebox{4.50in}{!}{\rotatebox{90}{\includegraphics[0,0][350,700]
{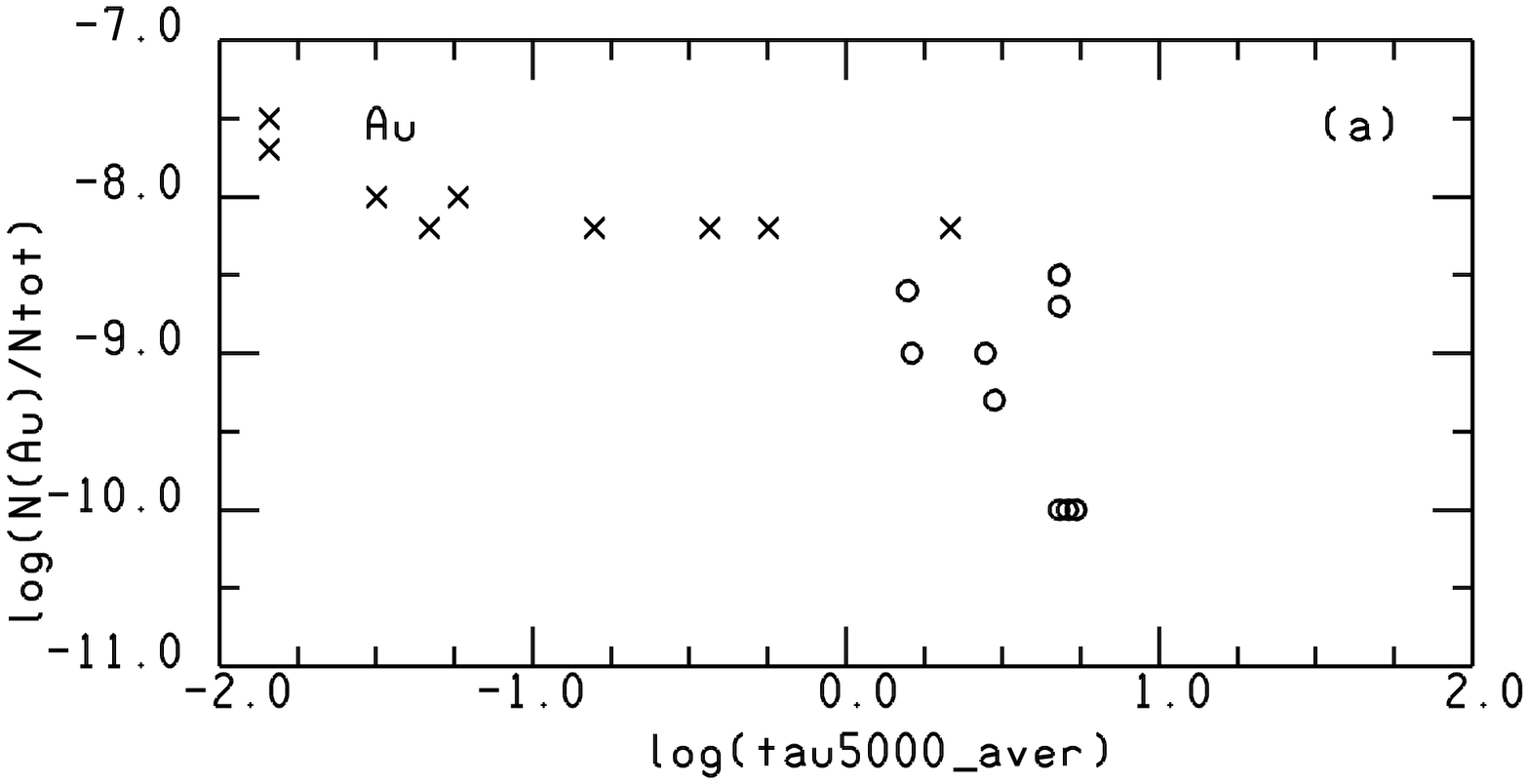}}}
\centering
\resizebox{4.50in}{!}{\rotatebox{90}{\includegraphics[0,0][250,700]
   {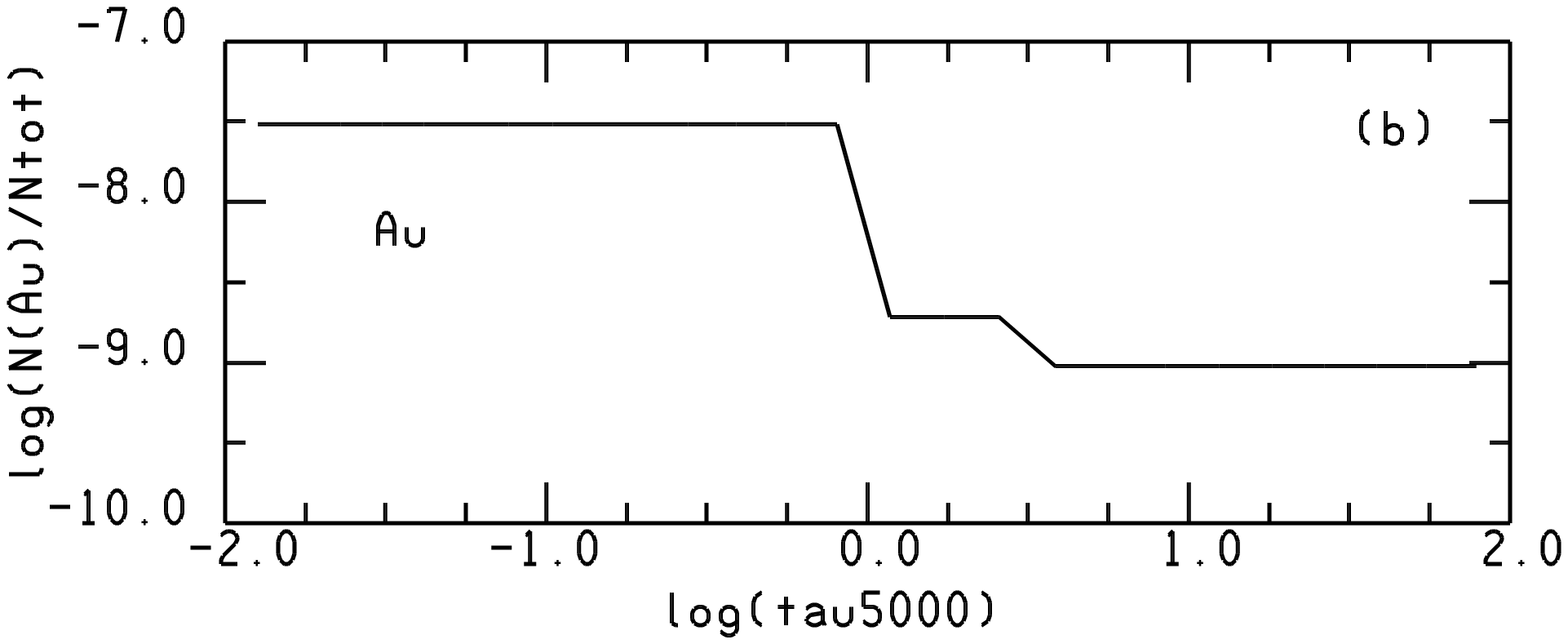}}}
\vskip -1.0cm
\caption{(a): Gold abundance versus $\log\tau_{5000}$(aver);
crosses are for \ion{Au}{ii}, circles for \ion{Au}{iii}.
(b): Vertical abundance
distrbution as a function of $\log\tau_{5000}$ obtained by  trial and error.}
\end{figure}

\section {Conclusions}

   This paper has extended to the ultraviolet the abundance analysis
   of HR\,6000 performed  by Castelli \& Hubrig (2007) and Castelli et al. (2009) 
  on high-resolution spectra  in the optical region.
   In this way we obtained  a panoramic view of the spectrum and  
   abundances of all the species observed in  this
    highly anomalous, ultrasharp-lined,  non-magnetic, main-sequence, 
   chemically peculiar star.

Thanks to  ultraviolet analysis we increased the number of the elements
for which abundances can be assigned, because some elements 
are only observable  in the ultraviolet. Also  some other
elements, which were  observed
in only one  ionization degree in the optical region, are present 
in more ionization degrees in the ultraviolet. In particular, ions not observed 
in the visible are : 
\ion{B}{ii}, \ion{N}{i}, \ion{N}{ii},
\ion{Al}{ii}, \ion{Al}{iii}, \ion{Si}{iii}, \ion{Si}{iv}, \ion{P}{i}, 
\ion{Sc}{iii}, \ion{Ti}{iii}, \ion{V}{ii}, \ion{Cr}{iii}, \ion{Mn}{iii},
\ion{Fe}{iii}, \ion{Ni}{iii}, \ion{Cu}{ii}, \ion{Zn}{ii}, \ion{Ga}{ii}, 
\ion{Ga}{iii}, \ion{Ge}{ii}, \ion{As}{ii}, \ion{Y}{iii}, \ion{Zr}{iii},
\ion{Cd}{ii}, \ion{In}{ii}, \ion{Sn}{ii}, \ion{Xe}{i}, and
\ion{Au}{iii}. 
 
The comparison of the average abundances of HR\,6000 with those of other
HgMn stars taken from the Ghazaryan \& Alecian (2016) compilation has pointed out
the strong underabundance of silicon, already found in other studies,
followed by the underabundances determined from \ion{Co}{ii}, \ion{Cu}{ii}, \ion{Sr}{ii},
and \ion{Zr}{iii}. These elements  put HR\,6000  outside of the group of HgMn stars
according to the abundances collected by Ghazaryan \& Alecian (2016).
At the lower underabundance limit of the Ghazaryan \& Alecian (2016)
compilation there  is carbon, while phosphorous and iron are  at the upper
overabundance limit.

We found in this paper that the striking silicon underabundance ([$-$2.80$\pm$0.26])
is not constant with depth, but that  vertical abundance stratification
 accounts for most of the abundance anomalies 
observed for this element in the ultraviolet. It was modeled with an empirical
step function with abundances ranging from $-$5.35\,dex at the bottom of the
atmosphere to $-$9.35\,dex at the top of the atmosphere.
We examined the possibility of abundance stratification also for some other elements
which either do not meet the ionization balance requirement, such as nitrogen, phosphorous, manganese,
and gold, or show strong broad profiles that cannot be fitted by an
unique abundance, such as \ion{C}{ii} at $\lambda\lambda$ 1335.663, 1335.708\,\AA.
 For all these elements, except for  carbon and phosphorous, we found both a dependence of the
abundance on the optical depth and an empirical abundance profile, which reduces the
disagreement between the observed and computed spectra when it is adopted in the synthetic spectrum computation.
For carbon, we found an empirical stratified abundance profile which roughly
fits both the core and  wings of \ion{C}{ii} at 
$\lambda\lambda$ 1335.663, 1335.708\,\AA, reduces the disagreement between a few
observed and computed \ion{C}{i} profiles, but still does not explain 
other discrepancies,  in particular that for \ion{C}{ii} at 1323.9\,\AA.
For phosphorous, we pointed out a weak dependence of the \ion{P}{ii} abundance
on the optical depth, but the presence at the same depths of \ion{P}{i} 
and \ion{P}{ii} lines with abundances differing by 1\,dex, has hampered any
determination of some empirical abundance stratification profile.

 Other elements could be affected by abundance stratification, such as, for
instance, yttrium and gallium, which were not analyzed here because there
were too few observed lines or their $\log\,gf$ values were uncertain.
There are other elements 
that do not show any sign of vertical stratification. This is true for iron
in spite of its high overabundance.

In this paper we estimated  abundance stratification profiles for
\ion{C}, \ion{N}, \ion{Si}, \ion{Mn}, and \ion{Au}\, in an empirical way by using
the trial and error method. We are aware that  a more rigorous approach should be used, taking into
account either an empirical abundance stratification on the temperature-pressure structure of
the model atmosphere (Nesvacil et al. 2013) or including the diffusion velocities in the
model atmosphere computations (Alecian \& Stift 2010; Stift \& Alecian 2012). 
Last but not least, NLTE effects on the HR\,6000 abundances should be investigated.
We note that while numerous NLTE analyses were performed for stars
hotter than 15000\,K and stars cooler than 10000\,K, very few and sparse work
on NLTE has been carried out  in the temperature range covered by the HgMn stars
and, in particular, for the ultraviolet region.

\begin{acknowledgements}

We thank grant GO-13346, from the Space Telescope Science Institute, for providing support
for the ASTRAL Large Treasury Project.

\end{acknowledgements}

\begin{appendix}

\section{Element-by-element description of the ultraviolet spectrum of HR\,6000 }

A short description of the elements observed in the ultraviolet spectrum of
HR\,6000 but not considered in the main paper is given in this Appendix.

{\bf Boron (B) Z=5:} We derived the underabundance  [$-$0.75]
from the \ion{B}{ii} resonance line at 1362.463\,\AA. 
It is the main component of a blend with \ion{Fe}{ii} 1362.451\,\AA.

{\bf Carbon (C) Z=6:} see the main paper.

{\bf Nitrogen (N) Z=7:} see the main paper.

{\bf Oxygen (O) Z=8:} Only the three lines of  the \ion{O}{i} UV multipet 2 at $\lambda\lambda$
1302.2, 1304.8, and 1306.0\,\AA\
can be observed in the spectrum. The abundance $-$3.68\,dex derived by Castelli et al. (2009)
from the optical
region fits the lines at 1304.858\,\AA\ and 1306.020\,\AA. The
resonance line at 1302.168\,\AA\ is blended with a strong  interstellar
or circumstellar component that is blueshifted by about 0.05\,\AA, so that only the
red part of the profile is  closely predicted.
Although oxygen is marginally underabundant ([$-$0.3]), there are only two stars in the
Ghazaryan \& Alecian (2016) sample with  even lower oxygen abundance ([$-$0.4]).

{\bf Magnesium (Mg) Z=12:} Castelli et al. (2009) obtained the abundance 
of $-$5.66\,dex from \ion{Mg}{ii} at 4481\,\AA. 

The \ion{Mg}{ii} abundance from the ultraviolet is, on average, 0.25\,dex 
larger, in that it ranges from $-$5.36\,dex to $-$5.46\,dex depending on
the lines  considered. We used the \ion{Mg}{ii} lines at 2928.633\,\AA\ and 2936.510\,\AA\
(mult.\,2), those at 2790.777\,\AA\ and 2797.998\,\AA\ (mult.\,3), and the
lines at $\lambda\lambda$ 1734.852, 1737.628, and 1753.474\,\AA. 
The lines of UV multiplet 1 at 2795.528\,\AA\ and 2802.705\,\AA\ cannot be used
for abundance purposes because they both are contaminated by a very strong 
interstellar or circumstellar component.
Averaging all the abundances from the visual and ultraviolet regions we 
obtain $-$5.45$\pm$0.09\,dex, i.e, an underabundance of [$-$1.0] that is fully 
compatible with that of the other HgMn stars of the  
Ghazaryan \& Alecian (2016) sample.

The observed \ion{Mg}{i} resonance line at 2852.126\,\AA\ is very sharp and much 
stronger than that computed for $-$5.36\,dex,which is the largest abundance 
derived from the \ion{Mg}{ii} lines.
It is probably blended with a strong line of of interstellar or cicumstellar origin.

{\bf Aluminium (Al) Z=13:} 
Castelli et al. (2009) fixed an upper limit 
for the aluminium abundance of $-$7.30\,dex from the absence of
\ion{Al}{i} and \ion{Al}{ii} lines in the UVES spectrum.
In the ultraviolet, the resonance line of \ion{Al}{ii} at 1670.787\,\AA\ is 
blended with a blue interstellar (circumstellar) component, so that 
it cannot be used. Other  lines of \ion{Al}{ii} are well reproduced  by
the average abundance of $-$7.80$\pm$0.3\,dex.
\ion{Al}{iii} lines at 1854.716\,\AA\ and 1862.790\,\AA\ are better
fitted by the 0.1\,dex higher abundance of  $-$7.70\,dex.
No lines of \ion{Al}{i} were clearly detectable.
We note that all the lines considered
are more or less blended, except perhaps for  \ion{Al}{ii}
at 1862.311\,\AA. 

The average abundance from all the \ion{Al}{ii} and \ion{Al}{iii} lines listed in
table\,A.1 is $-$7.80$\pm$0.3\,dex, corresponding to  an underabundance
of [$-$2.19].
This value is compatible with the underabundances of the  HgMn stars of the  
Ghazaryan \& Alecian (2016) sample.

{\bf Silicon (Si) Z=14:} see the main paper.

{\bf Phosphous (P) Z=15:} see the main paper

{\bf Sulfur (S) Z=16:} A sulfur abundance of $-$6.26\,dex was obtained by Castelli et al. (2009)
from the weak unblended \ion{S}{ii} line at 4162.665\,\AA.
Using the sulfur ultraviolet lines we updated the abundance to $-$6.36\,dex,  
a value that agrees well with both the ultraviolet \ion{S}{i} unblended line 
at 1425.03\,\AA\ and the \ion{S}{ii} line of UV mult.\,1 at 1250.584\,\AA.
The other two lines of \ion{S}{ii} mult.\,1 at
1253.811\,\AA\ and  1259.519\,\AA\
are too blended with other strong lines, and probably also with some interstellar or
circumstellar component,  to be used for abundance purposes.
All the three strong \ion{S}{ii} lines of UV mult.\,1 lie in a spectral region affected 
by several uncertainties mostly
related with the continuum position and   numerous unidentified lines.
The sulfur underabundance of [$-$1.44] puts HR\,6000 among the HgMn stars 
with a very low sulfur abundance according to the Ghazaryan \& Alecian (2016)
compilation.

{\bf Chlorine (Cl) Z=17:} The ground-state line of \ion{Cl}{i} at 1379.528\,\AA\ is 
absent in the STIS spectrum. For this reason, we adopted as  chlorine abundance 
the upper limit  $-$7.74\,dex ([$-$1.20]) inferred by Castelli et al. (2009)
from the absence in the UVES spectrum of the line at 4794.556\,\AA.
Other \ion{Cl}{i} lines, if present in the UVES spectrum, are heavy blended
with stronger components.
 Only two stars in the Ghazaryan \& Alecian (2016) sample have
 chlorine abundance. These are $\chi$\,LupiA and HD\,46866 with  underabundance
 [$-$1.40] and  [$-$0.36], respectively.
 The  underabundance of HR\,6000, which is $\le$ [$-$1.20], puts the star in between 
$\chi$LupiA and HD\,46866.

{\bf Calcium (Ca) Z=20:} The average calcium abundance obtained from a few 
ultraviolet unblended lines lying in the region 2110-2210\,\AA\
is almost solar (Table\,A.1). The underabundance of [-0.02]$\pm$0.09  agrees 
with the calcium abundance 
of the HgMn stars of the Ghazaryan \& Alecian (2016) sample,
which is more or less scattered around the solar value.

{\bf Scandium (Sc) Z=21:} Only weak lines of \ion{Sc}{iii} were observed in the STIS spectrum.
From the
unblended line of \ion{Sc}{iii} at 1603.064\,\AA\  an abundance of $-$10.2\,dex
was deduced, corresponding to the underabundance of ([$-$1.11]).
There are four stars in the
Ghazaryan \& Alecian (2016) sample with a scandium underabundance less
than that of HR\,6000. Therefore,  HR\,6000 is similar to other HgMn stars
as far as the scandium abundance is concerned. 

{\bf Titanium (Ti) Z=22:} Castelli et al. (2009) derived
a titanium  abundance  equal to $-$6.47$\pm$0.13\,dex ([+0.64]) from the equivalent widths of
numerous \ion{Ti}{ii} lines observed in the UVES spectrum. This value was confirmed
in the ultraviolet by the analysis of several \ion{Ti}{ii} lines, in 
particular those at $\lambda\lambda$ 1909.207, 1909.662, and 1910.954\,\AA. 
 Several \ion{Ti}{iii}
lines of the UV multiplets 1, 3, 4, 5, 6, and 7 were examined. There is a large spread
in the abundances, which range from $-$6.0\,dex to $-$7.7\,dex (Table\,A.1).
Most lines are blended. In addition, the only source for the $\log\,gf$ values is the
Kurucz line list so that we do not have a critical compilation of the \ion{Ti}{iii}
oscillator strengths.
The average abundance of \ion{Ti}{ii} agrees within the error limits with
the average \ion{Ti}{iii} abundance $-$6.37$\pm$0.41\,dex.
According to the Ghazaryan \& Alecian (2016) compilation,
the titanium overabundance [$+$0.64] of HR\,6000 is a typical value for HgMn stars.

{\bf Vanadium (V) Z=21:} The \ion{V}{ii} lines of multiplets 1, 10, and 12 
at 2900\,\AA\ are either very weak or  not even
observed. From three lines at $\lambda\lambda$ 2908.817, 2924.019, and 2924.641\,\AA,
we derived  the average abundance of $-$9.23$\pm$0.12\,dex.
In the Ghazaryan \& Alecian (2016) compilation, there are two stars, $\Phi$Phe and HD\,71066,
with a vanadium underabundance that is lower than the [$-$1.12] value of HR\,6000.
 
{\bf Chromium (Cr) Z=24:} The lines of \ion{Cr}{ii} and \ion{Cr}{iii} are very numerous in HR\,6000.
From analyses of the \ion{Cr}{ii} lines at $\lambda\lambda$ 2055, 2061, 2653,
2666, 2971, and 2989\,\AA,  which are also adopted by Smith \& Dworetsky (1993), and of
some other unblended lines (Table\,A.1), we derived  an average abundance
of $-$5.9$\pm$0.2\,dex. 
This value agrees with the value of $-$6.10$\pm$0.09\,dex  obtained  from the optical region by
Castelli et al. (2009). The \ion{Cr}{iii} abundance 
was obtained from a few unblended lines that were extracted
by analyzing all  the \ion{Cr}{iii} lines listed in the NIST database for the
1250-3000\,\AA\ interval. The average abundance for \ion{Cr}{iii} is $-$6.22$\pm$0.25\,dex.
The final chromium abundance $-$6.10$\pm$0.25\,dex was obtained 
averaging the abundances of all the \ion{Cr}{ii} and \ion{Cr}{iii} lines
examined in  both STIS and UVES spectra.
According to the Ghazaryan \& Alecian (2016) compilation, the chromium overabundance [$+$0.3]
of HR\,6000 is a typical value for the HgMn stars.

{\bf Manganese (Mn) Z=25:} see the main paper.

{\bf Iron (Fe) Z=26:} Pratically all the lines 
of \ion{Fe}{ii} and \ion{Fe}{iii} listed in the NIST database are present
in the ultraviolet spectrum of HR\,6000.
A few lines of \ion{Fe}{i}, mostly those arising from the ground level,
were also identified. The iron abundance  $-$3.65$\pm$0.09\,dex
was carefully determined by Castelli et al. (2009) from several \ion{Fe}{i} and
\ion{Fe}{ii} lines observed in the UVES spectrum. We checked that this
value  closely predicts the ultraviolet \ion{Fe}{ii} lines used by Smith \& Dworetsky (1998) in their
analyses of HgMn and related stars (Table\,A.1).
The average abundance from the ultraviolet \ion{Fe}{ii} lines is 
$-$3.68$\pm$0.05\,dex. This  is close to the value of $-$3.7 $\pm$0.05\,dex obtained 
by  Smith \& Dworetsky (1998) for HR\,6000.

We extracted a set of unblended lines from the numerous \ion{Fe}{iii} lines observed in the ultraviolet
spectrum (Table\,A.1).
The average abundance  $-$3.78$\pm$0.14\,dex agrees within the error limits 
with the \ion{Fe}{ii} abundance. We finally adopted
the value $-$3.65$\pm$0.09\,dex deduced by Castelli et al. (2009) from 
the optical spectrum for the iron abundance of HR\,6000.
The corresponding iron overabundance, [+0.92], 
is an upper limit  according to the Ghazaryan \& Alecian (2016) compilation.

{\bf Cobalt (Co) Z=27}: Castelli et al. (2009) estimated an upper limit $-$8.42\,dex
(underabundance [$-$1.3])
for the cobalt abundance from the unobserved  \ion{Co}{ii} line 
at 3501.708\,\AA\ and  from the blended \ion{Co}{ii} line at 4160.657\,\AA. 
In the ultraviolet we examined the lines used by  Smith \& Dworetsky (1993)
to derive the cobalt abundance in normal late-B and HgMn stars.
They are the lines at $\lambda\lambda$ 2286.159, 2307.86, 2324.321, and 2580.326\,\AA.
We added a few lines listed in Table\,A.1.
While the  line at 2307.86\,\AA\ is well
predicted by $-$8.42\,dex, the other lines, although predicted
for this abundance, are not observed.
Assuming that the absorption at 2307.86\,\AA\ is due to some other
unidentified element, we lowered the  upper limit of the cobalt abundance
to $-$10.12\,dex (underabundance [$-$3.0]), which suppresses the predicted
unobserved lines.
Smith \& Dworetsky (1993) assigned the upper limit  $<$-9.0$\pm$0.5\,dex for 
the cobalt abundance in HR\,6000. The cobalt underabundance $\le$[$-$3.0] 
of HR\,6000 is lower than the lower limit given in the  
Ghazaryan \& Alecian (2016) compilation.

{\bf Nickel (Ni) Z=28:} Castelli et al. (2009) obtained the abundance $-$6.24\,dex
from the equivalent width of the \ion{Ni}{ii} line at 4067.031\,\AA.
Smith \& Dworetsky (1993) derived the nickel abundance in the HgMn stars 
of their sample from the 
\ion{Ni}{ii} lines at $\lambda\lambda$ 2165.550, 2184.602, 2270.212, and
2287.081\,\AA. For HR\,6000 they determined  $-$6.34\,dex.
In the STIS spectrum the above lines are well fitted by $-$6.24\,dex.
The corresponding underabundance is [$-$0.40].
There are several \ion{Ni}{iii} lines in the spectrum, but they are
all blended except for $\lambda\lambda$ 1774.896, 1829.986, and 1830.060\,\AA.
The average abundance from the three lines quoted above is $-$6.64$\pm$0.14.
We adopted as nickel abundance $-$6.24\,dex from \ion{Ni}{ii}, mostly 
because \ion{Ni}{ii} is the dominant ion. According to the Ghazaryan \& Alecian (2016) compilation, 
the nickel underabundance [$-$0.40] of HR\,6000 is a typical value in HgMn stars.

{\bf Copper (Cu) Z=29:} Castelli et al. (2009) derived an upper limit for the copper
abundance equal  to $-$7.83\,dex (solar value) from the
\ion{Cu}{ii} unobserved lines at 4909.734\,\AA\ and 4931.698\,\AA.
In the ultraviolet, the abundance  $-$10.53\,dex was derived from the \ion{Cu}{ii}
resonance line at 1358.773\,\AA. The corresponding  underabundance is [$-$2.7].
This value is well below the lower limit of [$-$0.89] for 112\,Her\,A of
the Ghazaryan \& Alecian (2016) compilation.

{\bf Zinc (Zn) Z=30:} The zinc abundance of $-$8.84\,dex was derived from the
very weak \ion{Zn}{ii} line at 2064.227\,\AA.
In fact, the two strong resonance lines at
2025.502\,\AA\ and 2062.004\,\AA\ are not useful for this purpose.
They both seem to be blueshifted by 0.03\,\AA\ and the second 
line is computed too strong, as compared with the observed line, for  $-$8.84\,dex. 
We  argue that the two resonance lines are blended with an interstellar or
circumstellar component that affects the profiles in an unpredictable way.
We assumed the abundance of $-$8.84\,dex as an upper limit. This abundance corresponds to
the underabundance $\le$[$-$1.36], which is consistent with the zinc abundance
of the HgMn stars in the Ghazaryan \& Alecian (2016) sample.

{\bf Gallium (Ga) Z=31:}
All the \ion{Ga}{ii} lines are blended.
The abundance  $-$8.85\,dex was derived
from the \ion{Ga}{ii} resonance line  at 1414.399\,\AA,
which is the main component in a blend.
The \ion{Ga}{ii} resonance line is compatible with
that from other subordinate \ion{Ga}{ii} lines.
The resonance lines of \ion{Ga}{iii} at
1495.045\,\AA\ and 1534.462\,\AA\ give the lower abundance  $-$8.15\,dex,
so that there is a discrepancy of $-$0.65\,dex between the \ion{Ga}{ii}
and \ion{Ga}{iii} abundances. Because \ion{Ga}{ii} is the dominant
state, we assumed for gallium the final abundance of $-$8.85\,dex.
The overabundance [+0.17] puts HR\,6000
below the lowest overabundance limit equal to [$+$0.81]$\pm$0.30 provided by  46\,Aql,
according to the Ghazaryan \& Alecian (2016) compilation. 

{\bf Germanium (Ge) Z=32}: \ion{Ge}{ii} lines were not observed in the spectrum.
The solar germanium abundance was decreased to to $-$9.64\,dex  to fit 
the spectrum at the position
of the ground-configuration line of \ion{Ge}{ii} at 1649.194\,\AA.
Because the line at 1261.905 is still predicted as a strong line for this
abundance,  the germanium abundance was  further lowered to $-$10.64\,dex.

{\bf Arsenic (As) Z=33}:
The \ion{As}{ii} line at 1375.07\,\AA\ is predicted for  a solar arsenic abundance
as minor component in a blend with \ion{Ti}{ii} which closely fits the observed 
spectrum. Because other \ion{As}{ii} lines
were neither observed nor predicted, we adopted the  solar abundance
$-$9.74\,dex for arsenic.

{\bf Strontium (Sr) Z=38}:
No \ion{Sr}{ii} lines were observed in the whole spectrum from ultraviolet
to the optical regions. We adopted the upper limit  $-$10.07\,dex derived from
the absence of \ion{Sr}{ii} at 4077.709\,\AA. The corresponding underabundance, $\le$[$-$0.9],
puts HR\,6000 among the most deficient strontium stars of the Ghazaryan \& Alecian (2016) sample. 

{\bf Yttrium (Y) Z=39}:
Castelli et al. (2009) obtained the abundance equal to $-$8.60\,dex from the profiles 
of \ion{Y}{ii} at $\lambda\lambda$ 3950.349, 4883.682, and 4900.120\,\AA. 
However, the strong \ion{Y}{iii}
lines at $\lambda\lambda$ 2367.228, 2414.643, 2817.023, and 2945.995\,\AA\ require 
the higher abundance of $-$7.6\,dex
to be fitted. Yttrium overabundance is usual among HgMn stars, with values  closer
to what we found from  \ion{Y}{iii}  than from \ion{Y}{ii}.

{\bf Zirconium (Zr) Z=40}:
An upper limit of -10.24\,dex was derived from the unobserved \ion{Zr}{iii} line at 1941.053\,\AA.
The corresponding underabundance $\le$[$-$0.8] contrasts with the  usual 
zirconium overabundance in HgMn stars, as is evident from the 
Ghazaryan \& Alecian (2016) compilation.

{\bf Cadmium (Cd) Z=48}:
The resonance line of \ion{Cd}{ii} at 2144.393\,\AA, although blended, 
is well predicted by the abundance  $-$7.00\,dex. The almost unblended
\ion{Cd}{ii} line at 2265.018\,\AA\ and  two other lines at 2312.766  
and 2572.93\,\AA\ have intensities that are consistent with this abundance.
However, the observed \ion{Cd}{ii} line at 2748.549\,\AA\ is fitted by $-$7.4\,dex.
It is also redshifted by 0.007\,\AA, indicating that the energy levels 
for this line should be better determined. 
\ion{Cd}{i} at 2288.728\,\AA\ is observed, but blended with several features.
The most important is \ion{Fe}{ii}  2288.022\,\AA.
The cadmium overabundance of [$+$3.33] is a new finding for HR\,6000
abundance analyses.  
In the Ghazaryan \& Alecian (2016) sample
only  $\chi$Lupi\,A is quoted for cadmium, which is [+0.56] overabundant
in this star.

{\bf Indium (In) Z=49:}
The line of \ion{In}{ii} at 1586.331\,\AA\ is blended with two stronger
\ion{Fe}{ii} lines. The  abundance  $-$10.24\,dex is not in conflict with
the observed blended profile, but \ion{In}{iii} at 1625.295\,\AA\ is
predicted as a strong line that is not observed.

{\bf Tin (Sn) Z=50}:
The lines of \ion{Sn}{ii} are fitted by abundances
ranging from $-$7.0\,dex for the  line  
at 2151.514\,\AA\ to $-$8.7\,dex for the lines at $\lambda\lambda$1400.440, 1474.997, and
1757.905\,\AA. However, there are other lines,  for example, those
at 1312.274\,\AA\ and 1899.881\,\AA\ which are
predicted to be rather strong for $-$8.7\,dex, but are not observed at all.
We suspect that incorrect energy levels for \ion{Sn}{ii} may be
responsible. The average abundance is $-$8.23$\pm$0.64\,dex.
The corresponding overabundance is [+1.77]$\pm$0.774\,dex.
The only star with Sn abundance included in the Ghazaryan \& Alecian (2016)
compilation is $\chi$Lupi\,A. The Sn overabundance of this star is $<$ [+1.38], which  was derived only 
from the line at 2151.514\,\AA\ (Leckrone et al. 1999).

{\bf Xenon (Xe) Z=54:}
The numerous \ion{Xe}{ii} lines observed in the optical region yielded
the abundance of  $-$5.25$\pm$0.17\,dex (Y\"uce et al. 2011). There are no \ion{Xe}{ii}
lines in the ultraviolet spectrum, but two \ion{Xe}{i} lines were
observed at  1469.612\,\AA\ and 1295.588\,\AA. They give the abundance
 $-$5.55$\pm$0.25\,dex, in agreement with the \ion{Xe}{ii} value.
The xenon overabundance of HR\,6000 is similar to
that of most HgMn stars of the Ghazaryan \& Alecian (2016) sample. 

{\bf Gold (Au) Z=79:} see the main paper.

{\bf Mercury (Hg) Z=80}:
Castelli et al. (2009) derived an abundance of $-$8.20\,dex from a weak structure
observed at 3983.9\,\AA\ that they interpreted as due to \ion{Hg}{ii} affected by
an isotopic anomaly.
However, this abundance cannot be confirmed by the clearly observable
\ion{Hg}{ii} lines at 1649.937\,\AA\ and 1942.273\,\AA, which both yield 
the abundance of $-$9.6\,dex, i.e., an overabundance of [+1.27].
The  lower \ion{Hg}{ii} abundance is not in conflict with the abundances from a few
\ion{Hg}{iii} lines that are predicted as minor components of blends.
An example is \ion{Hg}{iii} at 1647.471\,\AA.   

In the Ghazaryan \& Alecian (2016) compilation only 53\,Tau has an overabundance
of mercury that is lower than that of HR\,6000.

\begin{table*}[!hbp]
  \caption[ ]{Lines analyzed in the STIS spectrum of HR 6000. For \ion{Si}{ii}, \ion{P}{ii}
    \ion{P}{iii}, and \ion{S}{ii}, lines observed in the UVES spectrum by Castelli \& Hubrig (2007)
    are added. The successive columns list: element, wavelength and multiplet number, if available, $\log\,gf$ value, its source,
    excitation potential of the lower and upper levels, abundance $\log(N_{elem}/N_{tot})$. In the last column some
    notes are  added.}
\font\grande=cmr7
\grande
\begin{flushleft}
\begin{tabular}{llllrrllllllll}
\hline\noalign{\smallskip}
\multicolumn{1}{c}{Elem}&
\multicolumn{1}{c}{$\lambda$, mult.}&
\multicolumn{1}{c}{$\log\,gf$}&
\multicolumn{1}{c}{source$^{a}$}&
\multicolumn{1}{c}{$\chi_{low}$(cm$^{-1}$)}&
\multicolumn{1}{c}{$\chi_{up}$(cm$^{-1}$)}&
\multicolumn{1}{c}{abund}&
\multicolumn{1}{l}{Notes}\\
\hline\noalign{\smallskip}
\ion{B}{ii}& 1362.463 (1)& $-$0.001& NIST5&  0.000 &  73396.510 & $-$10.1 & blend \ion{Fe}{ii}\\
\ion{C}{i} & 1277.245 (7)& $-$1.035& NIST5&  0.000  & 78293.490 & $-$5.8 &\\
\ion{C}{i} & 1277.283 (7)& $-$0.683& NIST5& 16.417 & 78307.630 & $-$5.8 & \\
\ion{C}{i} & 1277.513 (7)& $-$1.169& NIST5& 16.417 & 78293.490 & $-$5.3 & \\
\ion{C}{i} & 1277.550 (7)& $-$0.409& NIST5& 43.414 & 78318.250 & $-$6.0 & \\
\ion{C}{i} & 1277.723 (7)& $-$1.113& NIST5& 43.414 & 78306.619 &$-$5.5: & blend \ion{Fe}{ii}\\
\ion{C}{i} & 1277.954 (7)& $-$2.360& NIST5& 43.414 & 78293.501 &$-$4.9  & single\\
\ion{C}{i} & 1279.891 (5)&  $-$1.376& NIST5& 16.417 & 78148.089 & $-$5.10: & blend Cr III,Fe II\\
\ion{C}{i} & 1280.135 (5)&  $-$1.583& NIST5&  0.000 & 78116.748 & $-$5.00 & blend Mn II\\
\ion{C}{i} & 1280.333 (5)&  $-$1.106& NIST5& 43.414 & 78148.089 & $-$5.60 & single\\
\ion{C}{i} & 1280.404 (5)&  $-$1.859& NIST5& 16.417 & 78116.748 & $-$5.50 & blend Mn II\\
\ion{C}{i} & 1280.597 (5)&  $-$1.664& NIST5& 16.417 & 78104.967 & $-$5.10 & blend Mn II\\
\ion{C}{i} & 1280.847 (5)&  $-$1.573& NIST5& 43.414 & 78116.748 & $-$5.10 & single\\
\ion{C}{i} & 1328.833 (4) &  $-$1.236& NIST5&  0.000 & 75253.983 & $-$5.2 & blend Fe II\\
\ion{C}{i} & 1329.085 (4) &  $-$1.231& NIST5& 16.417 & 75256.153 & $-$5.5 & \\
\ion{C}{i} & 1329.100 (4) &  $-$1.147& NIST5& 16.417 & 75255.276 & $-$5.5 & \\
\ion{C}{i} & 1329.123 (4) &  $-$1.355& NIST5& 16.417 & 75253.983 & $-$5.5 &  blend Fe III,Fe II\\
\ion{C}{i} & 1329.578 (4) &  $-$0.662& NIST5& 43.414 & 75255.276 & $-$6.0 & blend Fe II\\
\ion{C}{i} & 1329.600 (4) &  $-$1.136& NIST5& 43.414 & 75253.983 & $-$6.0 & blend Fe II\\
\ion{C}{i} & 1560.309 (3)& $-$1.145 & NIST5&  0.000 & 64089.863 &$-$5.5: & blend Fe II\\
\ion{C}{i} & 1560.682 (3)& $-$0.793 & NIST5& 16.417 & 64090.969 &$-$5.8 & blend  Fe II\\
\ion{C}{i} & 1560.709 (3)& $-$1.271 & NIST5& 16.417 & 64089.863 &$-$5.8 & blend  Fe II\\
\ion{C}{i} & 1561.340 (3)& $-$1.271 & NIST5& 43.414 & 64090.969 &$-$6.0 & \\
\ion{C}{i} & 1561.367 (3)& $-$2.448 & NIST5& 43.414 & 64089.863 &$--$   & blend, not pred \\
\ion{C}{i} & 1561.438 (3)& $-$0.522 & NIST5& 43.414 & 64086.951 & $-$6.3&  \\
\ion{C}{i} & 1656.267 (2)& $-$0.746 & NIST5& 16.417 & 60393.148 & $-$6.7 & \\
\ion{C}{i} & 1656.928 (2)& $-$0.844 & NIST5&  0.000 & 60352.639 & $-$5.6 &\\
\ion{C}{i} & 1657.008 (2)& $-$0.271 & NIST5& 43.414 & 60393.148 & $-$6.5 & \\
\ion{C}{i} & 1657.379 (2)& $-$0.971 & NIST5& 16.417 & 60352.639 & $-$5.5 & blend\\
\ion{C}{i} & 1657.907 (2)& $-$0.845 & NIST5& 16.417 & 60333.429 & $-$5.9 & blend\\
\ion{C}{i} & 1658.121 (2)& $-$0.748 & NIST5& 43.414 & 60352.639 & $-$6.0\\
\ion{C}{ii}& 1323.862 (11)& $-$1.284  & NIST5&74930.100&  150466.690 & $-$5.5 :& no fit\\
\ion{C}{ii}& 1323.906 (11)& $-$0.337  & NIST5&74932.620&  150466.690 & $-$5.5 :& no fit\\
\ion{C}{ii}& 1323.951 (11)& $-$0.144  & NIST5&74930.100&  150461.580 &$-$5.5 :& no fit\\
\ion{C}{ii}& 1323.995 (11)& $-$1.288  & NIST5&74932.620&  150461.580 &$-$5.5 :& no fit \\
\ion{C}{ii}& 1334.532 (1)& $-$0.589  & NIST5&    0.000 & 74932.620 &$-$5.0 ? & wings, blend interstellar comp.\\
\ion{C}{ii}& 1335.663 (1)& $-$1.293  & NIST5&    63.420 & 74932.620&$-$5.0 ?  & wings  \\
\ion{C}{ii}& 1335.708 (1)& $-$0.335  & NIST5&    63.420 & 74930.100&$-$5.0 ?  & wings\\
\ion{N}{i}& 1310.540 (13)& $-$0.926& NIST5& 28839.306 & 105143.710 & $-$5.6& blend Fe II\\
\ion{N}{i}& 1310.943 (13)& $-$1.205& NIST5& 28838.920 & 105119.880 & $-$5.6& blend Fe III \\
\ion{N}{i}& 1310.950 (13)& $-$1.743& NIST5& 28839.306 & 105119.880 & $-$5.6& blend Fe III \\
\ion{N}{i}& 1411.931 (10)& $-$1.273& NIST5 & 28838.920 & 99663.912 & $-$5.9\\
\ion{N}{i}& 1411.939 (10)& $-$1.916& NIST5 & 28839.306 & 99663.912 & $-$5.9\\
\ion{N}{i}& 1411.948 (10)& $-$1.019& NIST5 & 28839.306 & 99663.427 & $-$5.9\\
\ion{N}{i}& 1492.625 (4)& $-$0.381& NIST5 & 19224.464 & 86220.510 & $-$6.8\\
\ion{N}{i}& 1492.820 (4)& $-$1.360& NIST5 & 19233.177 & 86220.510 & $--$ & not obs\\
\ion{N}{i}& 1494.675 (4)& $-$0.634& NIST5 & 19233.117 & 86137.350 & $-$6.8 \\
\ion{N}{ii}& 1275.038 ( )& $-$1.206 & NIST5 & 92237.200 & 170666.230 & $-$5.3\\
\ion{N}{ii}& 1275.251 ( )& $-$1.944 & NIST5 & 92250.300 & 170666.230 & $-$5.2\\
\ion{N}{ii}& 1276.201 ( )& $-$1.478 & NIST5 & 92250.300 & 170607.890 & $-$5.2\\
\ion{N}{ii}& 1276.225 ( )& $-$1.948 & NIST5 & 92251.800 & 170607.890 & $-$5.2\\
\ion{O}{i}& 1302.168 (2) &$-$0.585& NIST5& 0.000 & 76794.978 & $--$ & blue interstellar comp.\\
\ion{O}{i}& 1304.858 (2) &$-$0.808& NIST5& 158.265 & 76794.978 &$-$3.68 &  \\
\ion{O}{i}& 1306.029 (2)& $-$1.285& NIST5& 226.977 & 76794.978 &$-$3.68 &  \\
\hline
\noalign{\smallskip}
\end{tabular}
\end{flushleft}
\end{table*}

\setcounter{table}{0}

\begin{table*}[!hbp]
\caption[ ]{Cont.}
\font\grande=cmr7
\grande
\begin{flushleft}
\begin{tabular}{llllrrllllllll}
\hline\noalign{\smallskip}
\multicolumn{1}{c}{Elem}&
\multicolumn{1}{c}{$\lambda$}&
\multicolumn{1}{c}{$\log\,gf$}&
\multicolumn{1}{c}{source$^{a}$}&
\multicolumn{1}{c}{$\chi_{low}$(cm$^{-1}$)}&
\multicolumn{1}{c}{$\chi_{up}$(cm$^{-1}$)}&
\multicolumn{1}{l}{abund}&
\multicolumn{1}{l}{Notes}\\
\hline\noalign{\smallskip}
\ion{Mg}{i} & 2852.126 (1) & $+$0.255 & NIST5&    0.00  & 35051.264& $--$ & strong interstellar comp.\\ 
\ion{Mg}{ii} & 1734.852 ( ) & $-$1.111& NIST5& 35669.31 & 93311.11& $-$5.36 & blend\\
\ion{Mg}{ii} & 1737.628 ( ) & $-$0.859& NIST5& 35760.88 & 93310.59& $-$5.46 & blend\\
\ion{Mg}{ii} & 1753.474 ( ) & $-$1.133& NIST5& 35760.88 & 92790.51& $-$5.46: & blend \ion{Fe}{iii}\\
\ion{Mg}{ii}& 2790.777 (3)& $+$0.273 & NIST5& 35669.31 & 71491.06 & $-$5.36\\
\ion{Mg}{ii}& 2795.528 (1) & $+$0.085 & NIST5&     0.00 & 35760.88 &$--$& broad violet comp.\\
\ion{Mg}{ii}& 2797.998 (3)& $+$0.528 & NIST5& 35760.88 & 71490.19 & $-$5.36\\
\ion{Mg}{ii}& 2802.705 (1) & $-$0.218 & NIST5&     0.00 & 35669.31&$--$& broad violet comp.\\
\ion{Mg}{ii}& 2928.633 (2) &  $-$0.529& NIST5& 35669.31 & 69804.95& $-$5.46\\
\ion{Mg}{ii}& 2936.510 (2) &  $-$0.225& NIST5& 35760.88 & 69804.95& $-$5.46\\
\ion{Al}{ii} &1670.787 (2)& $+$0.248& NIST5& 0.000 & 59852.02& $-$7.80: & strong blue comp\\
\ion{Al}{ii}& 1719.442 (6)& $-$0.060& NIST5& 37393.03 & 95551.44&  $-$7.30& hfs, blend \\
\ion{Al}{ii}& 1721.271 (6)& $+$0.292 & NIST5& 37577.79 & 95549.42& $-$7.80& hfs \\
\ion{Al}{ii}& 1724.982 (6)& $+$0.565& NIST5& 37453.91 & 95550.51 & $-$7.80:&hfs, blend Fe II\\
\ion{Al}{ii}& 1763.869 (5)& $-$0.371& NIST5& 37453.91 & 94147.46& $-$7.80 & blend\\    
\ion{Al}{ii}& 1763.952 (5)& $+$0.331& NIST5& 37577.79 & 94268.68& $-$8.30 & bl Mn II\\
\ion{Al}{ii}& 1767.732 (5)& $-$0.143& NIST5& 37577.79 & 94147.46& $-$7.50 & bl Mn II\\
\ion{Al}{ii}& 1855.926 (4)& $-$0.886& NIST5& 37393.03 & 91274.50& $-$7.80: & hfs, blend P II \\
\ion{Al}{ii}& 1858.025 (4)& $-$0.412& NIST5& 37453.91 & 91274.50& $--$ & too blended Mn II, FeII\\
\ion{Al}{ii}& 1862.311 (4)& $-$0.197& NIST5& 37577.79 & 91274.50& $-$8.10& hfs, single\\
\ion{Al}{iii}& 1854.716 (1)& $+$0.050& NIST5&  0.00    & 53916.60& $-$7.70 & blend \\
\ion{Al}{iii}& 1862.790 (1)& $-$0.253& NIST5&  0.00    & 53682.93& $-$7.70 & blend\\    
\ion{Si}{ii} & 1260.422 (4) & $+$0.387& NIST5&  0.000 & 79338.50 & $--$ & strong blue component\\
\ion{Si}{ii} & 1264.738 (4) & $+$0.639& NIST5& 287.240& 79355.02 & $-$8.35 core, $-$6.65 wings & \\ 
\ion{Si}{ii} & 1265.002 (4) & $-$0.345& NIST5& 287.240& 79338.50 & $-$8.35 core, $-$6.65 wings&\\
\ion{Si}{ii} & 1304.370 (3) & $-$0.731& NIST5&  0.000 & 76665.35 &$<$ $-$8.35 & strong blue component\\
\ion{Si}{ii} & 1305.592 (13.04)& $+$0.474& K16 & 55325.18& 131918.800 & $-$6.40   & \\
\ion{Si}{ii} & 1309.276 (3) & $-$0.495& NIST5&287.240 & 76665.35 & $-$8.35 & bl \ion{Mn}{iii}\\   
\ion{Si}{ii} & 1309.453 (13.04)& $+$0.318& K16 & 55309.35& 131677.100 & $-$6.45   & \\
\ion{Si}{ii} & 1346.884 (7) & $-$0.389& NIST5& 42932.62& 117178.06 & $-$6.65 & bl \ion{Mn}{iii}\\
\ion{Si}{ii} & 1348.543 (7) & $-$0.437& NIST5& 42824.29& 116976.38 & $-$6.65:  & blend, blue unident.\\
\ion{Si}{ii} & 1350.072 (7)& $-$0.057& NIST5& 43107.91& 117178.06 & $-$8.00 ? & bl \ion{Fe}{ii}\\
\ion{Si}{ii} & 1350.516 (7) & $-$0.754& NIST5& 42932.62& 116078.38 & $-$6.65  & blend\\
\ion{Si}{ii} & 1350.656 (7)& $-$0.975& NIST5& 42824.29& 116862.38 & $-$6.65  & blend\\
\ion{Si}{ii} & 1352.635 (7) & $-$0.474& NIST5& 42932.62& 116862.38 & $-$7.00 ?& blend \ion{Fe}{iii}\\
\ion{Si}{ii} & 1353.721 (7) & $-$0.451& NIST5& 43107.91& 116978.38 & $---$  & too blended \\
\ion{Si}{ii} & 1403.784 (13.03)& $-$0.961& K16 & 55309.35& 126545.400 & $-$6.00   & \\
\ion{Si}{ii} & 1404.482 (13.03)& $-$0.826& K16 & 55325.18& 126525.800 & $-$6.00   & \\
\ion{Si}{ii} & 1409.053 (13.02)& $-$0.638& NIST5 & 55309.35& 126279.000 & $-$6.25 & \\
\ion{Si}{ii} & 1410.214 (13.02)& $-$0.383& NIST5 & 55325.18& 126326.400 & $-$6.25 & \\
\ion{Si}{ii} & 1509.092 (11.01)& $-$0.410& NIST5& 55325.18& 121590.19 & $-$6.65  & single \\
\ion{Si}{ii} & 1526.707 (2)& $-$0.575& NIST5&  0.000 & 65500.47 & $---$ & strong blue component\\
\ion{Si}{ii} & 1533.431 (2)& $-$0.274& NIST5 &287.24  & 65000.47 & $-$8.35 & bl \ion{Fe}{iii}\\ 
\ion{Si}{ii} & 1808.013 (1)& $-$2.203& NIST5&   0.000 & 55309.35&  $--$ & blend \ion{Fe}{iii}+blue comp.?\\
\ion{Si}{ii} & 1816.928 (1)& $-$2.103& NIST5& 287.240 & 55325.18&$\le$$-$8.35& not obs !!!!\\
\ion{Si}{ii} & 1817.451 (1)& $-$3.194& NIST5& 287.240 & 55309.35& $--$       & not obs, not pred for -7.35 \\
\ion{Si}{ii} & 2072.015 (9)& $-$0.430& NIST5& 55309.35 & 103556.16&$-$7.35: & blend \ion{Fe}{ii}\\
\ion{Si}{ii} & 2072.700 (9)& $-$0.290& NIST5& 55325.18 & 103556.030&$-$7.35: & bl \ion{Mn}{iii}\\
\ion{Si}{ii} & 3853.665 (vis 1)& $-$1.341& NIST5& 55309.35& 81251.32 & $-$7.00  & single\\
\ion{Si}{ii} & 3856.018 (vis 1)& $-$0.406& NIST5& 55325.18& 81251.32 & $-$7.65  & blend unknown ?\\
\ion{Si}{ii} & 3862.595 (vis 1)& $-$0.757& NIST5& 55309.35& 81191.34 & $-$7.35  & single\\
\ion{Si}{ii} & 4128.054 (vis 3)& $+$0.359& NIST5& 79338.500& 103556.160 & $-$7.35 ? & blend\\
\ion{Si}{ii} & 4130.894 (vis 3)& $+$0.552& NIST5& 79355.020& 103556.030 & $-$7.40  & \\
\ion{Si}{ii} & 5041.024 (vis 5)& $+$0.029& NIST5& 81191.34.020& 101023.05 & $-$7.35  & weak \\
\ion{Si}{ii} & 6347.109 (vis 2)& $+$0.149& NIST5& 65500.47& 81251.32 & $\le$ $-$8.00  & not obs ?\\
\ion{Si}{ii} & 6371.371 (vis 2)& $-$0.082& NIST5& 65500.47& 81191.34 & $\le$ $-$8.00  & not obs\\
\hline
\noalign{\smallskip}
\end{tabular}
\end{flushleft}
\end{table*}

\setcounter{table}{0}

\begin{table*}[!hbp]
\caption[ ]{Cont.}
\font\grande=cmr7
\grande
\begin{flushleft}
\begin{tabular}{llllrrllllllll}
\hline\noalign{\smallskip}
\multicolumn{1}{c}{Elem}&
\multicolumn{1}{c}{$\lambda$}&
\multicolumn{1}{c}{$\log\,gf$}&
\multicolumn{1}{c}{source$^{a}$}&
\multicolumn{1}{c}{$\chi_{low}$(cm$^{-1}$)}&
\multicolumn{1}{c}{$\chi_{up}$(cm$^{-1}$)}&
\multicolumn{1}{c}{abund}&
\multicolumn{1}{l}{Notes}\\
\hline\noalign{\smallskip}

\ion{Si}{iii} & 1294.545 (4) & $-$0.173 & NIST5& 52853.28 & 130100.52 & $---$    & bl \ion{Fe}{iii}\\
\ion{Si}{iii} & 1296.726 (4) & $-$0.270 & NIST5& 52724.69 & 129841.97 & $-$7.35       & bl uniden.\\     
\ion{Si}{iii} & 1298.892 (4) & $-$0.396 & NIST5& 52853.28 & 129841.97 & $-$7.85    & blend\\
\ion{Si}{iii} & 1298.946 (4) & $+$0.302 & NIST5& 53115.01 & 130100.52 & $-$7.85:    & blend\\
\ion{Si}{iii} & 1301.149 (4) & $-$0.272 & NIST5& 52853.28 & 129708.45 & $-$7.75       & blend \ion{Fe}{ii}\\
\ion{Si}{iii} & 1303.323 (4) & $-$0.177 & NIST5& 53155.01 & 129841.97 & $-$8.35  & blend Fe II\\ 
\ion{Si}{iii} & 1417.237 (9) & $-$0.184  & NIST5& 82884.41 & 153444.23 & $-$6.45 & single \\  
\ion{Si}{iii} & 1312.591 (10) & $-$0.765  & NIST5& 82884.41 & 159069.61 & $---$  & blend\\
\ion{Si}{iv} & 1393.755 (1) & $+$0.011     & NIST5&     0.00 & 71784.64  & $-$6.65 & blend\\
\ion{Si}{iv} & 1402.770 (1) & $-$0.292    & NIST5&     0.00 & 71287.54  & $-$6.65 & single !\\  
\ion{P}{i} & 1381.476 (2)& $+$0.081&K16 &0.000 & 72386.347 & $-$5.45 \\
\ion{P}{i} & 1671.671 (2)& $-$2.805&K16 &0.000      & 59820.371& $--$ & not obs\\
\ion{P}{i} & 1674.595 (2)& $-$2.453& K16  &   0.000 & 59715.921& $--$ & not obs \\
\ion{P}{i} & 1679.697 (2)& $-$2.190& K16  &   0.000 & 59534.549& $\le$ $-$5.55 & not obs \\
\ion{P}{i} & 1685.975 (6)& $+$0.006& K16 &11376.630 & 70689.504&  $--$& blend \ion{Fe}{ii} \\
\ion{P}{i} & 1694.031 (6)& $-$0.149& K16 &11361.020 & 70391.801&  $--$& blend, wrong \ion{Fe}{ii} \\
\ion{P}{i} & 1774.949 (1)& $-$0.211& NIST5 &  0.000 & 56339.656&  $-$5.60 & bad fit, blue comp ? \\
\ion{P}{i} & 1782.829 (1)& $-$0.389& NIST5 &  0.000 & 56090.626&  $-$5.60 & blend\\
\ion{P}{i} & 1787.648 (1)& $-$0.690& NIST5 &  0.000 & 55939.421&  $-$5.50 & \\
\ion{P}{i} & 1858.871 (5)& $-$1.142& NIST5 &11361.020 & 65157.126&  $-$5.60 & \\
\ion{P}{i} & 1858.901 (5)& $-$0.279& NIST5 &11361.020 & 65156.242&  $-$5.60 & \\
\ion{P}{i} & 1859.410 (5)& $-$0.086& NIST5 &11376.630 & 65157.126&  $-$5.60 & blend \ion{Fe}{ii} \\
\ion{P}{i} & 1859.441 (5)& $-$1.209& K16 &11376.630 & 65156.242&  $--$ & too blended \\
\ion{P}{i} & 2135.469 (4)& $-$1.076& K16 &11361.020 & 58174.366&$\le$ $-$5.50 & not obs    \\
\ion{P}{i} & 2136.182 (4)& $-$0.114& NIST5 &11376.630 & 58174.366& $-$5.50 & single\\
\ion{P}{i} & 2149.142 (4)& $-$0.211& K16 &11361.020 & 57876.574& $-$5.35 & single  \\
\ion{P}{ii} & 1301.874 (2)& $-$1.42& NIST5 &   0.000 & 76812.330 & $-$4.45: & blend \\
\ion{P}{ii} & 1304.492 (2)& $-$1.42& NIST5 & 164.900 & 76823.110 & $-$4.45:  & blend \\
\ion{P}{ii} & 1304.675 (2)& $-$1.55& NIST5 & 164.900 & 76812.330 & $-$4.45: & blend \\
\ion{P}{ii} & 1305.497 (2)& $-$1.31& NIST5 & 164.900 & 76764.060 & $-$4.45: & blend \\
\ion{P}{ii} & 1309.874 (2)& $-$2.215& K16 & 469.120 & 76812.330 & $-$4.45: & blend \\
\ion{P}{ii} & 1310.703 (2)& $-$0.85& NIST5 & 469.120 & 76764.060 & $-$4.45: & blend \\
\ion{P}{ii} & 1452.900 ( )& $-$1.331& HI  &8882.310 & 77710.19  & $-$4.55 & blend \ion{Fe}{ii} \\
\ion{P}{ii} & 1485.496 ( )& $-$0.825& K16 &21575.63 & 88893.22 & $-$4.55  & from red wing \\
\ion{P}{ii} & 1532.533 (1)& $-$2.122& K16 &   0.000 & 65251.450 &  $-$4.45 & blend \\
\ion{P}{ii} & 1535.923 (1)& $-$1.765& K16 & 164.900 & 65272.350 & $-$4.45 & blend \\
\ion{P}{ii} & 1536.416 (1)& $-$2.274& K16 & 164.900 & 65251.450 & $-$4.45 & blend \\
\ion{P}{ii} & 1542.304 (1)& $-$1.501& K16 & 469.120 & 65307.170 & $-$4.45 & blend \\
\ion{P}{ii} & 1543.051 ( )& $-$1.645& K16 & 65251.45 & 130058.110 & $-$4.80 & \\
\ion{P}{ii} & 1543.133 (1)& $-$2.295& K16 & 469.120 & 65272.350 & $-$4.45 & blend \\
\ion{P}{ii} & 1543.631 (1)& $-$3.496& K16 & 469.120 & 65251.450 &  $-$4.45 & blend \\
\ion{P}{ii} & 1685.850 ( )& $-$0.301& K16 & 88192.130 & 147509.39 & $-$4.70  &  \\
\ion{P}{ii} & 1858.209 ( )& $+$0.048& K16 & 88192.130 & 142007.39 & $-$4.70  & blend \\
\ion{P}{ii} & 2280.991 (6)& $+$0.182& K16 & 87804.100 & 121631.16 & $--$ & too blended \\
\ion{P}{ii} & 2285.105 (7)& $+$0.466& K16 & 88192.130 & 131940.29 & $-$4.60  & single \\
\ion{P}{ii} & 2497.372 (5)& $-$1.089& K16 & 65272.35 & 105302.37 & $-$4.45 & single \\
\ion{P}{ii} & 4044.576 (30)& $+$0.669& K16 & 107360.25 & 132077.74 & $-$4.60  & single,4044.595 ? \\
\ion{P}{ii} & 4062.149 (17)& $-$0.637& K16 & 103339.14 & 127949.70 & $-$4.75  & single \\
\ion{P}{ii} & 4072.289 (16)& $-$1.074& K16 & 103339.14 & 127888.42 & $-$4.60  & single \\
\ion{P}{ii} & 4127.559 (16)& $+$0.006& K16 & 103667.86 & 127888.42 & $-$4.65  & single \\
\ion{P}{ii} & 4160.623 (31)& $-$0.760& K16 & 103339.14 & 127367.23 & $-$4.65  & single \\
\ion{P}{ii} & 4244.622 (30)& $-$0.412& K16 & 130912.84 & 107360.25 & $-$4.65  & single \\
\ion{P}{ii} & 4288.606 (33)& $-$1.118& K16 & 101635.69 & 124946.73 & $-$4.65: & blend Fe II \\
\ion{P}{ii} & 4420.712 ( ) & $-$0.395& K16 & 88893.22 & 111597.66 & $-$4.50 & single \\

\hline
\noalign{\smallskip}
\end{tabular}
\end{flushleft}
\end{table*}

\setcounter{table}{0}

\begin{table*}[!hbp]
\caption[ ]{Cont.}
\font\grande=cmr7
\grande
\begin{flushleft}
\begin{tabular}{llrrrcclcrr}
\hline\noalign{\smallskip}
\multicolumn{1}{c}{Species}&
\multicolumn{1}{c}{$\lambda$($\AA$)}&
\multicolumn{1}{c}{$\log\,gf$}&
\multicolumn{1}{c}{source$^{a}$}&
\multicolumn{1}{c}{$\chi_{low}$}&
\multicolumn{1}{c}{$\chi_{up}$}&
\multicolumn{1}{c}{$log(N_{Z})/N_{\rm tot}$)}&
\multicolumn{1}{c}{Notes}
\\
\hline\noalign{\smallskip}
\ion{P}{ii} & 4452.472 (31)& $-$0.083& K16 & 105302.37 & 127755.500 & $-$4.60 & single \\
\ion{P}{ii} & 4463.027 (25)& $+$0.164& K16 & 105549.67 & 127949.700 & $-$4.60 & single \\
\ion{P}{ii} & 4466.140 (24)& $-$0.483& K16 & 105549.67 & 127934.090 & $-$4.65 & single \\
\ion{P}{ii} & 4468.000 (25)& $+$0.010& K16 & 105224.060 & 127599.160 & $-$4.65 & blend \\
\ion{P}{ii} & 4475.270 (24)& $+$0.535& K16 & 105549.670 & 127888.420 & $-$4.65 & single \\
\ion{P}{ii} & 4483.693 (25)& $-$0.644& K16 & 105302.370 & 127599.160 & $-$4.50 & single  \\
\ion{P}{ii} & 4499.230 ( ) & $+$0.614& K16 & 107922.930 & 130142.720 & $-$4.60 & single  \\
\ion{P}{ii} & 4927.197 (13)& $-$0.799& K16 & 103165.610 & 123455.460 & $-$4.50 & single  \\
\ion{P}{ii} & 4943.497 (13)& $+$0.082& K16 & 103667.860 & 123890.810 & $-$4.40 & single  \\
\ion{P}{ii} & 4954.367 (13)& $-$0.501& K16 & 103165.610 & 123344.190 & $-$4.40 & single  \\
\ion{P}{ii} & 4969.701 (13)& $-$0.174& K16 & 103339.140 & 123455.460 & $-$4.40 & single  \\
\ion{P}{ii} & 5152.221 (7) & $-$1.395& K16 & 86597.550 & 106001.250 & $-$4.60 & single  \\
\ion{P}{ii} & 5191.393 (7) & $-$0.675& K16 & 86743.960 & 106001.250 & $-$4.40 & single  \\
\ion{P}{ii} & 5296.077 (7) & $-$0.053& K16 & 87124.600 & 106001.250 & $-$4.25 & single  \\
\ion{P}{ii} & 5316.055 (6) & $-$0.291& K16 & 86743.960 & 105549.670 & $-$4.30 & bl  \\
\ion{P}{ii} & 5344.729 (6) & $-$0.280& K16 & 86597.550 & 105302.370 & $-$4.25 &   \\
\ion{P}{ii} & 5386.895 (6) & $-$0.257& K16 & 86743.960 & 105302.370 & $-$4.30 &   \\
\ion{P}{ii} & 5409.722 (6) & $-$0.312& K16 & 86743.960 & 105224.060 & $-$4.30 &   \\
\ion{P}{ii} & 5425.880 (6) & $+$0.288& K16 & 87124.600 & 105549.670 & $-$4.30 & core not fitted   \\
\ion{P}{ii} & 5450.709 (23)& $+$0.083& K16 &105549.670 & 123890.810 & $-$4.45 &   \\
\ion{P}{ii} & 5499.697 (6) & $-$0.441& K16 & 87124.600 & 105302.370 & $-$4.35 &   \\
\ion{P}{ii} & 5507.174 (23)& $-$0.559& K16 &105302.370 & 123455.460 & $-$4.45 & bl   \\
\ion{P}{ii} & 5541.139 (23)& $-$0.515& K16 &105302.370 & 123344.190 & $-$4.45 &    \\
\ion{P}{ii} & 5583.235 (23)& $-$0.508& K16 &105549.670 & 123455.460 & $-$4.50 &    \\
\ion{P}{ii} & 6024.178 (5) & $+$0.198& K16 & 86743.960 & 103339.140 & $-$4.20 & core not fitted   \\
\ion{P}{ii} & 6034.039 (5) & $-$0.151& K16 & 86597.550 & 103165.610 & $-$4.20 & core not fitted   \\
\ion{P}{ii} & 6043.084 (5) & $+$0.442& K16 & 87124.600 & 103667.860 & $-$4.25 & core not fitted   \\
\ion{P}{ii} & 6087.837 (5) & $-$0.381& K16 & 86743.960 & 103165.610 & $-$4.30 &  \\
\ion{P}{ii} & 6165.598 (5) & $-$0.412& K16 & 87124.600 & 103339.140 & $-$4.30 &   \\
\ion{P}{iii} & 1334.813 (1)& $-$1.168& K16 & 0.000   & 74916.85 & $--$ & blend \\
\ion{P}{iii} & 1344.326 (1)& $-$0.931& K16 & 559.140 & 74945.86 & $--$ & blend \\
\ion{P}{iii} & 1344.850 (1)& $-$1.958& K16 & 559.140 & 74916.85 & $-$4.80 & from the red wing, blend \\
\ion{P}{iii} & 1379.912 (7)& $-$2.170& K16 & 74916.85 &147385.26 &$\le$ $-$5.54 & not obs \\
\ion{P}{iii} & 1380.464 (7)& $-$1.012& K16 & 74945.86 &147385.26 & $-$5.00  & blend  \\
\ion{P}{iii} & 1381.095 (7)& $-$1.225& K16 & 74916.86 &147323.19 &$\le$ $-$5.54  & blend \\
\ion{P}{iii} & 1381.648 (7)& $-$2.046& K16 & 74945.86 &147323.19 &$--$ & too blended \\
\ion{P}{iii} & 4059.340 (1)& $-$0.236& K16 & 116885.87 &141513.63 & $-$4.70 &  \\
\ion{P}{iii} & 4080.100 (1)& $-$0.494& K16 & 116874.56 &141376.91 & $-$4.70 &  \\
\ion{P}{iii} & 4222.232 (3)& $+$0.218& K16 & 117835.95 &141513.63 & $-$4.55 &  \\
\ion{P}{iii} & 4246.750 (3)& $-$0.086& K16 & 117835.95 &141376.91 & $-$4.80 & single \\
\ion{S}{i}& 1425.030 (5)& $-$0.094& NIST5 &    0.000 & 70173.968 & $-$6.26 & single\\
\ion{S}{i}& 1433.278 (5)& $-$0.365& NIST5 &  396.055 & 70166.195 &  $-$6.90 & \\
\ion{S}{i}& 1436.967 (5)& $-$0.716& NIST5 &  573.64 & 70164.658 & $>$ $--$ & $\lambda$?,bl unknown ?, bl \ion{Fe}{ii} \\
\ion{S}{i}& 1448.229 ( )& $-$0.326& NIST5 & 9238.069 & 78288.44 &$--$ & blend Fe II \\
\ion{S}{i}& 1472.971 (4)& $-$1.071& NIST5&  0.000 & 67890.016 & $--$   & not obs\\
\ion{S}{i}& 1473.995 (3)& $-$0.349& NIST5&   0.000& 67842.967& $--$   & bl \ion{Fe}{ii}\\
\ion{S}{i}& 1474.378 (3)& $-$1.10 & NIST5&   0.000& 67825.188& $--$   & bl \ion{Fe}{ii}\\
\ion{S}{i}& 1474.571 (3)& $-$2.278& NIST5&   0.000& 67816.351& $--$ & not obs\\
\ion{S}{i}& 1481.663 (4)& $-$1.536& K16&  396.055 & 67887.805 & $--$   & not obs\\
\ion{S}{i}& 1483.038 (3)& $-$0.624& NIST5&  396.055 & 67825.188 & $-$6.36 & weak, bl \ion{Fe}{iii}\\
\ion{S}{i}& 1483.233 (3)& $-$1.101& NIST5&  396.055 & 67816.315 & $--$   & bl unknown ?  \\
\ion{S}{i}& 1485.622 (4)& $-$2.638& NIST5& 573.64 & 67885.535 & $--$   & not obs\\
\ion{S}{i}& 1487.150 (3)& $-$0.979& NIST5&  573.640 & 67813.351 & $-$6.90   & bl \ion{Au}{iii}\\
\ion{S}{i}& 1666.687 (11)&$-$0.021& NIST5   & 9238.609 & 69237.886 & $--$ & bl \ion{Fe}{ii}\\
\ion{S}{i}& 1807.318 (2) & $-$0.319& NIST5&   0.000& 55330.811 &$>$ $-$6.26 ? & bl \ion{Mn}{ii}\\
\ion{S}{i}& 1820.343 (2) & $-$0.593& NIST5& 396.055& 55330.811 & $--$ &bl \ion{Mn}{ii}\\
\ion{S}{i}& 1826.245 (2) & $-$1.073& NIST5& 373.640& 55330.811 & $--$ & bl \ion{Fe}{iii}\\

\hline
\noalign{\smallskip}
\end{tabular}
\end{flushleft}
\end{table*}

\setcounter{table}{0}

\begin{table*}[!hbp]
\caption[ ]{Cont.}
\font\grande=cmr7
\grande
\begin{flushleft}
\begin{tabular}{lrlllrrlllllll}
\hline\noalign{\smallskip}
\multicolumn{1}{c}{Elem}&
\multicolumn{1}{c}{$\lambda$}&
\multicolumn{1}{c}{$\log\,gf$}&
\multicolumn{1}{c}{source$^{a}$}&
\multicolumn{1}{c}{$\chi_{low}$(cm$^{-1}$)}&
\multicolumn{1}{c}{$\chi_{up}$(cm$^{-1}$)}&
\multicolumn{2}{c}{abund}&
\multicolumn{1}{l}{Notes}\\
\hline\noalign{\smallskip}
\ion{S}{ii}& 1250.584 (1)&$-$1.618& NIST5   & 0.000 & 79962.61 & $-$6.36 & blend\\
\ion{S}{ii}& 1253.811 (1)&$-$1.315& NIST5   & 0.000 & 79756.83 & $>$ $-$6.26 & blend \ion{Mn}{ii}\\
\ion{S}{ii}& 1259.519 (1)&$-$1.138& NIST5   & 0.000 & 79395.39 & $-$6.26 & bl \ion{Fe}{ii}\\
\ion{S}{ii}& 1363.031 ( )&$-$2.638& NIST5   &24524.83 & 97890.74 &$<$$-$5.70  & blend \ion{Fe}{ii}\\
\ion{S}{ii}& 1363.376 ( )&$-$2.357& NIST5   &24571.54 & 97918.86 &$-$5.70  & blend \ion{Fe}{ii}\\
\ion{S}{ii}& 4162.665 (1)&$+$0.780& NIST5   & 128599.16 & 152615.46 & $-$6.36  & single\\
\ion{Cl}{i}& 1335.572 (2) &$-$0.755 & K16&  0.00 & 74865.667 & $--$ & bl \ion{C}{ii}\\
\ion{Cl}{i}& 1347.240 (2) &$-$0.341 & NIST5&  0.00 & 74225.846 & $-$6.85 & blend\\
\ion{Cl}{i}& 1351.656 (2) &$-$0.486 & K16& 882.352 & 74865.667 & $-$7.75 & blend\\
\ion{Cl}{i}& 1363.447 (2) &$-$0.835 & K16& 882.352 & 74225.846 & $-$7.75 ? & blend\\
\ion{Cl}{i}& 1379.528 (1) &$-$1.722 & K16& 0.00 & 72488.568 & $--$ & not obs\\
\ion{Cl}{i}& 1389.957 (1) &$-$3.010 & NIST5& 882.352 & 72827.038 & $--$ & not obs\\
\ion{Cl}{i}& 1396.527 (1) &$-$2.629 & NIST5& 882.352 & 72488.568 & $--$ & not obs\\
\ion{Ca}{ii}& 2112.757 (9) &$-$0.409 & K16& 25414.40 & 72730.93 & $-$ 5.80 & single \\
\ion{Ca}{ii}& 2113.146 (9) &$-$1.364 & K16& 25414.40 & 72722.23 & $-$ 5.80 & single\\
\ion{Ca}{ii}& 2131.505 (3) &$-$2.310 & NIST5& 13710.88 & 60611.28 & $-$ 5.68 & blend \\
\ion{Ca}{ii}& 2197.787 (8) &$-$1.303 & K16  & 25191.51 & 70677.62 & $-$ 5.60 & single \\
\ion{Ca}{ii}& 2208.611 (8) &$-$1.004 & K16  & 25414.40 & 70677.62 & $-$ 5.60 & single \\
\ion{Sc}{iii}& 1598.001 (1) & $-$1.239 & K16& 0.00 & 62578.18 & $\le$$-$10.2  \\
\ion{Sc}{iii}& 1603.064 (1)& $-$0.285 & K16& 197.64 & 62578.18 & $-$10.2 & single \\
\ion{Sc}{iii}& 1610.194 (1)& $-$0.543 & K16& 197.64 & 62104.30 & $-$9.6  \\
\ion{Sc}{iii}& 2699.066 ( )& $+$0.080 & K16& 25539.32 & 62578.18 & $-$10.2& blend  \\
\ion{Ti}{ii}& 1909.207 ( )& $-$0.321& K16 &  94.114 & 52471.895 & $-$6.46 & blend\\
\ion{Ti}{ii}& 1909.662 ( )& $-$0.132& K16 &  94.114 & 52459.395 & $-$6.46 & blend\\
\ion{Ti}{ii}& 1910.954 ( )& $-$0.407& NIST5 &  0.000 & 52329.889 & $-$6.46 & blend\\
\ion{Ti}{iii}& 1295.884 (1)& $-$0.439& K16 &   0.000 & 77167.43 & $-$6.46 & blend\\
\ion{Ti}{iii}& 1298.633 (1)& $-$0.906& K16 & 420.400 & 77424.45 & $-$6.46 & blend\\
\ion{Ti}{iii}& 1298.697 (1)& $-$0.271& K16 & 0.000 & 77000.23 & $-$6.46 & blend\\
\ion{Ti}{iii}& 1298.996 (1)& $-$0.210& NIST5  & 184.900 & 77167.43 & $-$6.46 & blend\\
\ion{Ti}{iii}& 1327.609 (4)& $-$0.538& K16 & 8473.500 & 83796.86 & $-$6.46 & blend Fe II\\
\ion{Ti}{iii}& 1455.195 (5)& $+$0.233& K16 &14397.600 & 83116.93 & $-$6.1  & blend\\
\ion{Ti}{iii}& 1498.695 (3)& $-$0.238& K16 & 8473.500 & 75198.21 & $-$6.0  & blend Fe II\\
\ion{Ti}{iii}& 2374.99 (10)& $-$0.003& K16 &41704.270 & 83796.86 & $-$6.46  & blend Fe II\\
\ion{Ti}{iii}& 2527.845 (7)& $+$0.138& K16 &38198.95  & 77746.94 & $-$6.0  & blend Mn II \\
\ion{Ti}{iii}& 2540.048 (7)& $-$0.065& K16 &38064.35  & 77421.86 & $-$6.26  & blend Fe II \\
\ion{Ti}{iii}& 2565.408 (6)& $-$0.146& K16 &38198.950 & 77167.43 & $-$6.0  & single \\
\ion{Ti}{iii}& 2567.556 (6)& $-$0.118& K16 &38064.350 & 77000.23 & $-$6.1   & $\lambda$=2567.54 ?,single \\
\ion{Ti}{iii}& 2576.463 (6)& $-$0.521& K16 &38198.950 & 77000.23 & $-$6.3   & blend \ion{Fe}{ii} \\
\ion{Ti}{iii}& 2984.744 (8)& $+$0.171& K16 &41704.270 & 75198.21 & $-$7.7   & blend \ion{Fe}{ii} \\
\ion{V}{ii}& 2908.817 (12)& $+$0.310&  K16 &3162.966 & 37531.132 & $-$9.14 & blend\\
\ion{V}{ii}& 2924.019 (10)& $+$0.420&  K16 &3162.966 & 37352.464 & $-$9.40 & blend\\
\ion{V}{ii}& 2924.641 (10)& $+$0.810&  K16 &2968.389 & 37150.615 & $-$9.14 & blend\\
\ion{Cr}{ii} & 2055.599 (1)  &   $-$0.186 &K16 &  0.000  & 48632.059 & $-$6.1& single   \\       
\ion{Cr}{ii} & 2061.577 (1) &   $-$0.312 & K16&  0.000  & 48491.057 & $-$6.1& bl \ion{Fe}{iii} \\       
\ion{Cr}{ii} & 2653.581 (8)&   $-$0.617 & K16 & 12032.545 & 49706.261 & $-$6.1& bl \ion{Fe}{ii},\ion{Mn}{ii} \\       
\ion{Cr}{ii} & 2666.014 (8) &   $-$0.106 & K16& 12147.771 & 49645.806 & $-$5.9 & blend & \\       
\ion{Cr}{ii} & 2858.910 (5)&   $-$0.224 & K16 & 12496.457 & 47464.557 & $-$5.9 & single & \\       
\ion{Cr}{ii} & 2860.931 (5)&   $-$0.449 & K16 & 11961.746 & 46905.137 & $-$5.7 & single & \\       
\ion{Cr}{ii} & 2862.569 (5)&   $-$0.070 & K16 & 12303.820 & 47227.291 & $-$5.8 & single & \\       
\ion{Cr}{ii} & 2971.901 (80)&   $+$0.523 & K16 & 30391.831 & 54030.505 & $-$5.7 & single & \\       
\ion{Cr}{ii} & 2989.190 (80)&   $+$0.237 & K16 & 30156.734 & 63600.861 & $-$5.8 & single & \\       
\hline
\noalign{\smallskip}
\end{tabular}
\end{flushleft}
\end{table*}

\setcounter{table}{0}

\begin{table*}[!hbp]
\caption[ ]{Cont.}
\font\grande=cmr7
\grande
\begin{flushleft}
\begin{tabular}{llllrrllllllll}
\hline\noalign{\smallskip}
\multicolumn{1}{c}{Elem}&
\multicolumn{1}{c}{$\lambda$}&
\multicolumn{1}{c}{$\log\,gf$}&
\multicolumn{1}{c}{source$^{a}$}&
\multicolumn{1}{c}{$\chi_{low}$(cm$^{-1}$)}&
\multicolumn{1}{c}{$\chi_{up}$(cm$^{-1}$)}&
\multicolumn{1}{c}{abund}&
\multicolumn{1}{l}{Notes}\\
\hline\noalign{\smallskip}
\ion{Cr}{iii} & 1261.865 (20) & $-$0.649 & K16 & 20851.87 & 100099.66 & $-$5.90 & \\
\ion{Cr}{iii} & 1263.611 (20) & $-$0.765 & K16 & 20702.45 & 99840.73 & $-$6.60 &\\
\ion{Cr}{iii} & 1268.021 (5) & $-$1.309 & K16 & 17850.13 & 96713.15 & $-$6.6 & single\\
\ion{Cr}{iii} & 1696.642 (71) & $+$0.835 & K16 & 94375.18 & 153315.13 &$\le$ $-$6.6 & not obs\\
\ion{Cr}{iii} & 1827.336 (46) & $-$0.846 & K16 & 56650.51 & 111374.97 & $-$6.3 & single\\
\ion{Cr}{iii} & 2203.228 (47) & $+$0.236 & K16 & 63420.87 & 108794.65 & $-$6.2 & single\\
\ion{Cr}{iii} & 2226.679 (39)& $+$0.631 & K16 & 50409.28 & 95305.25 & $-$5.9 & single\\
\ion{Cr}{iii} & 2277.483 (67) & $+$0.384 & K16 & 71676.22 & 115570.79 & $-$6.5 & single\\
\ion{Cr}{iii} & 2319.074 (44) & $+$0.414 & K16 & 56992.24 &100099.66 & $-$5.9 & single\\
\ion{Cr}{iii} & 2483.073 (43) & $+$0.192 & K16 & 57422.53 & 97683.060 & -6.0 & single\\
\ion{Cr}{iii} & 2531.023 (42) & $-$0.484 & K16 & 56650.51 & 96148.35 & $-$6.3  & single\\
\ion{Cr}{iii} & 2537.756 (42) & $-$0.509 & K16 & 56992.24 & 96385.29 & $-$6.1  & single\\
\ion{Cr}{iii} & 2544.373 (42) & $-$0.710 & K16 & 57422.53 & 96713.15 & $-$6.1  & single\\
\ion{Cr}{iii} & 2564.774 (39) & $-$0.401 & K16 & 74787.89 & 113766.00 & $-$6.2   & single\\
\ion{Cr}{iii} & 2640.737 (65) & $-$0.443 & K16 & 71676.22 & 109533.15 & $-$6.1   & single\\
\ion{Mn}{ii} & 1382.301 (14)& $-$1.568 & K16 & 14593.835 &  86936.98 &  $-$5.40  & hfs, single \\
\ion{Mn}{ii} & 1383.049 (14)& $-$1.842 & K16 & 14593.835 &  86897.932 & $-$5.55  & hfs, single \\
\ion{Mn}{ii} & 1385.890 (14)& $-$1.753 & K16 & 14781.205 &  86936.98 & $-$5.30  & hfs, blend \\
\ion{Mn}{ii} & 1853.266 (12)& $-$0.230 & K16 & 14325.866 &  68284.664 &  $-$5.80 & hfs, single & \\
\ion{Mn}{ii} & 1868.585 (  )& $-$1.492 & K16 & 14901.203 &  68417.697 &  $-$5.30 & hfs,bl \ion{Fe}{iii}& \\
\ion{Mn}{ii} & 1911.409 (10)& $-$0.407 & NIST5 & 14325.866   & 66643.296 & $-$5.30 & hfs, blend & \\
\ion{Mn}{ii} & 1923.339 (11)& $-$1.053 & NIST5  & 14901.203   & 66894.13  & $-$5.30 & hfs,single& \\
\ion{Mn}{ii} & 1925.511 (11)& $-$1.069 & NIST5  & 14959.876   & 66894.130 & $-$5.30 & hfs, bl \ion{Fe}{iii}&\\
\ion{Mn}{ii} & 1926.577 (10)& $-$0.700 & NIST5  & 14781.205   & 66686.739 & $-$5.30 & hfs, single  \\
\ion{Mn}{ii} & 1926.945 (10)& $-$1.324 & NIST5  & 14781.205   & 66676.833 & $-$5.40 & hfs, bl \ion{Fe}{iii}  \\
\ion{Mn}{ii} & 2535.977 (  )& $-$1.045 & NIST5  & 27588.534   & 67002.217 & $-$5.30 & hfs, single  \\
\ion{Mn}{ii} & 2576.104 (1)&   $+$0.399 &NIST5  &      0.000  & 38806.691 & $-$5.30 &hfs, bl \ion{Mn}{ii}, core &  \\       
\ion{Mn}{ii} & 2593.721 (1)&   $+$0.290 &NIST5 &      0.000  & 38543.122 & $-$5.30 &hfs, bl \ion{Fe}{ii},core& \\       
\ion{Mn}{ii} & 2605.680 (1)&   $+$0.137 &NIST5&      0.000  & 38366.232 & $-$5.30 &hfs, bl \ion{Mn}{ii}, core& \\
\ion{Mn}{ii} & 2701.698 (18)&  $+$0.608 & K16    &  27547.260  & 64550.040 & $-$5.30  &hfs,bl \ion{Fe}{ii},\ion{Cr}{ii},  core & \\
\ion{Mn}{ii} & 2705.730 (18)&  $+$0.479 & K16    &  27571.250  & 64518.890 & $-$5.30  &hfs,single,core & \\
\ion{Mn}{ii} & 2708.449 (18)&  $+$0.219 & K16    &  27583.590  & 64494.140 & $-$5.30  &hfs, bl \ion{Fe}{ii}, core & \\
\ion{Mn}{ii} & 2710.334 (18)&  $+$0.213 & K16    &  27588.534  & 64473.421 & $-$5.30  &hfs,single, core & \\
\ion{Mn}{ii} & 2711.562 (18)&  $-$0.525 & K16    &  27588.534  & 64456.720 & $-$5.30  &hfs,single, core & \\
\ion{Mn}{ii} & 2711.623 (18)&  $+$0.100 & K16    &  27589.360  & 64456.720 & $-$5.30  &hfs,single, core & \\
\ion{Mn}{ii} & 2933.054 (5)&   $-$0.102 & NIST5    &   9472.993  & 43557.175 & $-$5.30 & hfs,single, core & \\
\ion{Mn}{ii} & 2933.785 ( )&   $-$1.458 & NIST5    & 32818.440  & 66894.130 & $-$5.30 & hfs, single & \\
\ion{Mn}{ii} & 2935.362 ( )&   $-$2.002 & NIST5   &  32836.740  & 66894.130 & $-$5.30 & hfs, single & \\
\ion{Mn}{ii} & 2939.308 (5)&   $+$0.108 & NIST5    &   9472.993  & 43484.664 & $-$5.30 & hfs,single, core & \\
\ion{Mn}{ii} & 2949.204 (5)&   $+$0.253 & NIST5    &   9472.993  & 43370.527 & $-$5.30 & hfs, bl \ion{Fe}{ii},core &\\
\ion{Mn}{iii} & 1283.580 (9)&$-$0.251& K16& 43573.160  & 121480.240 & $-$5.9 & $\lambda$ 1283.564 ?  \\
\ion{Mn}{iii} & 1287.584 (9)&  $-$0.429 &K16 & 43602.500  & 121267.320 & $-$5.9 & blend  \\
\ion{Mn}{iii} & 1972.869 (12)& $-$0.603 &K16 &  62747.500  & 113676.530 & $-$5.90 & single \\
\ion{Mn}{iii} & 1975.451 (12)& $-$0.614 &K16 & 62456.990  & 113078.340 & $-$6.00  & blend  \\
\ion{Mn}{iii} & 1986.853 (12)& $-$0.807 &K16 & 62747.500  & 113078.340 & -5.55  & blend \ion{Fe}{iii} \\
\ion{Mn}{iii} & 2013.519 (17)& $-$0.552 &K16 & 71831.980  & 121480.240 & $-$6.2  & blend \ion{Mn}{iii} \\
\ion{Mn}{iii} & 2016.333 (17)& $-$0.522 &K16 & 71395.270  & 120974.250 & -6.4   & single \\
\ion{Mn}{iii} & 2018.455 (17)& $-$0.304 &K16 &  71564.210  & 121091.080 & -6.2  & single \\
\ion{Mn}{iii} & 2026.938 (11)& $-$0.465 &K16 & 62456.990  & 111776.600 & -5.55 & blend \\
\ion{Mn}{iii} & 2027.106 (11)& $-$0.736 &K16 & 62568.080  & 111883.600 & -5.55 & single,$\lambda$=2027.125 ? \\
\ion{Mn}{iii} & 2027.872 (17)& $+$0.345 &K16 & 71823.330  & 121480.240 & $-$6.10 & blend \\
\ion{Mn}{iii} & 2029.428 (17)& $-$0.317 &K16 & 71831.980  & 121091.080 & -6.30 & single  \\
\ion{Mn}{iii} & 2031.515 (11)& $-$0.260 &K16 & 62568.080  & 111776.600 & -5.55 &blend  \\
\ion{Mn}{iii} & 2036.670 (17)& $-$0.376 &K16 & 72183.330  & 121267.320 & -6.30  &single  \\
\ion{Mn}{iii} & 2049.682 (  )& $+$0.331 &K16 & 63285.370  & 112057.800 & -4.90 & single\\  
\ion{Mn}{iii} & 2077.369 (10)& $+$0.410 &K16 & 62988.920  & 111111.390 & -5.30 & bl \ion{Fe}{ii}\\  

\hline
\noalign{\smallskip}
\end{tabular}
\end{flushleft}
\end{table*}

\setcounter{table}{0}

\begin{table*}[!hbp]
\caption[ ]{Cont.}
\font\grande=cmr7
\grande
\begin{flushleft}
\begin{tabular}{llllrrllllllll}
\hline\noalign{\smallskip}
\multicolumn{1}{c}{Elem}&
\multicolumn{1}{c}{$\lambda$}&
\multicolumn{1}{c}{$\log\,gf$}&
\multicolumn{1}{c}{source$^{a}$}&
\multicolumn{1}{c}{$\chi_{low}$(cm$^{-1}$)}&
\multicolumn{1}{c}{$\chi_{up}$(cm$^{-1}$)}&
\multicolumn{1}{c}{abund}&
\multicolumn{1}{l}{Notes}\\
\hline\noalign{\smallskip}

\ion{Mn}{iii} & 2094.773 (10)& $-$0.026 &K16 & 62988.920  & 110711.620 & -5.30 & single\\  
\ion{Mn}{iii} & 2106.012 (  )& $-$0.830 &K16 & 62568.080  & 110036.140 & -5.80 & single\\  
\ion{Mn}{iii} & 2215.233 (16)& $+$0.153 &K16 & 71564.210  & 116692.150 & -6.00 & single\\  
\ion{Mn}{iii} & 2220.558 (  )& $+$0.330 &K16 & 71831.980  & 116851.690 & -5.60 & blend\\  
\ion{Mn}{iii} & 2220.743 (  )& $-$0.494 &K16 & 71564.210  & 116580.170 & -5.90 & single\\  
\ion{Mn}{iii} & 2354.663 (15)& $-$0.834 &K16 & 71564.210  & 116692.150 & -5.60 & single \\  
\ion{Mn}{iii} & 2374.314 (15)& $-$0.171 &K16 & 72183.330  & 114287.910 & -5.55 & single, $\lambda$=2374.33 ?\\  
\ion{Mn}{iii} & 2408.086 (14)& $-$0.551 &K16 & 71564.210  & 113078.340 & $-$5.90 & single\\  
\ion{Mn}{iii} & 2423.720 (14)& $-$0.323 &K16 & 71831.980  & 113078.340 & $-$5.40  & single, $\lambda$=2423.735 ?\\  
\ion{Fe}{i}   & 2966.898 (1) & $-$0.404 & NIST5    &  0.00    & 33695.397 & $-$3.74 \\
\ion{Fe}{i}   & 2973.132 (1) & $-$0.901 & NIST5    &  704.007    & 34328.752 & $-$3.74 \\
\ion{Fe}{i}   & 2973.235 (1) & $-$0.660 & NIST5    &  415.933    & 34039.516 & $-$3.74 \\
\ion{Fe}{ii}   & 2598.369 (1) & $-$0.062 &NIST5 & 384.787  &38858.97 & $-$3.74 &  SM\\
\ion{Fe}{ii}   & 2607.087 (1) & $-$0.152 &NIST5 & 667.683  &39013.216 & $-$3.74 &  SM\\
\ion{Fe}{ii}   & 2611.873 (1) & $-$0.009 &NIST5 & 384.787  &38660.054 & $-$3.74 &  SM\\
\ion{Fe}{ii}   & 2613.824 (1) & $-$0.362 &NIST5 & 862.612  &39109.316 & $-$3.74 &  SM\\
\ion{Fe}{ii}   & 2617.617 (1) & $-$0.522 &NIST5 & 867.683  &38858.97  & $-$3.64 &  SM\\
\ion{Fe}{ii}   & 2621.669 (1) & $-$0.938 &NIST5 & 977.050  &39109.316  & $-$3.64 &  SM\\
\ion{Fe}{ii}   & 2628.293 (1) & $-$0.441 &NIST5 & 977.050  &39013.216  & $-$3.64 &  SM\\
\ion{Fe}{ii}   & 2730.734 (62)& $-$0.900 &NIST5 & 8680.471  &45289.825  & $-$3.64 &  SM\\
\ion{Fe}{ii}   & 2743.197 (62)& $-$0.051 &NIST5 & 8846.784  &45289.825  & $-$3.65 & blend\\
\ion{Fe}{ii}   & 2755.736 (62)& $+$0.389 &NIST5 & 7955.319  &44232.540  & $-$3.64  & blend\\
\ion{Fe}{iii}   & 1468.982 ( )& $-$1.471 &K16 & 50275.840  &118350.20  & $-$3.00 & single ?\\
\ion{Fe}{iii}   & 1854.395 ( )& $-$0.564 &K16 & 83647.000  &137572.94  & $-$3.90  & single\\
\ion{Fe}{iii}   & 1882.369 ( )& $+$0.160 &K16 & 89783.550  &142908.100  & $-$3.90  & single\\
\ion{Fe}{iii}   & 1883.208 ( )& $-$0.913 &K16 & 87901.860  &141002.750  & $-$3.64  & single\\
\ion{Fe}{iii}   & 1904.265 ( )& $+$0.157 &K16 & 109570.800  &162084.49  & $-$3.90  & single\\
\ion{Fe}{iii}   & 1924.125 ( )& $+$0.098 &K16 & 97041.300  &149012.98  & $-$3.90  & single\\
\ion{Fe}{iii}   & 1945.344 ( )& $+$0.359 &K16 & 69836.890  &121241.69  & $-$3.80  & single\\
\ion{Fe}{iii}   & 1946.766 ( )& $+$0.345 &K16 & 114351.930  &165719.17  & $-$3.80  & single\\
\ion{Fe}{iii}   & 1951.331 ( )& $-$0.702 &K16 & 70694.170  &121941.23  & $-$3.90  & single\\
\ion{Fe}{iii}   & 1958.744 ( )& $-$0.030 &K16 & 89697.470  &140750.59  & $-$3.64  & single\\
\ion{Fe}{iii}   & 1961.014 ( )& $+$0.112 &K16 & 90472.370  &141466.39  & $-$3.80  & single\\
\ion{Fe}{iii}   & 1961.727 ( )& $-$0.471 &K16 & 90423.540  &141399.04  & $-$3.90  & single\\
\ion{Fe}{iii}   & 1994.087 ( )& $+$0.351 &K16 & 63487.080  &113635.34  & $-$3.90  & single\\
\ion{Fe}{iii}   & 1994.377 ( )& $-$1.669 &K16 & 63494.380  &113635.34  &  $-$3.40  & single\\
\ion{Fe}{iii}   & 2053.527 ( )& $-$0.768 &K16 & 76956.760  &125637.87  & $-$3.70  & single\\
\ion{Fe}{iii}   & 2057.936 ( )& $+$0.184 &K16 & 97041.300  &145681.13  & $-$3.70  & single\\
\ion{Fe}{iii}   & 2087.145 ( )& $+$0.189 &K16 & 76956.760  &124853.86  & $-$3.64  & single\\
\ion{Fe}{iii}   & 2103.818 ( )& $+$0.175 &K16 & 70728.930  &118246.49  & $-$3.8  & single\\
\ion{Fe}{iii}   & 2107.337 ( )& $+$0.149 &K16 & 70725.220  &118163.44  & $-$3.9  & single\\
\ion{Co}{ii}      & 2011.516 ( ) & $-$0.480 & NIST5 &   0.000 &49697.683 & $\le$ $-$10.12 & not obs\\
\ion{Co}{ii}      & 2022.354 ( ) & $-$0.490 & NIST5 & 950.324 &50381.724 & $\le$ $-$10.12 & not obs\\
\ion{Co}{ii}      & 2025.759 ( ) & $-$0.950 & NIST5 &   0.000 &49308.304 & $\le$ $-$10.12 & not obs\\
\ion{Co}{ii}      & 2286.159 (9) & $+$0.530 & NIST5 &3350.494 &47078.491 & $\le$ $-$10.12 & not obs\\
\ion{Co}{ii}      & 2307.860 (9) & $+$0.360 & NIST5 &4028.988 &47345.842 & $-$8.42 & blend\\
\ion{Co}{ii}      & 2324.321 (8) & $-$0.350 & NIST5 &4028.988 &47039.102 &$\le$ $-$10.12 & blend\\
\ion{Co}{ii}      & 2580.326 (14)& $+$0.360 & NIST5 &9812.859 &48556.049 & $-$10.12 & blend\\
\ion{Ni}{ii}      & 2165.550 (12)& $+$0.253 & K16 & 8393.900 & 54557.05 & $-$6.24 & \\
\ion{Ni}{ii}      & 2184.602 (  )& $-$0.031 & K16 & 10663.890 &56424.49 & $-$6.24 & \\
\ion{Ni}{ii}      & 2270.212 (  )& $+$0.120 & K16 & 9330.04 &53365.17 & $-$6.24 & blend \ion{Fe}{ii}\\
\ion{Ni}{ii}      & 2287.081 (  )& $+$0.020 & K16 & 14995.57 &58705.95 & $-$6.24 & blend\\
\ion{Ni}{ii}      & 2312.917 (  )& $+$0.521 & K16 & 32499.53 &75721.68 & $-$6.24 & single\\
\hline
\noalign{\smallskip}
\end{tabular}
\end{flushleft}
\end{table*}

\setcounter{table}{0}

\begin{table*}[!hbp]
\caption[ ]{Cont.}
\font\grande=cmr7
\grande
\begin{flushleft}
\begin{tabular}{llllrrllllllll}
\hline\noalign{\smallskip}
\multicolumn{1}{c}{Elem}&
\multicolumn{1}{c}{$\lambda$}&
\multicolumn{1}{c}{$\log\,gf$}&
\multicolumn{1}{c}{source$^{a}$}&
\multicolumn{1}{c}{$\chi_{low}$(cm$^{-1}$)}&
\multicolumn{1}{c}{$\chi_{up}$(cm$^{-1}$)}&
\multicolumn{1}{c}{abund}&
\multicolumn{1}{l}{Notes}\\
\hline\noalign{\smallskip}
\ion{Ni}{iii}     & 1794.896 (  )& $-$0.103 & K16 & 54657.83 &110371.35 & $-$6.44 & single\\
\ion{Ni}{iii}     & 1829.986 (  )& $+$0.207 & K16 & 61338.58 &116191.93 & $-$6.74 & single\\
\ion{Ni}{iii}     & 1830.060 (  )& $+$0.054 & K16 & 63471.93 &118114.95 & $-$6.74 & single\\
\ion{Cu}{ii}      & 1358.773 (3) & $-$0.174 & K16 &     0.00 &73595.813& $-$10.53 & single\\ 
\ion{Zn}{ii}      & 2064.227 (4) & $+$0.070 & NIST5 &48481.077 &96909.893& $-$8.84 & single\\ 
\ion{Ga}{ii}      & 1414.399 (2) & $+$0.248 & NIST5 &   0.00 & 70701.427 & $-$8.85 & blend\\  
\ion{Ga}{iii}     & 1495.045 (  )& $+$0.033 & NIST5 &   0.00 & 66887.630 & $-$8.16 & blend\\  
\ion{Ga}{iii}     & 1534.462 (  )& $-$0.281 & NIST5 &   0.00 & 65169.400 & $-$8.16 & blend\\  
\ion{Ge}{ii}      & 1261.905 (4)& $+$0.500 & NIST5 & 1767.357 &81012.598 &$\le$ $-$10.64 & not obs\\  
\ion{As}{ii}      & 1375.074 (4)& $+$0.000 & BMQ   & 10095.082& 82819.214 & $-$9.74 & blend\\
\ion{Y}{iii}      & 2414.643 (1)& $-$0.290 & Bie   &   0.00  & 41401.46 & $-$7.6&  single\\  
\ion{Y}{iii}      & 2945.995 (3)& $-$0.150 & Bie   & 7467.10 & 41401.46 & $-$7.6&  single\\  
\ion{Zr}{iii}     & 1941.053 (18)& $-$0.036 & NIST5 &8839.97  & 60358.40 &$\le$ $-$10.24& not obs\\  
\ion{Cd}{i}       & 2288.728 (  )& $+$0.100 & NIST5 &   0.00  & 43692.389 & $--$ & blend\\
\ion{Cd}{ii}      & 2144.393 (  )& $-$0.110 & NIST5  &   0.00  & 46618.532 & $-$7.00 & blend\\
\ion{Cd}{ii}      & 2265.018 (  )& $-$0.340 & NIST5  &   0.00  & 44136.080 & $-$7.00 & single\\
\ion{Cd}{ii}      & 2572.930 (  )& $-$0.470 & NIST5  &44136.08 & 82990.660 & $-$7.00 & blend\\
\ion{Cd}{ii}      & 2748.549 (  )& $-$0.200 & NIST5  &46618.55 & 82990.660 & $-$7.40 & single, $\lambda$=2748.556\,\AA\ \\
\ion{In}{ii}      & 1586.331 (  )& $+$0.160 & NIST5 &   0.000  & 63038.546 & $-$10.24& blend\\
\ion{Sn}{ii}      & 1290.875 (  )& $+$0.470 & NIST5 & 4251.494 & 81718.300 & $-$7.80 & single\\
\ion{Sn}{ii}      & 1316.581 (  )& $+$0.050 & NIST5 &   0.000 & 75954.300  & $-$8.50 & blend\\
\ion{Sn}{ii}      & 1400.440 (  )& $+$0.380 & NIST5 &   0.000  & 71406.142 & $-$8.70 & blend\\
\ion{Sn}{ii}      & 1474.997 (  )& $+$0.580 & NIST5 & 4251.494  & 72048.26 & $-$8.70 & blend\\
\ion{Sn}{ii}      & 1757.905 (  )& $-$0.550 & NIST5 &   0.000  & 56885.895 & $-$8.70 & single\\
\ion{Sn}{ii}      & 2151.514 (  )& $-$2.530 & NIST5 &   0.000  & 46464.29 & $-$7.00 & single\\
\ion{Xe}{i}       & 1295.588 (  )& $-$0.730 & NIST5 &   0.000  & 77185.041 & $-$6.00 & single\\ 
\ion{Xe}{i}       & 1469.612 (  )& $-$0.564 & NIST5 &   0.000  & 68045.156 & $-$5.10 & blend\\ 
\ion{Au}{ii}& 1469.142 ( ) & $-$1.470& Fiv& 17640.616& 85707.570 & $-$8.20  & blend \ion{Fe}{iii}\\
\ion{Au}{ii}& 1673.587 ( )& $-$0.130& Fiv& 15039.572& 74791.477 & $-$8.00  & blend \ion{Fe}{iii}\\
\ion{Au}{ii}& 1740.475 ( )& $+$0.230& Fiv& 15039.572& 72495.129 &  $-$7.70& single .\\
\ion{Au}{ii}& 1749.756 ( )& $-$0.330& Fiv& 17640.616& 74791.477 & $-$8.00& blend\\
\ion{Au}{ii}& 1756.098 ( )& $-$0.290& Fiv& 29621.249& 86565.667 & $-$8.20& blend Fe II\\
\ion{Au}{ii}& 1783.199 ( )& $+$0.100& Fiv& 29621.249& 85700.201 & $-$8.20 &\\
\ion{Au}{ii}& 1793.297 ( ) & $-$0.050& RW97&17640.616& 73403.839 & $-$8.20 &strong, blend \ion{Fe}{ii}\\
\ion{Au}{ii}& 1800.579 ( )& $-$0.060& Fiv &17640.616& 73178.291 & $-$6.30 & blended  unknown ?\\
\ion{Au}{ii}& 1823.220 ( )& $-$0.670& Fiv &27765.758& 82613.781 & $--$ & too blended Mn II\\
\ion{Au}{ii}& 1921.651 ( )& $-$0.620& Fiv &29621.249& 81659.828 & $-$8.20& blend\\
\ion{Au}{ii}& 2000.792 ( )& $-$0.350& Fiv &15039.572& 65003.594 & $-$7.50& single\\
\ion{Au}{ii}& 2082.074 (1)& $-$0.090& Fiv &15039.572& 63053.318 &  $-$7.50& single\\
\ion{Au}{iii}&1278.521 ( )& $-$0.610 &YBD & 40345.900& 118561.300& $<$$-$10.00 & very weak \\
\ion{Au}{iii}&1290.028 ( )& $-$0.940 &YBD & 44425.900& 121943.600& $-$8.50 & \\
\ion{Au}{iii}&1291.983 ( )& $-$0.610 &YBD & 44425.900& 121826.300& $\le$$-$10.0 & not obs \\
\ion{Au}{iii}&1314.833 ( )& $-$0.740 &YBD & 29754.00 & 105809.100& $\le$$-$10.0&  blend P II wrong\\
\ion{Au}{iii}&1336.707 ( )& $+$0.180 &YBD & 29754.00 & 104564.600& $--$ & blend \ion{Fe}{ii}\\
\ion{Au}{iii}&1348.879 ( )& $-$0.600 &YBD & 44425.900& 118561.300& $-$8.70 & weak single\\
\ion{Au}{iii}&1350.307 ( )& $-$0.560 &YBD & 38822.400& 112879.700& $\le$$-$10\\
\ion{Au}{iii}&1353.206 ( )& $-$0.690 &YBD & 44425.900& 118324.500& $-$8.50 & $\lambda$=1353.990 ?\\
\ion{Au}{iii}&1365.382 ( )& $+$0.500 &YBD & 29754.000& 102993.000& $-$9.00 & $\lambda$=1365.375 ?\\
\ion{Au}{iii}&1385.768 ( )& $+$0.190 &YBD & 38822.400& 110984.600& $-$9.00 & $\lambda$=1385.78 ?\\
\ion{Au}{iii}&1487.130 ( )& $-$0.140 &YBD & 35076.900& 102320.500& $-$9.30 & \\
\ion{Au}{iii}&1746.057 ( )& $-$0.220 &YBD & 38822.400&  96097.700& $-$8.60 & weak single\\
\ion{Hg}{ii} &1649.937 (1)& $+$0.290& NIST5& 0.00& 60608.362 & $-$9.60  &\\
\ion{Hg}{ii} &1942.273 (1)& $-$0.418& Proff& 0.00& 51486.070 & $-$9.60  & hfs\\
\hline
\noalign{\smallskip}
\end{tabular}
\end{flushleft}
$^{a}$ NIST5= NIST database: http://www.nist.gov/PHYSRefData/ASD/lines-form.html\\
K16= Kurucz (2016): http://kurucz.harvard.edu/linelists/gfnew\\
BMQ=Biemont et al. (1998); Bie=Biemont et al. (2011); Fiv=Fivet wt al. (2006);\\
Hi= Hibbert (1988); RW97=Rosberg et al. (1997); YBD=Enzonga Yoca et al. (2008);\\
Proff=Proffitt et al. (1999).\\
\end{table*}

\end{appendix}


\begin{thebibliography}{}


\bibitem[Alecian \& Artru(1988)]{1988IAUS..132..235A} Alecian, G., \& Artru, M.~C.\ 1988, The Impact of Very High S/N Spectroscopy on Stellar Physics, 132, 235 

\bibitem[Alecian \& Stift(2010)]{2010A&A...516A..53A} Alecian, G., \& Stift, M.~J.\ 2010, \aap, 516, A53 

\bibitem[Alonso-Medina et al.(2005)]{2005PhyS...71..154A} Alonso-Medina, A., Col{\'o}n, C., \& Rivero, C.\ 2005, \physscr, 71, 154 




 \bibitem[Andersen \& Jaschek(1984)]{1984A&AS...55..469A} Andersen, J., \& Jaschek, M.\ 1984, \aaps, 55, 469 

 \bibitem[Andersen et al.(1984)]{1984A&A...132..354A} Andersen, J., Jaschek, M., \& Cowley, C.~R.\ 1984, \aap, 132, 354 

\bibitem[Ansbacher et al.(1986)]{1986CaJPh..64.1365A} Ansbacher, W., Pinnington, E.~H., Kernahan, J.~A., \& Gosselin, R.~N.\ 1986, Canadian Journal of Physics, 64, 1365 




   
\bibitem[Asplund et al.(2009)]{2009ARA&A..47..481A} Asplund, M., Grevesse, N., Sauval, A.~J., \& Scott, P.\ 2009, \araa, 47, 481 


\bibitem[Ayres(2010)]{2010ApJS..187..149A} Ayres, T.~R.\ 2010, \apjs, 187, 149 

\bibitem[Ayres \& The ASTRAL I  Science Teams(2014)]{2014AAS...22325437A} Ayres, T.~R., \& The ASTRAL I Science Teams, I.\ 2014, American Astronomical Society Meeting Abstracts \#223, 223, 254.37 



\bibitem[Babel(1994)]{1994IAUS..162..333B} Babel, J.\ 1994, Pulsation; Rotation; and Mass Loss in Early-Type Stars, 162, 333 

\bibitem[Bailey \& Landstreet(2013)]{2013A&A...551A..30B} Bailey, J.~D., \& Landstreet, J.~D.\ 2013, \aap, 551, A30 
  
 \bibitem[Bessell \& Eggen(1972)]{1972ApJ...177..209B} Bessell, M.~S., \& Eggen, O.~J.\ 1972, \apj, 177, 209 

\bibitem[Biemont et al.(1998)]{1998MNRAS.297..713B} Bi{\'e}mont, E., Morton, D.~C., \& Quinet, P.\ 1998, \mnras, 297, 713 

\bibitem[Bi{\'e}mont et al.(2007)]{2007MNRAS.380.1581B} Bi{\'e}mont, {\'E}., Blagoev, K., Fivet, V., et al.\ 2007, \mnras, 380, 1581 

\bibitem[Bi{\'e}mont et al.(2011)]{2011MNRAS.414.3350B} Bi{\'e}mont, {\'E}., Blagoev, K., Engstr{\"o}m, L., et al.\ 2011, \mnras, 414, 3350 







\bibitem[Castelli(2005)]{2005MSAIS...8...44C} Castelli, F.\ 2005, Memorie della Societa Astronomica Italiana Supplementi, 8, 44 




 \bibitem[Castelli et al.(1981)]{1981LIACo..23..149C} Castelli, F., Cornachin, M., \& Hack, M.\ 1981, Liege International Astrophysical Colloquia, 23, 149 


 \bibitem[Castelli et al.(1985)]{1985A&AS...59....1C} Castelli, F., Cornachin, M., Morossi, C., \& Hack, M.\ 1985, \aaps, 59, 1 





\bibitem[Castelli \& Hubrig(2007)]{2007A&A...475.1041C} Castelli, F., \& Hubrig, S.\ 2007, \aap, 475, 1041 

\bibitem[Castelli \& Kurucz(2010)]{2010A&A...520A..57C} Castelli, F., \& Kurucz, R.~L.\ 2010, \aap, 520, A57 


\bibitem[Castelli et al.(2009)]{2009A&A...508..401C} Castelli, F., Kurucz, R., \& Hubrig, S.\ 2009, \aap, 508, 401 

\bibitem[Castelli et al.(2015)]{2015A&A...580A..10C} Castelli, F., Kurucz, R.~L., \& Cowley, C.~R.\ 2015, \aap, 580, A10 


\bibitem[Catanzaro et al.(2016)]{2016MNRAS.460.1999C} Catanzaro, G., Giarrusso, M., Leone, F., et al.\ 2016, \mnras, 460, 1999 

\bibitem[Catanzaro et al.(2004)]{2004A&A...425..641C} Catanzaro, G., Leone, F., \& Dall, T.~H.\ 2004, \aap, 425, 641

\bibitem[Cugier \& Hardorp(1988)]{1988A&A...197..163C} Cugier, H., \& Hardorp, J.\ 1988, \aap, 197, 163 




  
\bibitem[Curtis et al.(2000)]{2000PhRvA..62e2513C} Curtis, L.~J., Matulioniene, R., Ellis, D.~G., \& Fischer, C.~F.\ 2000, \pra, 62, 052513 



  
\bibitem[Dunlop(1829)]{1829Mem. RAS...3...257} Dunlop, J.\ 1829, Mem. RAS, 3, 257


\bibitem[Eggen(1975)]{1975PASP...87...37E} Eggen, O.~J.\ 1975, \pasp, 87, 37 

\bibitem[Enzonga Yoca et al.(2008)]{2008PhyS...78b5303E} Enzonga Yoca, S., Bi{\'e}mont, {\'E}., Delahaye, F., Quinet, P., \& Zeippen, C.~J.\ 2008, \physscr, 78, 025303 





\bibitem[Fivet et al.(2006)]{2006JPhB...39.3587F} Fivet, V., Quinet, P., Bi{\'e}mont, {\'E}., \& Xu, H.~L.\ 2006, Journal of Physics B Atomic Molecular Physics, 39, 3587 

\bibitem[Fuhr \& Wiese (2005)]{Fuhr2005} Fuhr, J.R. \& Wiese, W.L. 1998, CRC Handbook of Chemistry and Physics, 10-88-10-146 (ed. by D.R. Lide, CRC Press, Boca Raton, Fl)  

\bibitem[Gavrila(1967)]{1967PhRv..163..147G} Gavrila, M.\ 1967, Physical Review, 163, 147 



  

\bibitem[Ghazaryan \& Alecian(2016)]{2016MNRAS.460.1912G} Ghazaryan, S., \& Alecian, G.\ 2016, \mnras, 460, 1912 




\bibitem[Grevesse et al.(2015)]{2015A&A...573A..27G} Grevesse, N., Scott, P., Asplund, M., \& Sauval, A.~J.\ 2015, \aap, 573, A27 

\bibitem[Haris et al.(2014)]{2014PhyS...89k5403H} Haris, K., Kramida, A., \& Tauheed, A.\ 2014, \physscr, 89, 115403 


\bibitem[Hempel \& Holweger(2003)]{2003A&A...408.1065H} Hempel, M., \& Holweger, H.\ 2003, \aap, 408, 1065 

 
\bibitem[Hibbert(1988)]{1988PhyS...38...37H} Hibbert, A.\ 1988, \physscr, 38, 37 


  
\bibitem[Holt et al.(1999)]{1999MNRAS.306..107H} Holt, R.~A., Scholl, T.~J., \& Rosner, S.~D.\ 1999, \mnras, 306, 107 





\bibitem[Hubrig et al.(1999)]{1999A&A...341..190H} Hubrig, S., Castelli, F., \& Mathys, G.\ 1999, \aap, 341, 190 





\bibitem[James et al.(2006)]{2006A&A...446..971J} James, D.~J., Melo, C., Santos, N.~C., \& Bouvier, J.\ 2006, \aap, 446, 971 

\bibitem[J{\"o}nsson \& Andersson(2007)]{2007JPhB...40.2417J} J{\"o}nsson, P., \& Andersson, M.\ 2007, Journal of Physics B Atomic Molecular Physics, 40, 2417 




  
\bibitem[Khalack et al.(2007)]{2007A&A...466..667K} Khalack, V.~R., Leblanc, F., Bohlender, D., Wade, G.~A., \& Behr, B.~B.\ 2007, \aap, 466, 667 

\bibitem[Khalack et al.(2008)]{2008A&A...477..641K} Khalack, V.~R., Leblanc, F., Behr, B.~B., Wade, G.~A., \& Bohlender, D.\ 2008, \aap, 477, 641 

\bibitem[Kramida et al.(2015)]{ } Kramida, A.,
Ralchenko, Yu., Reader, J., \& NIST ADS Team. 2015, NIST Atomic Spectra Database,
(version 5.3), National Institute of Standards and Technology, Gaithersburg, MD 

\bibitem[Kurtz \& Marang(1995)]{1995MNRAS.276..191K} Kurtz, D.~W., \& Marang, F.\ 1995, \mnras, 276, 191 

\bibitem[Kurucz(1970)]{1970SAOSR.309.....K} Kurucz, R.~L.\ 1970, SAO Special Report, 309, 309 


																	

\bibitem[Kurucz(2005)]{2005MSAIS...8...14K} Kurucz, R.~L.\ 2005, Memorie della Societa Astronomica Italiana Supplementi, 8, 14
  

\bibitem[Kurucz(2016)]{2016 ASOS} Kurucz, R.~L.\ 2016, 21 ASOS Colloquium, San Paulo 

\bibitem[LeBlanc et al.(2015)]{2015MNRAS.453.3766L} LeBlanc, F., Khalack, V., Yameogo, B., Thibeault, C., \& Gallant, I.\ 2015, \mnras, 453, 3766 


\bibitem[Leckrone et al.(1999)]{1999AJ....117.1454L} Leckrone, D.~S., Proffitt, C.~R., Wahlgren, G.~M., Johansson, S.~G., \& Brage, T.\ 1999, \aj, 117, 1454 

\bibitem[Leone(1998)]{1998CoSka..27..285L} Leone, F.\ 1998, Contributions of the Astronomical Observatory Skalnate Pleso, 27, 285 

\bibitem[Leone et al.(1997)]{1997A&A...325.1125L} Leone, F., Catalano, F.~A., \& Malaroda, S.\ 1997, \aap, 325, 1125 

\bibitem[Leone \& Lanzafame(1997)]{1997A&A...320..893L} Leone, F., \& Lanzafame, A.~C.\ 1997, \aap, 320, 893 






\bibitem[Ljung et al.(2006)]{2006A&A...456.1181L} Ljung, G., Nilsson, H., Asplund, M., \& Johansson, S.\ 2006, \aap, 456, 1181 


\bibitem[Michaud(1970)]{1970ApJ...160..641M} Michaud, G.\ 1970, \apj, 160, 641 





\bibitem[Morton(2000)]{2000ApJS..130..403M} Morton, D.~C.\ 2000, \apjs, 130, 403 


\bibitem[Nesvacil et al.(2013)]{2013A&A...552A..28N} Nesvacil, N., Shulyak, D., Ryabchikova, T.~A., et al.\ 2013, \aap, 552, A28 




\bibitem[Nielsen et al.(2005)]{2005AJ....130.2312N} Nielsen, K.~E., Wahlgren, G.~M., Proffitt, C.~R., Leckrone, D.~S., \& Adelman, S.~J.\ 2005, \aj, 130, 2312 




\bibitem[Oliver \& Hibbert(2010)]{2010JPhB...43g4013O} Oliver, P., \& Hibbert, A.\ 2010, Journal of Physics B Atomic Molecular Physics, 43, 074013 



\bibitem[Peterson \& Kurucz(2015)]{2015ApJS..216....1P} Peterson, R.~C., \& Kurucz, R.~L.\ 2015, \apjs, 216, 1 

\bibitem[Proffitt et al.(1999)]{1999ApJ...512..942P} Proffitt, C.~R., Brage, T., Leckrone, D.~S., et al.\ 1999, \apj, 512, 942 


\bibitem[Raassen \& Uylings(1998)]{1998A&A...340..300R} Raassen, A.~J.~J., \& Uylings, P.~H.~M.\ 1998, \aap, 340, 300 


\bibitem[Reader \& Acquista(1997)]{1997PhyS...55..310R} Reader, J., \& Acquista, N.\ 1997, \physscr, 55, 310 



  

\bibitem[Rosberg \& Wyart(1997)]{1997PhyS...55..690R} Rosberg, M., \& Wyart, J.-F.\ 1997, \physscr, 55, 690 








\bibitem[Ryabchikova \& Smirnov(1988)]{1988ATsir1534...21R} Ryabchikova, T.~A., \& Smirnov, J.~M.\ 1988, Astronomicheskij Tsirkulyar, 1534, 21 




\bibitem[Ryabchikova \& Smirnov(1994)]{1994ARep...38...70R} Ryabchikova, T.~A., \& Smirnov, Y.~M.\ 1994, Astronomy Reports, 38, 70 

\bibitem[Ryabchikova et al.(2003)]{2003IAUS..210..301R} Ryabchikova, T., Wade, G.~A., \& LeBlanc, F.\ 2003, Modelling of Stellar Atmospheres, 210, 301 

\bibitem[Sansonetti \& Reader(2001)]{2001PhyS...63..219S} Sansonetti, C.~J., \& Reader, J.\ 2001, \physscr, 63, 219 

\bibitem[Scott et al.(2015)]{2015A&A...573A..25S} Scott, P., Grevesse, N., Asplund, M., et al.\ 2015a, \aap, 573, A25 

\bibitem[Scott et al.(2015)]{2015A&A...573A..26S} Scott, P., Asplund, M., Grevesse, N., Bergemann, M., \& Sauval, A.~J.\ 2015b, \aap, 573, A26 


\bibitem[Shirai et al.(2007)]{2007JPCRD..36..509S} Shirai, T., Reader, J., Kramida, A.~E., \& Sugar, J.\ 2007, Journal of Physical and Chemical Reference Data, 36, 509 




\bibitem[Sigut(2001)]{2001ApJ...546L.115S} Sigut, T.~A.~A.\ 2001, \apjl, 546, L115 






\bibitem[Smith(1993)]{1993A&A...276..393S} Smith, K.~C.\ 1993, \aap, 276, 393 

\bibitem[Smith(1994)]{1994A&A...291..521S} Smith, K.~C.\ 1994, \aap, 291, 521 

\bibitem[Smith(1997)]{1997A&A...319..928S} Smith, K.~C.\ 1997, \aap, 319, 928 

\bibitem[Smith \& Dworetsky(1993)]{1993A&A...274..335S} Smith, K.~C., \& Dworetsky, M.~M.\ 1993, \aap, 274, 335 


\bibitem[Stift \& Alecian(2012)]{2012MNRAS.425.2715S} Stift, M.~J., \& Alecian, G.\ 2012, \mnras, 425, 2715 




\bibitem[Thiam et al.(2010)]{2010MNRAS.405.1384T} Thiam, M., LeBlanc, F., Khalack, V., \& Wade, G.~A.\ 2010, \mnras, 405, 1384 

\bibitem[Townley-Smith et al.(2016)]{2016MNRAS.461...73T} Townley-Smith, K., Nave, G., Pickering, J.~C., \& Blackwell-Whitehead, R.~J.\ 2016, \mnras, 461, 73 


\bibitem[Uylings \& Raassen(1997)]{1997A&AS..125..539U} Uylings, P.~H.~M., \& Raassen, A.~J.~J.\ 1997, \aaps, 125,  

\bibitem[van den Ancker et al.(1996)]{1996A&A...313..517V} van den Ancker, M.~E., de Winter, D., \& The, P.~S.\ 1996, \aap, 313, 517 

\bibitem[Wiese \& Martin(1980)]{1980wtpa.book.....W} Wiese, W.~L., \& Martin, G.~A.\ 1980, Wavelengths and transition probabilities for atoms and atomic ions: Part 2.~Transition probabilities, NSRDS-NBS Vol.~68., 68,  




\bibitem[Wyart et al.(1996)]{1996PhyS...53..174W} Wyart, J.-F., Joshi, Y.~N., Tchang-Brillet, L., \& Raassen, A.~J.~J.\ 1996, \physscr, 53, 174 

\bibitem[Y{\"u}ce et al.(2011)]{2011A&A...528A..37Y} Y{\"u}ce, K., Castelli, F., \& Hubrig, S.\ 2011, \aap, 528, A37 







\end{thebibliography}
\end{document}